\documentclass[11pt]{scrartcl}

\usepackage{ifthen}
\usepackage[symbol]{footmisc}

\newcounter{DisplayAuthors}
\newcommand{\twooptions}[2]{\ifthenelse{\value{DisplayAuthors}=0}{#1}{#2}}

\title{A Dichotomy for Terminal Cyclability{\twooptions{}{\footnotemark}}}

\author{%
 \twooptions{\textcolor{BostonUniversityRed}{An onymous}}{%
 \parbox{\textwidth}{\centering
 {Matthias Bentert\footnotemark \qquad
     Dario Cavallaro\footnotemark[2] \qquad
 Sebastian Wiederrecht\footnotemark}
 }%
}%
}

\date{}

\usepackage[utf8]{inputenc}
\usepackage[T1]{fontenc}
\usepackage{lmodern}
\usepackage{alphabeta}

\usepackage{titling}
\usepackage{xspace}
\usepackage{soul} 

\usepackage[letterpaper,top=1in,bottom=1.5in,left=1in,right=1in]{geometry}

\DeclareSectionCommand[%
level=4,
indent=0pt,
beforeskip=1ex plus 1ex minus .2ex,
afterskip=-1em,
font={},
tocindent=7em,
tocnumwidth=4.1em,
counterwithin=subsubsection
]{paragraph}

\usepackage[inline]{enumitem}

\usepackage{dsfont}
\usepackage{setspace}

\usepackage{bm}
\usepackage{microtype}
\usepackage{amsmath}
\usepackage{amssymb}
\usepackage{amsfonts}
\usepackage{mathtools}
\usepackage[mathscr]{euscript}

\usepackage[hyphens]{url}
\usepackage{amsthm}
\usepackage{stackengine}

\usepackage{tikz}
\usepackage{xcolor}
\usepackage{subcaption}
\usepackage{graphicx}
\usepackage{wrapfig}

\usepackage{hyperref}
\usepackage{zref-clever}
\usepackage{thm-restate}

\usepackage{nicefrac}
\usepackage[textwidth=1.5in,disable]{todonotes}

\usepackage{ifthen}

\newcommand{\cref}[1]{\zcref{#1}}
\renewcommand{\autoref}[1]{\zcref{#1}}

\colorlet{myGreen}{green!50!black}
\colorlet{myLightgreen}{green}
\colorlet{myRed}{red!90!black}
\definecolor{myBlue}{rgb}{0.25, 0.0, 1.0}
\definecolor{myLightBlue}{rgb}{0.39, 0.58, 0.93}
\colorlet{myViolet}{myBlue!55!myRed}
\definecolor{myOrange}{rgb}{1.0, 0.66, 0.07}

\definecolor{CornflowerBlue}{rgb}{0.39, 0.58, 0.93}
\definecolor{DarkGoldenrod}{rgb}{0.72, 0.53, 0.04}
\definecolor{BritishRacingGreen}{rgb}{0.0, 0.26, 0.15}
\definecolor{DarkMagenta}{rgb}{0.55, 0.0, 0.55}
\definecolor{AO}{rgb}{0.0, 0.5, 0.0}
\definecolor{BostonUniversityRed}{rgb}{0.8, 0.0, 0.0}
\definecolor{myRed}{rgb}{0.8, 0.0, 0.0}
\definecolor{DarkMidnightBlue}{rgb}{0.0, 0.2, 0.4}
\definecolor{DarkTangerine}{rgb}{1.0, 0.66, 0.07}
\definecolor{AppleGreen}{rgb}{0.55, 0.71, 0.0}
\definecolor{BrightUbe}{rgb}{0.82, 0.62, 0.91}
\definecolor{Amethyst}{rgb}{0.6, 0.4, 0.8}
\definecolor{DarkGray}{rgb}{0.52, 0.52, 0.51}
\definecolor{Gray}{rgb}{0.66, 0.66, 0.66}
\definecolor{BananaYellow}{rgb}{1.0, 0.88, 0.21}
\definecolor{Amber}{rgb}{1.0, 0.75, 0.0}
\definecolor{LightGray}{rgb}{0.83, 0.83, 0.83}
\definecolor{PrincetonOrange}{rgb}{1.0, 0.56, 0.0}
\definecolor{DeepCarrotOrange}{rgb}{0.91, 0.41, 0.17}
\definecolor{CarrotOrange}{rgb}{0.93, 0.57, 0.13}
\definecolor{MidnightBlue}{rgb}{0.1, 0.1, 0.44}
\definecolor{Magenta}{rgb}{0.50, 0.0, 0.50}
\definecolor{BrightPink}{rgb}{1.0, 0.0, 0.5}
\definecolor{BrilliantRose}{rgb}{1.0, 0.33, 0.64}
\definecolor{ChromeYellow}{rgb}{1.0, 0.65, 0.0}
\definecolor{HotMagenta}{rgb}{1.0, 0.11, 0.81}
\definecolor{Auburn}{rgb}{0.43, 0.21, 0.1}
\definecolor{BrightTurquoise}{rgb}{0.03, 0.91, 0.87}
\definecolor{DarkCyan}{rgb}{0.0, 0.55, 0.55}
\definecolor{DarkGray}{rgb}{0.66, 0.66, 0.66}
\definecolor{DimGray}{rgb}{0.41, 0.41, 0.41}
\definecolor{DarkBananaYellow}{RGB}{240,181,67}
\definecolor{LavenderFloral}{rgb}{0.71, 0.49, 0.86}
\definecolor{Persimmon}{rgb}{0.93, 0.35, 0.0}
\definecolor{LavenderMagenta}{rgb}{0.93, 0.51, 0.93}

\vfuzz \hfuzz
\setlist[itemize]{topsep=0pt,partopsep=0pt,itemsep=0pt,parsep=0pt}
\setlist[itemize,1]{label={\small\textbullet}}
\setlist[itemize,2]{label={\tiny\textbullet}}
\setlist[itemize,3]{label=$\cdot$}
\setlist[enumerate]{topsep=0pt,partopsep=0pt,itemsep=0pt,parsep=0pt}
\setlist[enumerate,1]{label=\roman*)}
\setlist[enumerate,2]{label=\alph*)}
\setlist[enumerate,3]{label=\arabic*)}

\hypersetup{
colorlinks=true,
linkcolor=AO!65!black,
citecolor=AO!65!black,
urlcolor=AppleGreen!65!black,
bookmarksopen=true,
bookmarksnumbered,
bookmarksopenlevel=2,
bookmarksdepth=3,
}

\newenvironment{claimproof}[1][Proof of claim]
  {\begin{proof}[#1]}
  {\end{proof}}

\theoremstyle{definition}

\newtheorem{environment}{Environment}[section]

\newtheorem{lemma}[environment]{Lemma}
\AddToHook{env/lemma/begin}{\zcsetup{countertype={environment=lemma}}}
\zcRefTypeSetup{lemma}{
Name-sg = Lemma ,
name-sg = Lemma ,
Name-pl = Lemmas ,
name-pl = Lemmas ,
}

\newtheorem*{lemma*}{Lemma}
\AddToHook{env/lemma*/begin}{\zcsetup{countertype={environment=lemma*}}}
\zcRefTypeSetup{lemma*}{
Name-sg = Lemma ,
name-sg = Lemma ,
Name-pl = Lemmas ,
name-pl = Lemmas ,
}

\newtheorem{proposition}[environment]{Proposition}
\AddToHook{env/proposition/begin}{\zcsetup{countertype={environment=proposition}}}
\zcRefTypeSetup{proposition}{
Name-sg = Proposition ,
name-sg = Proposition ,
Name-pl = Propositions ,
name-pl = Propositions ,
}

\AddToHook{env/corollary/begin}{\zcsetup{countertype={environment=corollary}}}
\zcRefTypeSetup{corollary}{
Name-sg = Corollary ,
name-sg = Corollary ,
Name-pl = Corollaries ,
name-pl = Corollaries ,
}

\newtheorem{theorem}[environment]{Theorem}
\AddToHook{env/theorem/begin}{\zcsetup{countertype={environment=theorem}}}
\zcRefTypeSetup{theorem}{
Name-sg = Theorem ,
name-sg = Theorem ,
Name-pl = Theorems ,
name-pl = Theorems ,
}

\newtheorem*{theorem*}{Theorem}
\AddToHook{env/theorem*/begin}{\zcsetup{countertype={environment=theorem*}}}
\zcRefTypeSetup{theorem*}{
Name-sg = Theorem ,
name-sg = Theorem ,
Name-pl = Theorems ,
name-pl = Theorems ,
}

\AddToHook{env/conjecture/begin}{\zcsetup{countertype={environment=conjecture}}}
\zcRefTypeSetup{conjecture}{
Name-sg = Conjecture ,
name-sg = Conjecture ,
Name-pl = Conjectures ,
name-pl = Conjectures ,
}

\newtheorem*{hypothesis*}{Hypothesis}
\AddToHook{env/hypothesis*/begin}{\zcsetup{countertype={environment=hypothesis*}}}
\zcRefTypeSetup{hypothesis*}{
Name-sg = Hypothesis ,
name-sg = Hypothesis ,
Name-pl = Hypotheses ,
name-pl = Hypotheses ,
}

\newtheorem{observation}[environment]{Observation}
\AddToHook{env/observation/begin}{\zcsetup{countertype={environment=observation}}}
\zcRefTypeSetup{observation}{
Name-sg = Observation ,
name-sg = Observation ,
Name-pl = Observations ,
name-pl = Observations ,
}

\AddToHook{env/example/begin}{\zcsetup{countertype={environment=example}}}
\zcRefTypeSetup{example}{
Name-sg = Example ,
name-sg = Example ,
Name-pl = Examples ,
name-pl = Examples ,
}

\AddToHook{env/remark/begin}{\zcsetup{countertype={environment=remark}}}
\zcRefTypeSetup{remark}{
Name-sg = Remark ,
name-sg = Remark ,
Name-pl = Remarks ,
name-pl = Remarks ,
}

\zcRefTypeSetup{equation}{
Name-sg = Equation ,
name-sg = Equation ,
Name-pl = Equations ,
name-pl = Equations ,
}

\zcRefTypeSetup{chapter}{
Name-sg = Chapter ,
name-sg = Chapter ,
Name-pl = Chapters ,
name-pl = Chapters ,
}

\zcRefTypeSetup{section}{
Name-sg = Section ,
name-sg = Section ,
Name-pl = Sections ,
name-pl = Sections ,
}

\zcRefTypeSetup{algorithm}{
Name-sg = Algorithm ,
name-sg = Algorithm ,
Name-pl = Algorithms ,
name-pl = Algorithms ,
}

\AddToHook{env/notation/begin}{\zcsetup{countertype={environment=notation}}}
\zcRefTypeSetup{notation}{
Name-sg = Notation ,
name-sg = Notation ,
Name-pl = Notations ,
name-pl = Notations ,
}

\AddToHook{env/question/begin}{\zcsetup{countertype={environment=question}}}
\zcRefTypeSetup{question}{
Name-sg = Question ,
name-sg = Question ,
Name-pl = Questions ,
name-pl = Questions ,
}

\AddToHook{env/problem/begin}{\zcsetup{countertype={environment=problem}}}
\zcRefTypeSetup{problem}{
Name-sg = Problem ,
name-sg = Problem ,
Name-pl = Problems ,
name-pl = Problems ,
}

\AddToHook{env/definition/begin}{\zcsetup{countertype={environment=definition}}}
\zcRefTypeSetup{definition}{
Name-sg = Definition ,
name-sg = Definition ,
Name-pl = Definitions ,
name-pl = Definitions ,
}

\zcRefTypeSetup{figure}{
Name-sg = Figure ,
name-sg = Figure ,
Name-pl = Figures ,
name-pl = Figures ,
}

\newtheoremstyle{beautypalour}
{3pt}
{3pt}
{}
{}
{\itshape}
{.}
{.5em}
{}
\theoremstyle{beautypalour}

\newtheorem{claim}{Claim}[environment]
\AddToHook{env/claim/begin}{\zcsetup{countertype={environment=claim}}}
\zcRefTypeSetup{claim}{
Name-sg = Claim ,
name-sg = Claim ,
Name-pl = Claims ,
name-pl = Claims ,
}

\usetikzlibrary{calc}
\usetikzlibrary{fit}
\usetikzlibrary{decorations}
\usetikzlibrary{decorations.pathmorphing}
\usetikzlibrary{decorations.text}
\usetikzlibrary{shapes,hobby,backgrounds}

\tikzset{
	position/.style args={#1:#2 from #3}{
		at=($(#3)+(#1:#2)$)
	}
}

\tikzset{
  v:main/.style = {draw, circle, scale=0.8, thick,fill=black,inner sep=0.7mm},
  v:ghost/.style = {inner sep=0pt,scale=1},
  >={latex},
  e:marker/.style = {line width=8.5pt,line cap=round,opacity=0.35,color=DarkGoldenrod},
  e:main/.style = {line width=1pt},
}

\tikzstyle{path} = [color=black!15,line cap=round, line join=round, line width=8pt]

\newcommand{\Coordinates}{%
    \def\R{6}

    \draw[step=.25, very thin, gray!40] (-\R,-\R) grid (\R,\R);

    \draw[step=1, thin, gray!70] (-\R,-\R) grid (\R,\R);

    \draw[->, thick] (-\R,0) -- (\R+0.35,0);
    \draw[->, thick] (0,-\R) -- (0,\R+0.35);

    \foreach \x in {-6,-5,...,6}{
        \draw (\x,0.08) -- (\x,-0.08);
        \ifnum\x=0\else
            \node[below,font=\scriptsize] at (\x,0) {\x};
        \fi
    }

    \foreach \y in {-6,-5,...,6}{
        \draw (0.08,\y) -- (-0.08,\y);
        \ifnum\y=0\else
            \node[left,font=\scriptsize] at (0,\y) {\y};
        \fi
    }

}

\newcommand{\SW}[1]{\todo[linecolor=red,bordercolor=white,color=Auburn,backgroundcolor=AppleGreen!40,textcolor=Auburn,size=\footnotesize]{\textsf{SW: #1}}}

\newcommand{\SWin}[1]{\todo[inline,bordercolor=white,color=Auburn,backgroundcolor=AppleGreen!40,textcolor=Auburn,size=\footnotesize]{\textsf{SW: #1}}}

\DeclareMathOperator{\poly}{\mathbf{poly}}
\DeclareMathOperator{\comb}{comb}
\DeclareMathOperator{\Comb}{Comb}
\newcommand{\pema}{\ensuremath{\mathsf{PM}}}
\newcommand{\Tc}{\textnormal{\textsc{Terminal Cyclability}}}

\newcommand{\LLL}{\mathcal{L}}

\newcommand{\bd}{\mathsf{bd}}
\newcommand{\sth}{\mathrel : }

\begin{document}

\footnotetext[1]{
Matthias Bentert was supported by the German Research Foundation (DFG) under research grant SPP 2378: ReNO-2 (project number 511099228).
Sebastian Wiederrecht was supported by the NRF South Korea (RS-2026-25605318) and by the Institute for Basic Science (IBS-R029-C1).

Emails of the authors:
\href{mailto:bentert@tu-berlin.de}{bentert@tu-berlin.de},
\href{mailto:d.cavallaro@tu-berlin.de}{d.cavallaro@tu-berlin.de},
\href{mailto:wiederrecht@kaist.ac.kr}{wiederrecht@kaist.ac.kr}.
}
\footnotetext[2]{Technische Universität Berlin, Berlin, Germany.}
\footnotetext[3]{School of Computing, KAIST, Daejeon, South Korea.}

\maketitle

\begin{abstract}
    Given an undirected graph~$G$ and a set~$R\subseteq V(G)$ of terminals, \Tc{} asks whether~$G$ contains a simple cycle visiting every vertex of~$R$.
    We establish a complexity dichotomy for this problem on red-minor-closed classes of annotated graphs.
    In particular, we show that \Tc{} is polynomial-time solvable on every such class excluding any annotated planar graph.
    The problem is \textsf{NP}-complete whenever the class contains all annotated planar graphs.

    More precisely, we give an~$|G|^{2^{\mathbf{poly}(b)}}$-time algorithm, where~$b$ is the bidimensionality of~$(G,R)$.
    Our main technical contribution is the introduction of a weak irrelevant vertex rule.
    Rather than removing irrelevant vertices until the treewidth becomes constant, we only remove irrelevant vertices until the treewidth is logarithmic in the graph size.
    Such a weakening is necessary as a parameter-only treewidth threshold for an irrelevant vertex rule in this setting is known to be impossible to realise.
    Combined with a single-exponential algorithm for \Tc{} parameterized by treewidth, this yields the polynomial-time algorithm for our dichotomy.
\end{abstract}
\let\sc\itshape
\thispagestyle{empty}

\newpage

\newpage
\thispagestyle{empty}
\tableofcontents

\newpage

\setcounter{page}{1}

\section{Introduction}\label{sec:Intro}
The \textsc{Hamiltonian Cycle} problem asks if a given graph contains a cycle spanning all of its vertices.
The problem belongs to Karp's original list of \textsf{NP}-complete problems \cite{Karp1972Reducibility} and is known to remain \textsf{NP}-complete even when restricted to planar graphs \cite{GareyJohnsonTarjan1976PlanarHC}.
It is one of the oldest and most studied computationally hard problems.
The fact that \textsc{Hamiltonian Cycle} is \textsf{NP}-complete on planar graphs gives rise to a neat delineation of the problem's tractability within minor-closed graph classes:
\textsc{Hamiltonian Cycle} is \textsf{NP}-complete in every minor-closed graph class containing all planar graphs.
On the other hand, the celebrated grid theorem of Robertson and Seymour \cite{RobertsonS1986GraphMinorsV} states that any minor-closed class that excludes some planar graph has bounded treewidth, implying the existence of polynomial-time algorithms for \textsc{Hamiltonian Cycle} within such classes (see for example~\cite{CyganNPPvRWJ2022SolvingConnectivity}).

\begin{figure}[tbp]
    \centering
    \begin{tikzpicture}
        \def\x{2.4}
        \def\y{-1.2}
                
        \foreach \i/\j/\w in {1/2/gride,1/3/guide,1/4/guile,1/6/quilt,1/7/quirt,2/1/bride,2/2/shade,2/3/shake,2/4/spake,2/5/spike,2/6/spice,2/7/quire,3/1/brede,3/3/share,3/4/spare,3/5/spire,3/7/quite,4/1/arede,4/2/serre,4/3/seare,4/4/shire,4/5/prods,4/6/brods,4/7/suite,5/1/serry,5/2/areae,5/3/woods,5/4/poods,5/5/broos,5/6/brook,5/7/groom,6/1/sorry,6/2/words,6/3/aread,6/4/tread,6/5/brood,6/6/broom,7/2/wordy,7/4/bread,7/5/broad,7/6/blood,7/7/bloom}{
            \node[rectangle,rounded corners,draw,thick,minimum height=.6cm,minimum width=1cm] at (\x*\j,\y*\i) (\w) {\w};
        }
        
        \foreach \i/\j/\w in {1/1/pride,1/5/guilt,3/2/shame,3/6/spite,6/7/gloom,7/1/worry,7/3/dread}{
            \node[rectangle,rounded corners,draw,thick,fill=red,minimum height=.6cm,minimum width=1cm] at (\x*\j,\y*\i) (\w) {\w};
        }
        
        \foreach \a/\b in {pride/gride,pride/bride,gride/guide,guide/guile,guile/guilt,guilt/quilt,quilt/quirt,quirt/quire,bride/brede,shade/shake,shade/shame,shake/spake,shake/share,spake/spike,spake/spare,spike/spice,spike/spire,spice/spite,quire/quite,brede/arede,shame/share,share/spare,share/seare,spare/spire,spire/spite,quite/suite,serre/seare,prods/brods,serry/sorry,woods/poods,broos/brook,broos/brood,brook/broom,groom/gloom,sorry/worry,words/wordy,aread/tread,aread/dread,tread/bread,brood/broom,brood/broad,gloom/bloom,worry/wordy,dread/bread,bread/broad,blood/bloom}{
            \draw[very thick] (\a) to (\b);
        }

        \foreach \a/\b in {shade/share,spike/spite,share/shire,spite/suite,arede/areae,areae/aread,broos/broom,aread/bread,brood/blood,broom/bloom}{
            \draw[very thick] ([xshift=-1.5pt, yshift=1.5pt]\a.south east) to ([xshift=1.5pt, yshift=-1.5pt]\b.north west);
        }

        \foreach \a/\b in {gride/bride,shake/shame,spice/spire,spare/seare,spire/shire,serre/serry,prods/poods,brods/broos,woods/words,brook/brood,groom/broom,tread/dread}{
            \draw[white,line width=4pt,rounded corners] (\a) -- (\b);
            \draw[very thick] ([xshift=1.5pt, yshift=1.5pt]\a.south west) to ([xshift=-1.5pt, yshift=-1.5pt]\b.north east);
        }
        
        \foreach \i/\j/\w in {1/2/gride,1/3/guide,1/4/guile,1/6/quilt,1/7/quirt,2/1/bride,2/2/shade,2/3/shake,2/4/spake,2/5/spike,2/6/spice,2/7/quire,3/1/brede,3/3/share,3/4/spare,3/5/spire,3/7/quite,4/1/arede,4/2/serre,4/3/seare,4/4/shire,4/5/prods,4/6/brods,4/7/suite,5/1/serry,5/2/areae,5/3/woods,5/4/poods,5/5/broos,5/6/brook,5/7/groom,6/1/sorry,6/2/words,6/3/aread,6/4/tread,6/5/brood,6/6/broom,7/2/wordy,7/4/bread,7/5/broad,7/6/blood,7/7/bloom}{
            \node[rectangle,rounded corners,draw,thick,minimum height=.6cm,minimum width=1cm] at (\x*\j,\y*\i) {\w};
        }
        
        \foreach \i/\j/\w in {1/1/pride,1/5/guilt,3/2/shame,3/6/spite,6/7/gloom,7/1/worry,7/3/dread}{
            \node[rectangle,rounded corners,draw,thick,fill=red,minimum height=.6cm,minimum width=1cm] at (\x*\j,\y*\i) {\w};
        }
    \end{tikzpicture}
    \caption{An instance of \textsc{Doublets} with a pre-selected set of words from the SOWPODS scrabble dictionary.
    The task is to find a cycle through the seven sins, that is, the given words are~$R \coloneqq \{\text{dread}, \text{guilt}, \text{gloom}, \text{pride}, \text{shame}, \text{spite}, \text{worry} \}$.
    The interested reader may find it amusing to search for an $R$-cycle. A solution can be found in \zcref{sec:Conclusion}.}
    \label{fig:sevensins}
\end{figure}
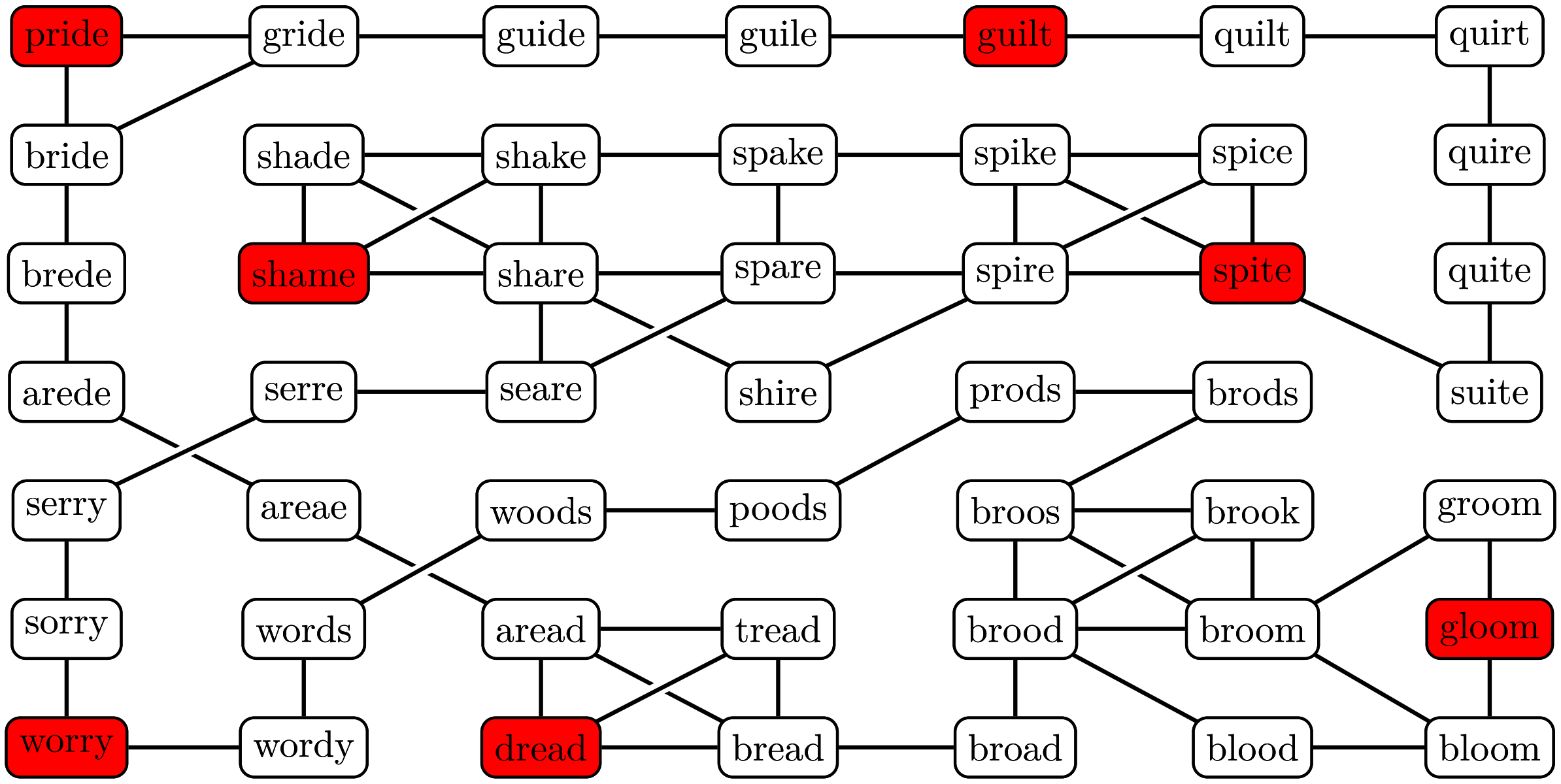

A particularly charming variant of \textsc{Hamiltonian Cycle} was considered by Lewis Carroll~\cite{Car78}.
In its necklace variant, one is given a list of English words, all of the same length~$n$.
The task is to arrange a sequence of English words of length $n$ in a cycle, changing exactly one letter at each step, including from the last word back to the first.
Every word on the given list must appear in the cycle and no word should appear more than once.
See \zcref{fig:sevensins} for an example of such a puzzle.

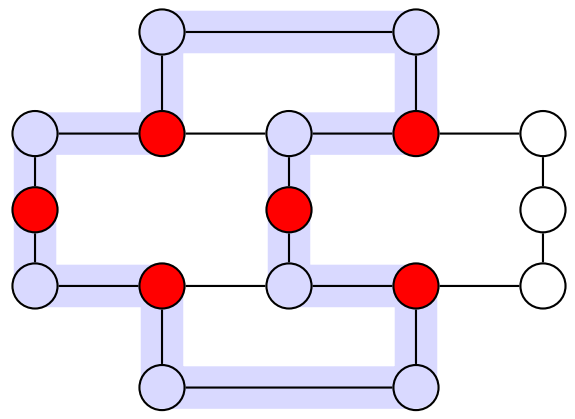
\begin{figure}[ht]
    \centering
    \begin{tikzpicture}[scale=.84]
        \node[circle,draw,inner sep=3pt] (uu1) at(-1,1.4) {};
        \node[circle,draw,inner sep=3pt] (uu2) at(1,1.4) {} edge(uu1);
        \node[circle,draw,inner sep=3pt] (u1) at(-2,.6) {};
        \node[circle,draw,fill=red,inner sep=3pt] (u2) at(-1,.6) {} edge(uu1) edge(u1);
        \node[circle,draw,inner sep=3pt] (u3) at(0,.6) {} edge(u2);
        \node[circle,draw,fill=red,inner sep=3pt] (u4) at(1,.6) {} edge(uu2) edge(u3);
        \node[circle,draw,inner sep=3pt] (u5) at(2,.6) {} edge(u4);
        \node[circle,draw,fill=red,inner sep=3pt] (m1) at(-2,0) {} edge(u1);
        \node[circle,draw,fill=red,inner sep=3pt] (m2) at(0,0) {} edge(u3);
        \node[circle,draw,inner sep=3pt] (m3) at(2,0) {} edge(u5);
        \node[circle,draw,inner sep=3pt] (l1) at(-2,-.6) {} edge(m1);
        \node[circle,draw,fill=red,inner sep=3pt] (l2) at(-1,-.6) {} edge(l1);
        \node[circle,draw,inner sep=3pt] (l3) at(0,-.6) {} edge(m2) edge(l2);
        \node[circle,draw,fill=red,inner sep=3pt] (l4) at(1,-.6) {} edge(l3);
        \node[circle,draw,inner sep=3pt] (l5) at(2,-.6) {} edge(m3) edge(l4);
        \node[circle,draw,inner sep=3pt] (ll1) at(-1,-1.4) {} edge(l2);
        \node[circle,draw,inner sep=3pt] (ll2) at(1,-1.4) {} edge(l4) edge(ll1);

        \begin{pgfonlayer}{background}
            \draw[path,blue!15] (uu1.center) -- (uu2.center) -- (u4.center) -- (u3.center) -- (l3.center) -- (l4.center) -- (ll2.center) -- (ll1.center) -- (l2.center) -- (l1.center) -- (u1.center) -- (u2.center) -- cycle;
        \end{pgfonlayer}
    \end{tikzpicture}
    \caption{An instance of \Tc. The vertices in~$R$ are red and an~$R$-cycle is highlighted.}
    \label{fig:intro-example}
\end{figure}

Returning to the world of graphs, this simple puzzle reveals a surprisingly rich algorithmic problem where one is given a graph~$G$ and a set~$R \subseteq V(G)$ of distinguished vertices called \emph{terminals} or \emph{red} vertices.
The task is then to find a simple cycle in~$G$ containing all vertices in $R$.
This question, known as \Tc{} or \textsc{$T$-Cycle}, is a fundamental problem in algorithmic graph structure theory and the focus of our work.
We call the pair~$(G,R)$ an \emph{annotated graph} and a solution an \emph{$R$-cycle}.
For an example, see~\cref{fig:intro-example}.

Note that a solution does not need to be a Hamiltonian cycle of $G$ in order to constitute a valid solution.
However, requiring the cycle to contain every vertex -- that is, ${R = V(G)}$ -- reveals that \Tc{} is in fact a generalisation of \textsc{Hamiltonian Cycle}.
In light of the dichotomy for \textsc{Hamiltonian Cycle} explained above, this gives rise to the central question motivating our work:
\begin{center}
    \textsl{Can we distinguish those classes of annotated graphs where \Tc{} is tractable from those where the problem is \textsf{NP}-complete?}
\end{center}

\paragraph{Our result.}
To start our quest, we draw further inspiration from the above dichotomy for \textsc{Hamiltonian Cycle} and restrict ourselves to classes of annotated graphs closed under an appropriate analogue of the minor relation for annotated graphs.
An annotated graph $(H,B)$ is a \emph{red-minor} of some annotated graph $(G,R)$ if $(H,B)$ can be obtained from $(G,R)$ by a sequence of (i) edge deletions, (ii) vertex deletions, (iii) edge contractions, where the resulting vertex is red if and only if at least one of the endpoints of the contracted edge was red, and (iv) colour erasures, that is, forgetting the red colour of some vertex.
Red-minors, sometimes also referred to as \textsl{rooted minors}, have been widely studied in the literature as a natural extension of the minor relation to annotated graphs and for their algorithmic properties.
Within the framework of red-minors, we provide the following full classification of the tractability landscape of \Tc{}:

\begin{restatable}{theorem}{maintheorem}
\label{thm:MainTheorem}
Let $\mathcal{R}$ be a red-minor-closed class of annotated graphs.
If $\mathcal{R}$ does not contain all annotated planar graphs, then \textsc{Terminal Cyclability} can be solved in polynomial time on $\mathcal{R}$.
Otherwise, \textsc{Terminal Cyclability} is \textsf{NP}-complete on $\mathcal{R}$.
\end{restatable}

One may see \zcref{thm:MainTheorem} as a natural extension of the treewidth-based dichotomy for \textsc{Hamiltonian Cycle} in minor-closed graph classes.
In fact any red-minor-closed graph class excluding some annotated graph~$(G,\emptyset)$ where $G$ is planar must have bounded treewidth.
Moreover, every red-minor-closed class $\mathcal{C}$ containing all planar annotated graphs contains in particular all planar annotated graphs of the form~$(G,V(G))$.
Hence, \Tc{} is \textsf{NP}-complete even when restricted to inputs from $\mathcal{C}$.
However, the novelty of \zcref{thm:MainTheorem} arises in the space between these two extremes:
For every planar annotated graph $(G,R)$ with~$R \neq \emptyset$, the class $\mathcal{C}_{(G,R)}$ of $(G,R)$-red-minor-free annotated graphs has unbounded treewidth and yet \Tc{} remains tractable on~$\mathcal{C}_{(G,R)}$.
\zcref{thm:MainTheorem} closes the remaining gap:
It completes previously known results by stating that excluding \textsl{any} annotated planar graph implies tractability.
This also answers an open question recently raised by Thilikos and Wiederrecht~\cite{ThilikosW2026Spanning} in the affirmative.

\paragraph{Our tools: Weak irrelevant vertices and bidimensionality.}
Red-minor-closed classes excluding an annotated planar graph are precisely those of bounded \emph{bidimensionality} \cite{ThilikosWiederrecht2025Bidimensionality,GorskyPW2026Quickly}.
Here the bidimensionality of an annotated graph $(G,R)$ is defined to be the largest $k$ such that $(G,R)$ contains an all-red $(k \times k)$-grid as a red-minor.
We show that every \Tc{} instance of bidimensionality $b$ can be reduced, by deleting irrelevant vertices, to an equivalent instance of treewidth $f(b)\cdot\log |G|$ for some computable function~$f$.
Combined with a single-exponential algorithm parameterized by treewidth, this yields polynomial-time solvability for every fixed $b$.
This \emph{weak irrelevant vertex rule} overcomes an obstruction identified by Thilikos and Wiederrecht~\cite{ThilikosW2026Spanning} who showed that there cannot be such a reduction in which the treewidth only depends on~$b$.
In fact, their examples show that treewidth~$\log |G|$ is necessary.
The logarithmic threshold is therefore both sufficient for tractability and necessary for this approach.
More broadly speaking, we reveal a way to extend the reach of the irrelevant vertex technique by matching the target treewidth to the dependence on treewidth of the final algorithm.

Our proof builds on a recent structure theorem for annotated graphs of bounded bidimensionality \cite{ProtopapasThilikosWiederrecht2025ColorfulMinors,GorskyPW2026Quickly}, analogous to Robertson and Seymour's theorem for $H$-minor-free graphs \cite{RobertsonS2003GraphMinorsXVI}.
It provides access to topological tools even though the input classes may contain arbitrary ordinary minors.
The central new argument shows that deep intrusions of a suitably chosen solution into a topologically controlled region force exponential branching: each additional constant amount of depth requires a constant-factor increase in the number of disjoint substructures.
Their depth is consequently $\mathbf{O}(\log |G|)$.
A second key argument then reroutes the solution through an insulated annulus using at most two crossing paths, independently of $b$.
The main technical challenge is to preserve the logarithmic threshold throughout these steps.
Even a polynomial loss in $\log |G|$ would defeat the polynomial-time conclusion, so the structural reductions and rerouting arguments must retain linear dependence on this scale.

\subsection{A closer look at our result}

Our main result rests on a relaxation of the irrelevant vertex technique.
The guiding principle is that, for a problem solvable in single-exponential time parameterized by treewidth, reducing the treewidth to a logarithmic function of the input size already suffices for polynomial-time solvability.
We show that this perspective extends the reach of irrelevant vertex methods beyond a known structural barrier for \Tc{}.

\paragraph{From bounded to logarithmic treewidth.}
The irrelevant vertex technique, pioneered by Robertson and Seymour
\cite{RobertsonSeymour1995DisjointPaths,RobertsonSeymour2009UniqueLinkages,RobertsonSeymour2012Irrelevant},
is a cornerstone of algorithmic graph minors.
Its usual implementation repeatedly deletes vertices whose removal preserves the answer until the remaining graph has treewidth bounded by a function only depending on the parameter.
An algorithm for bounded treewidth then solves the reduced instance.
This approach underlies many algorithms for routing and related problems
\cite{AdlerEtAl2017PlanarIrrelevant,CavallaroGKTW2026Optimal,KorhonenPilipczukStamoulis2024AlmostLinear,LokshtanovSaurabhZehavi2020Efficient,SauStamoulisThilikos2025CMSO}.

For polynomial-time solvability at each fixed parameter value, however, a parameter-only bound on the remaining treewidth is more than we need.
If a problem admits an algorithm with running time~$2^{\mathbf{O}(w)}\cdot \mathbf{poly}(|G|)$ on graphs of treewidth~$w$,
then a reduction to treewidth $f(k)\log |G|$ yields running time~$2^{\mathbf{O}(f(k)\log |G|)} \cdot \mathbf{poly}(|G|) = |G|^{\mathbf{O}(f(k))}$.
Thus, an efficiently computable irrelevant vertex rule with a logarithmic threshold can yield an XP-algorithm even when a parameter-only threshold is impossible.

\paragraph{A structural barrier for conventional irrelevant vertices.}
\label{subsec:IntroIrrelevant}
For \Tc{}, this distinction is essential.
Thilikos and Wiederrecht~\cite{ThilikosW2026Spanning} identified a parameter
$\mathsf{depth}_2$ that characterises the red-minor-closed classes on which the conventional irrelevant vertex approach applies.
Bounded $\mathsf{depth}_2$ permits a fixed-parameter algorithm that either bounds the treewidth by a function of the parameter or finds an irrelevant vertex.
Every red-minor-closed class of unbounded~$\mathsf{depth}_2$, however, contains a family $\mathcal{H} = \{(H_i,R_i)\}_{i\in\mathbb{N}}$ of annotated planar graphs with bounded bidimensionality and unbounded treewidth, but no irrelevant nonterminal vertices.
Each~$(H_i,R_i)$ has an $R_i$-cycle, and deleting any vertex outside $R_i$ destroys all such cycles.\SW{add reference to the counterexample illustration that will appear later}

These examples rule out a parameter-only irrelevant vertex threshold for \Tc{} parameterized by bidimensionality.
Crucially, they do not establish computational hardness.
Moreover, their size grows exponentially with their treewidth: they leave open the possibility of a logarithmic threshold.
Our result shows that this remaining possibility is sufficient to recover tractability throughout the entire bounded-bidimensionality regime.

\paragraph{The weak irrelevant vertex property.}
We formalise this relaxation as follows.
A parameterized graph problem with parameter~$k$ has the
\emph{weak irrelevant vertex property} if there exists a fixed-parameter algorithm that either determines that the input graph~$G$ has treewidth at most
$f(k)\log |G|$, for some computable function~$f$, or finds a non-empty set of vertices whose deletion preserves the answer to the instance.

For \Tc{}, a set $X\subseteq V(G)\setminus R$ is \emph{irrelevant} if
$(G-X,R)$ is a yes-instance if and only if $(G,R)$ is a yes-instance.
Our central algorithmic result establishes the weak irrelevant vertex property for \Tc{} parameterized by bidimensionality.
It also supplies the tree-decomposition needed to solve the reduced instance.

\begin{restatable}{theorem}{irrelevantvertex}
    \label{thm:MainIrrelevantVertex}
    There exists a function $\tau \colon \mathbb{N}^2 \to \mathbb{N}$ and an algorithm that takes as input an instance~$(G,R)$ of \Tc{} and an integer $b \geq 1$, and finds one of the following outcomes in~$2^{2^{\mathbf{poly}(b)}} \cdot \mathbf{poly}(|G|)$~time:
    \begin{enumerate}
        \item $\mathsf{bidim}(G,R) > b$,
        \item an induced subgraph~$G_0$ with $R \subseteq V(G_0)$ such that $V(G) \setminus V(G_0) \neq \emptyset$ is irrelevant and $G_0$ does not contain~$K_{6(b+1)^2 + 1}$ as a minor, or
        \item a tree-decomposition for~$G$ of width at most $\tau(b,|G|) \in 2^{\mathbf{poly}(b)} \cdot \log |G|$.
    \end{enumerate}
\end{restatable}

Since bidimensionality cannot increase under vertex deletion, we may repeatedly apply
\zcref{thm:MainIrrelevantVertex} until the remaining instance has logarithmic treewidth.
Our single-exponential algorithm parameterized by treewidth then gives an
$|G|^{2^{\mathbf{poly}(b)}}$-time algorithm, where
$b=\mathsf{bidim}(G,R)$.
Consequently, \Tc{} is polynomial-time solvable on every red-minor-closed class excluding an annotated planar graph.
The polynomial degree depends on the class, while both the number of terminals and the treewidth may be unbounded.
The obstruction family $\mathcal{H}$ explains why the logarithmic factor cannot simply be replaced by a constant depending only on~$b$.

\paragraph{Relation to earlier tractable regimes.}

Previous work on~\Tc{} establishes tractability through parameterizations
by the number of terminals~\cite{RobertsonSeymour1995DisjointPaths,GRZ25} and more general parameters that allow for unbounded numbers of terminals and instead restrict their distribution within the graph~\cite{JansenSwennenhuis2024SteinerTree,HodorLaMicekRambaud2026ApexForest,ProtopapasThilikosWiederrecht2025ColorfulMinors,ThilikosW2026Spanning}.
Our approach takes on a new point of view on this landscape which can be applied, for example, to the planar arguments of Gahlawat, Rathod, and
Zehavi~\cite{GRZ25} in order to derive another tractability result.
Although their algorithms are parameterized by the number
of terminals, we observe that their rerouting arguments
can be adapted to obtain an irrelevant vertex threshold
of $\mathbf{O}(\sqrt{f}\log |G|)$ when the terminals can
be covered by~$f$ faces.
Combined with a single-exponential algorithm parameterized
by treewidth, this gives an
$|G|^{\mathbf{O}(\sqrt{f})}$-time algorithm.
This consequence is not stated in their paper.
Since our main result subsumes this observation, although with worse bounds, we do not provide further details and leave the proof as an exercise for the interested reader.
Our main theorem extends tractability to all classes of
bounded bidimensionality, including classes of unbounded
$\mathsf{depth}_2$ and annotated planar graphs with bounded face-cover number.

For a concrete example beyond the bounded-$\mathsf{depth}_2$ regime, consider the class $\mathcal{D}$ defined as the red-minor closure of all planar annotated graphs admitting an embedding with a face~$F$ such that every terminal lies on~$F$ or has a neighbour on~$F$.
This class is red-minor closed and excludes the all-red $(5\times5)$-grid, yet has unbounded $\mathsf{depth}_2$.
Our theorem establishes polynomial-time solvability on this class despite the obstruction to conventional irrelevant vertex methods.

\paragraph{The remaining parameterized complexity challenge.}
Our dichotomy settles polynomial-time solvability on
red-minor-closed classes, but leaves a central question open:
is \Tc{} fixed-parameter tractable when parameterized by
bidimensionality, or can one establish $\mathsf{W}[1]$-hardness?
Our algorithm runs in $|G|^{2^{\mathbf{poly}(b)}}$ time,
so the degree of the polynomial depends on~$b$.
An FPT-algorithm would replace this bound by
$f(b)\cdot \mathbf{poly}(|G|)$, with an exponent independent
of~$b$.
The known obstructions to parameter-only irrelevant vertex
thresholds do not rule out such an algorithm: they constrain
a reduction technique rather than establishing computational
hardness.
Determining whether the parameter dependence can be removed
from the exponent, or proving a parameterized lower bound
that prevents this, is a crucial next challenge in understanding
how the structure of the terminal set governs spanning routing.

\paragraph{A broader algorithmic perspective.}
The weak irrelevant vertex property suggests a promising direction for algorithmic graph minors.
For problems admitting single-exponential algorithms parameterized by treewidth, a logarithmic irrelevant vertex threshold can already support polynomial-time algorithms on structurally restricted classes.
The central challenge is to obtain such a threshold while keeping every structural reduction linear in $\log |G|$.

\Tc{} demonstrates that this approach can succeed even in the presence of explicit obstructions to the conventional irrelevant vertex property.
Natural candidates for further investigation include \textsc{Spanning Disjoint Paths} and the annotated deletion problems discussed in our conclusion.
For spanning paths, prescribed connections prevent the direct use of our cycle rerouting argument as further expanded upon below, leaving \textsc{Spanning Disjoint Paths} as a natural yet challenging next test case.
Across the board, the weak irrelevant vertex property provides a concrete target for a possibly wide class of computational problems on annotated graphs: establish enough structural reduction for the final algorithm to run in polynomial time, even when reduction to parameter-bounded treewidth is impossible.

\subsection{Related work}\label{subsec:RelatedWork}

In the following, we give an overview of relevant related results in the known literature.

\begin{figure}[t]
 \centering
 \begin{tikzpicture}

 \pgfdeclarelayer{background}
		\pgfdeclarelayer{foreground}
			
		\pgfsetlayers{background,main,foreground}

 \begin{pgfonlayer}{background}
 \pgftext{\includegraphics[width=16cm]{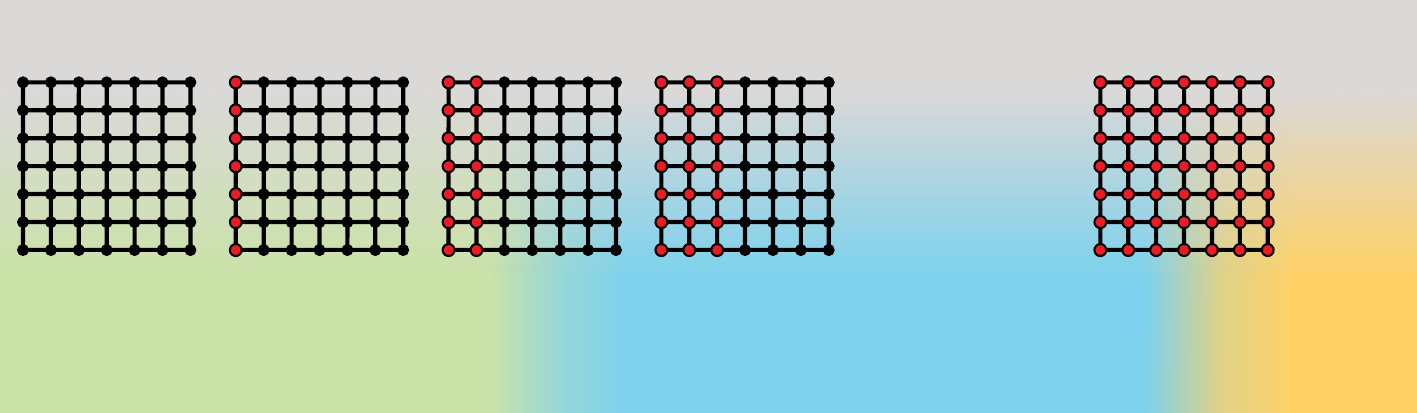}} at (C.center);
 \end{pgfonlayer}
			
 \begin{pgfonlayer}{main}
 \node (C) [v:ghost] {};
 
 \end{pgfonlayer}
 
 \begin{pgfonlayer}{foreground}

    \node (tw) [v:ghost] at (-6.75,1.8) {treewidth};

    \node (monodim) [v:ghost] at (-4.35,1.8) {monodim.\@};

    \node (bidim) [v:ghost] at (5.4,1.8) {bidim.\@};

    \node (depth_0) [v:ghost] at (-6.75,-1) {$\mathsf{depth}_0$};

    \node (depth_1) [v:ghost] at (-4.35,-1) {$\mathsf{depth}_1$};

    \node (depth_2) [v:ghost] at (-2,-1) {$\mathsf{depth}_2$};

    \node (depth_3) [v:ghost] at (0.4,-1) {$\mathsf{depth}_3$};

    \node (depth_dots) [v:ghost] at (2.8,-1) {$\dots$};

    \node (depth_infty) [v:ghost] at (5.4,-1) {$\mathsf{depth}_{\infty}$};

    \node (dots) [v:ghost] at (2.8,0.5) {$\mathbf{\dots}$};

    \node (sufficient_1) [v:ghost] at (-5.5,-1.8) {\textsf{FPT}};

    \node (boundary) [v:ghost] at (1.8,-1.8) {{This paper:} \textsf{XP}};

    \node (necessary_1) [v:ghost] at (7.1,-1) {\rotatebox{90}{\textsf{NP}-complete}};
 \end{pgfonlayer}
 \end{tikzpicture}
 \caption{The depth hierarchy of annotated grid parameters.
Known algorithms for \Tc{} are fixed-parameter tractable
parameterized by $\mathsf{depth}_2$, while our algorithm is
in \textsf{XP} parameterized by
$\mathsf{depth}_\infty=\mathsf{bidim}$.
A red-minor-closed class has bounded bidimensionality exactly
when it excludes an annotated planar graph; otherwise,
\Tc{} is $\mathsf{NP}$-complete on the class.}
 \label{fig:DepthHierarchy}
\end{figure}
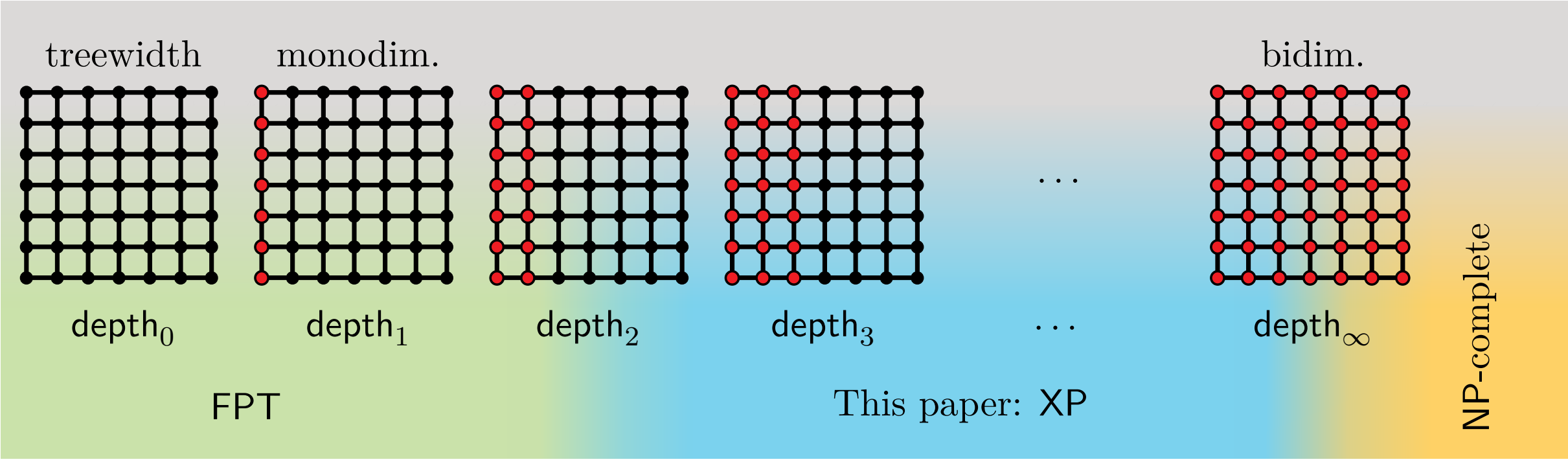

\paragraph{Annotated grids and structural parameters.}
The structural parameters underlying our result belong to the
\emph{depth hierarchy} introduced by Thilikos and Wiederrecht
\cite{ThilikosW2026Spanning}; see \zcref{fig:DepthHierarchy}.
This hierarchy extends the grid characterisation of treewidth:
Robertson and Seymour~\cite{RobertsonS1986GraphMinorsV} proved that
treewidth is functionally equivalent to the largest order of a square
grid minor.
For annotated graphs, prescribing which branch sets of the grid must
meet the terminal set gives a sequence of increasingly permissive
structural restrictions.

For integers $0\leq r\leq k$, the \emph{$r$-annotated
$(k\times k)$-grid} is the $(k\times k)$-grid with its $r$ leftmost
columns coloured red.
The parameter $\mathsf{depth}_r(G,R)$ is the largest~$k$ for which
this annotated grid is a red-minor of $(G,R)$.
At the endpoint of the hierarchy, $\mathsf{depth}_\infty(G,R)$
requires every vertex of the grid to be red and is precisely
$\mathsf{bidim}(G,R)$.
Thus, $\mathsf{depth}_0$ is functionally equivalent to treewidth,
while $\mathsf{depth}_1$ corresponds to \textsl{monodimensionality}, also
known as \textsl{torso treewidth}
\cite{MarxSeymourWollan2017RootedGrid,
ProtopapasThilikosWiederrecht2025ColorfulMinors,
DumasHMMRS2026Invitation}.

The hierarchy also reflects changes in the structural machinery
available.
The grid theorem and the rooted grid theorem of Marx, Seymour,
and Wollan~\cite{MarxSeymourWollan2017RootedGrid} provide
tree-decomposition-based descriptions at its first two levels.
Beyond these levels, known descriptions involve surfaces, apices,
and vortices, as in Robertson and Seymour's Graph Minor Structure
Theorem~\cite{RobertsonS2003GraphMinorsXVI}.
Such structure theorems are available for bounded
$\mathsf{depth}_2$~\cite{ThilikosW2026Spanning} and bounded
bidimensionality~\cite{ProtopapasThilikosWiederrecht2025ColorfulMinors,GorskyPW2026Quickly} where the latter makes up the structural foundation for our proof.
We explain its role in \zcref{subsec:Method}.

\paragraph{Disjoint Paths and spanning routing.}
The algorithmic study of \Tc{} is closely connected to
\textsc{Disjoint Paths}, which asks for vertex-disjoint paths
connecting $k$ prescribed pairs of terminals.
Guessing the cyclic order of the vertices in~$R$, and splitting
terminals appropriately, reduces \Tc{} to~$|R|!$ instances
of \textsc{Disjoint Paths}.
Robertson and Seymour's fixed-parameter algorithm for
\textsc{Disjoint Paths} therefore also establishes
fixed-parameter tractability of \Tc{}
\cite{RobertsonSeymour1995DisjointPaths}.

Recent advances improve different aspects of the running time.
Korhonen, Pilipczuk, and Stamoulis
\cite{KorhonenPilipczukStamoulis2024AlmostLinear}
obtained an $f(k)\cdot |\!|G|\!|^{1+o(1)}$-time algorithm.
Cavallaro, Gorsky, Kreutzer, Thilikos, and Wiederrecht
\cite{CavallaroGKTW2026Optimal} obtained a
$2^{2^{\mathbf{poly}(k)}}|G|^2$-time algorithm, with the more
refined bound~$2^{\mathbf{poly}(k)\cdot 2^{\mathbf{poly}(b)}}|G|^2$
when the terminals have bidimensionality~$b$.
Their quantitative bounds and irrelevant vertex machinery are
important ingredients of our proof.

Even closer to \Tc{} is the above-mentioned variant \textsc{Spanning Disjoint Paths}, where the prescribed paths must collectively visit every vertex of an additional set~$R$.
Thilikos and Wiederrecht~\cite{ThilikosW2026Spanning}
established its irrelevant vertex property when parameterized by the number of prescribed pairs and~$\mathsf{depth}_2(G,R)$.
Extending our weak irrelevant vertex rule to this problem remains open.
The key distinction is that our rerouting may change the order
in which a cycle visits the terminals, whereas spanning paths
must preserve their prescribed connections.

\paragraph{Terminal Cyclability.}
Cycles through prescribed vertices have a long graph-theoretic
history.
Dirac~\cite{Dirac1960Abstrakten} proved that any $k$ specified
vertices in a $k$-connected graph lie on a common cycle for any~$k \geq 2$.
We refer to Gould's survey~\cite{Gould2009Cycles} for extensions
and related results.

For algorithms parameterized by the number of terminals,
Kawarabayashi~\cite{Kawarabayashi2008CyclesThroughElements}
developed a direct irrelevant vertex argument for \Tc{},
improving substantially on the running time obtained through
\textsc{Disjoint Paths}.
A different approach uses algebraic methods:
Bj\"orklund, Husfeldt, and Taslaman
\cite{BjorklundHusfeldtTaslaman2012Shortest}
gave a randomised $2^k \cdot \mathbf{poly}(|G|)$-time algorithm,
and Wahlstr\"om~\cite{Wahlstrom2013Tutte} obtained another
single-exponential randomised algorithm together with a
randomised compression of size $\mathbf{O}(k^3)$ into an
algebraic problem.
The target problem is not known to be in $\mathsf{NP}$,
so this compression does not establish a polynomial kernel
for \Tc{}.
Indeed, \Tc{} is sometimes called the only \emph{natural} problem for which we know a polynomial compression, but no polynomial kernel~\cite{GRZ25}.

On the way towards a polynomial kernel for \Tc{} parameterized by the number of terminals, Gahlawat, Rathod, and Zehavi~\cite{GRZ25} recently
obtained the first polynomial kernel for planar graphs.
They also obtained a deterministic~$2^{\mathbf{O}(\sqrt{k}\log k)}\cdot |G|$-time algorithm for planar graphs.
Finally, Crespelle and Golovach~\cite{CrespelleFeghaliGolovach2022Cyclability} studied \Tc{} on restricted graph classes other than planar graphs and presented linear-time (unparameterized) algorithms for interval graphs, bipartite permutation graphs, and cographs.

\paragraph{Other variants of \textsc{Disjoint Paths}.}
\textsc{Disjoint Paths} is $\mathsf{NP}$-complete already
for two terminal pairs in directed graphs, and \textsc{Directed }\Tc{} is
$\mathsf{NP}$-complete already for two specified vertices
\cite{FHW80}.
Related work studies routing with congestion
\cite{BCHKKS26} and fixed-parameter algorithms for directed
routing in planar graphs and for edge-disjoint routing in
Eulerian digraphs~\cite{CyganMPP2013PlanarDigraph,CavallaroKK2024Eulerian}.
Another line imposes shortest-path requirements:
\textsc{Disjoint Shortest Paths} was introduced by
Eilam-Tzoreff~\cite{Eilam-Tzoreff98}, with XP-algorithms
developed by Lochet~\cite{Loc21} and Bentert, Nichterlein,
Renken, and Zschoche~\cite{BNRZ23}.
Bentert, Fomin, and
Golovach~\cite{BFG26} studied approximation algorithms, kernelization,
and parameterized algorithms for a variant where not all pairs of terminals must be connected.

\subsection{Our methods}\label{subsec:Method}
We prove the weak irrelevant vertex property for \Tc{} in
three stages.
Throughout this overview, let $n=|G|$ and
$b=\mathsf{bidim}(G,R)$.

\begin{enumerate}
    \item \textbf{Bounding vital instances.}
    We first show that a \emph{vital} instance, that is, an instance with a unique $R$-cycle visiting every vertex, has treewidth~$\mathbf{O}_b(\log n)$.
    The central argument shows that deep intrusions of this
    cycle into a topologically controlled region force
    exponential branching, which limits their depth.

    \item \textbf{Combing general solutions.}
    We then use this bound to control solutions in arbitrary
    instances.
    Inside a sufficiently large terminal-free annulus,
    a solution can be rerouted so that it crosses a central
    subannulus along at most two paths, so the number of paths does not depend on~$b$.

    \item \textbf{Finding an irrelevant vertex.}
    Finally, we use this constant-size interface to insulate
    a central region whose size depends only on~$b$.
    Existing irrelevant vertex machinery can then be applied
    within this region.
    The resulting vertex is chosen independently of any
    solution, even though the annulus witnessing its
    insulation may depend on the solution.
\end{enumerate}

A local rerouting lemma is the common tool behind the first
two stages.
It preserves a cycle through all terminals while allowing
the order of traversal to change.
This freedom drives both the branching argument and the
reduction to two crossing paths.
The following two constraints govern how we implement this strategy:
\begin{description}
    \item[(1) Structural requirements.]
    We use the structure theorem for excluding an annotated
    planar graph~\cite{GorskyPW2026Quickly}, variants of the
    Flat Wall Theorem~\cite{SauST2024More}, and
    homogenisation techniques~\cite{GorskySW2026Price}.
    Applying these tools may require an input wall
    substantially larger than the structure they produce.
    Such losses are usually compatible with a
    parameter-dependent treewidth threshold, but a polynomial
    loss in the wall order can be prohibitive at logarithmic
    scale.

    \item[(2) A logarithmic threshold.]
    Our irrelevant vertex rule must apply whenever the
    treewidth exceeds $f(b)\log n$, for some computable
    function~$f$.
    The final single-exponential treewidth algorithm then
    runs in polynomial time for each fixed~$b$.
    A threshold of $f(b)(\log n)^{1+\varepsilon}$, for any
    fixed~$\varepsilon>0$, would no longer yield this
    polynomial-time bound.
\end{description}

The challenge is therefore to retain linear dependence on
$\log n$ in the structural threshold throughout the proof.
This concerns the size of the structures required, rather
than polynomial factors in the running time needed to find
them.
It also dictates where we apply the more expensive tools:
we use a logarithmic-size outer region for insulation and
confine homogenisation to a central region
whose size depends only on~$b$.
This separation of scales allows us to use the graph minor
toolkit within the threshold required by our algorithm.

\paragraph{Excluding a clique minor.}
We first reduce to graphs excluding a clique minor whose
order depends only on~$b$.
This allows us to find large \emph{walls}, which are
grid-like subgraphs witnessing large treewidth, without
losing the logarithmic scale.
Indeed, directly using the general polynomial grid bound
of Chuzhoy and Tan~\cite{ChuzhoyT2021TowardsTight} would
require a superlogarithmic treewidth threshold to obtain
a wall of logarithmic order.
For $H$-minor-free graphs, however, treewidth
$\mathbf{poly}(|H|)\cdot r$ already forces a wall of
order~$r$~\cite{GorskySTW2026Catching}.
To reach this setting, we adapt Robertson and Seymour's
generic folio argument~\cite{RobertsonSeymour1995DisjointPaths}.
Given a sufficiently large clique minor of order
$\mathbf{poly}(b)$, we either certify that
$\mathsf{bidim}(G,R)>b$ or find an irrelevant vertex.
Combining this argument with the minor-testing algorithm
of Cavallaro, Gorsky, Kreutzer, Thilikos, and
Wiederrecht~\cite{CavallaroGKTW2026Optimal}, we can
repeatedly delete irrelevant vertices until the remaining
graph excludes $K_{\mathbf{poly}(b)}$ as a minor.
This preprocessing takes
$2^{2^{\mathbf{poly}(b)}}n^3$ time and makes the linear
grid bound available for the subsequent arguments.

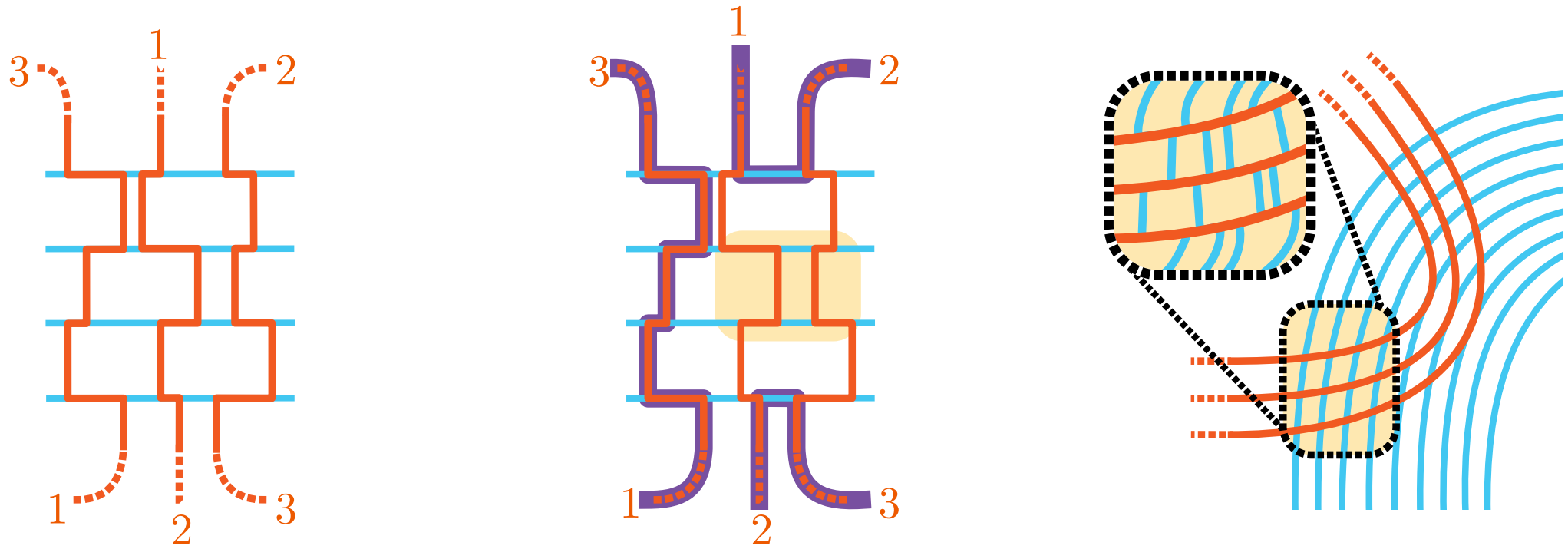
\begin{figure}[t]
 \centering
 \begin{tikzpicture}

 \pgfdeclarelayer{background}
		\pgfdeclarelayer{foreground}
			
		\pgfsetlayers{background,main,foreground}
			
 \begin{pgfonlayer}{main}
 \node (C) [v:ghost] {};

 \node(L) [v:ghost] at (-5,0) {
 \begin{tikzpicture}

 \pgfdeclarelayer{background}
		 \pgfdeclarelayer{foreground}
			
		 \pgfsetlayers{background,main,foreground}

 \begin{pgfonlayer}{background}
 \pgftext{\includegraphics[width=2.4cm]{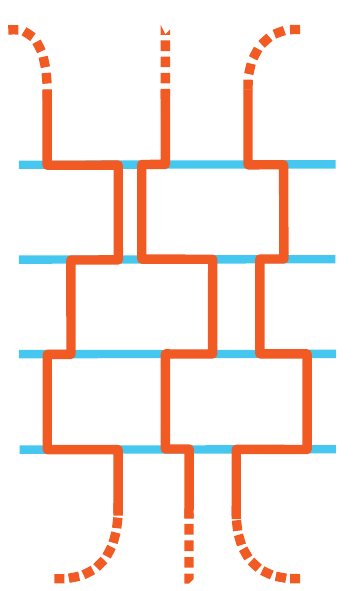}} at (C.center);
 \end{pgfonlayer}
 
 \begin{pgfonlayer}{main}
 \node (C) [v:ghost] {};
 
 \end{pgfonlayer}

 \end{tikzpicture}
 };

 \node(M) [v:ghost] at (0,0) {
 \begin{tikzpicture}

 \pgfdeclarelayer{background}
		 \pgfdeclarelayer{foreground}
			
		 \pgfsetlayers{background,main,foreground}

 \begin{pgfonlayer}{background}
 \pgftext{\includegraphics[width=2.4cm]{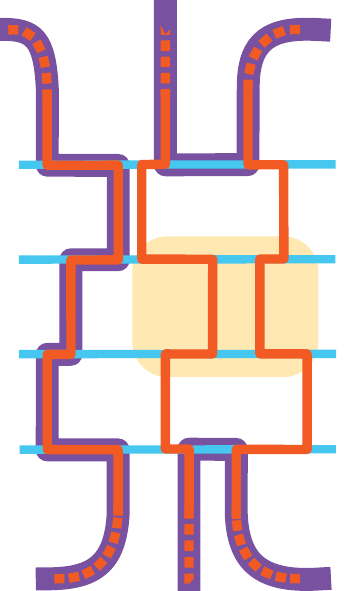}} at (C.center);
 \end{pgfonlayer}
			
 \begin{pgfonlayer}{main}
 \node (C) [v:ghost] {};
 
 \end{pgfonlayer}
 \end{tikzpicture}
 };

 \node(R) [v:ghost] at (5,0) {
 \begin{tikzpicture}

 \pgfdeclarelayer{background}
		 \pgfdeclarelayer{foreground}
			
		 \pgfsetlayers{background,main,foreground}

 \begin{pgfonlayer}{background}
 \pgftext{\includegraphics[width=4cm]{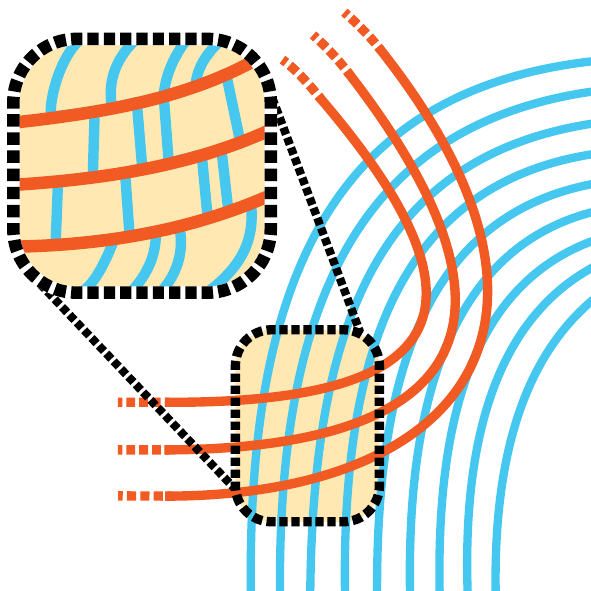}} at (C.center);
 \end{pgfonlayer}
			
 \begin{pgfonlayer}{main}
 \node (C) [v:ghost] {};
 
 \end{pgfonlayer}

 \end{tikzpicture}
 };

    \node (LU1) [v:ghost] at (-5.1,2) {\textcolor{Persimmon}{$1$}};
    \node (LU2) [v:ghost] at (-4,1.78) {\textcolor{Persimmon}{$2$}};
    \node (LU3) [v:ghost] at (-6.3,1.77) {\textcolor{Persimmon}{$3$}};

    \node (LD1) [v:ghost] at (-5.97,-2) {\textcolor{Persimmon}{$1$}};
    \node (LD2) [v:ghost] at (-4.9,-2.18) {\textcolor{Persimmon}{$2$}};
    \node (LD3) [v:ghost] at (-4,-2) {\textcolor{Persimmon}{$3$}};

    \node (RU1) [v:ghost] at (-0.1,2.2) {\textcolor{Persimmon}{$1$}};
    \node (RU2) [v:ghost] at (1.2,1.78) {\textcolor{Persimmon}{$2$}};
    \node (RU3) [v:ghost] at (-1.3,1.77) {\textcolor{Persimmon}{$3$}};

    \node (RD1) [v:ghost] at (-1.03,-1.95) {\textcolor{Persimmon}{$1$}};
    \node (RD2) [v:ghost] at (0.1,-2.18) {\textcolor{Persimmon}{$2$}};
    \node (RD3) [v:ghost] at (1.2,-1.95) {\textcolor{Persimmon}{$3$}};
 \end{pgfonlayer}
 \end{tikzpicture}
\caption{The local rerouting lemma.
Left: an $R$-cycle~$S$ intersects a terminal-free
$(4\times3)$-mesh in its three vertical paths.
Matching labels indicate their connections outside the
mesh in this example.
Middle: the rerouted cycle~$S'$ agrees with~$S$ outside
the mesh and avoids the vertices highlighted in
\textcolor{DarkBananaYellow}{yellow}.
Right: a configuration arising in the proof of
\zcref{thm:MainVitalSolution}.
The lemma does not require the ambient graph to be planar.}
 \label{fig:ReRoute}
\end{figure}

\paragraph{A local rerouting lemma.}
Our main routing tool is a local lemma inspired by the
planar rerouting argument of Gahlawat, Rathod, and
Zehavi~\cite{GRZ25}.
Their argument controls segments of a cheapest solution
within a family of concentric cycles.
Its proof uses planarity to restrict how the parts of the
cycle outside the local configuration can be connected.
We establish a rerouting statement that depends only on
a small mesh and its intersection with the solution,
without any assumption on the planarity of the ambient
graph or on how the solution is connected outside the mesh.
Let $M$ be a terminal-free $(4\times3)$-mesh with
horizontal paths $P_1,\dots,P_4$ and vertical paths
$Q_1,Q_2,Q_3$.
For a formal definition of meshes, we refer to \cref{sec:prelim}.
Suppose an $R$-cycle~$S$ intersects~$M$ precisely in
$Q_1\cup Q_2\cup Q_3$.
We show that there is an $R$-cycle~$S'$ that agrees
with~$S$ outside~$M$ and avoids every vertex of
$P_2\cap Q_2$.
See \zcref{fig:ReRoute}.
The lemma applies to any such cycle: it requires neither
minimality nor vitality, and the ambient graph may be
arbitrarily non-planar.

The freedom to change how the exterior portions of the
cycle are joined is essential.
In their work on \textsc{Spanning Disjoint Paths},
Thilikos and Wiederrecht~\cite{ThilikosW2026Spanning}
use the additional structure supplied by bounded
$\mathsf{depth}_2$ to obtain reroutings that preserve
the prescribed terminal connections.
That structure is unavailable under bounded
bidimensionality alone.
For \Tc{}, we instead exploit the freedom to change the
order of traversal while preserving a single cycle
through all terminals.
Consequently, our lemma need not preserve the prescribed
connections of a spanning linkage and in order to derive a weak irrelevant vertex rule for \textsc{Spanning Disjoint Paths}, new ideas seem necessary.

Our local rerouting statement in a mesh drives both main routing arguments of our proof.
In a vital instance, a mesh satisfying its hypotheses would permit a rerouting that avoids a vertex, contradicting vitality.
We use this contradiction to force branching among nested restrictions of the
solution to a disc.
Later, the same lemma allows us to reduce three crossings of a terminal-free annulus to one.
Repeated applications yield the bound of at most two
crossings needed to insulate the central region.

\paragraph{Bounding the treewidth of a vital instance.}
Following the vital-linkage approach of Robertson and
Seymour~\cite{RobertsonSeymour2009UniqueLinkages}, we first
bound the treewidth of \emph{vital} instances of \Tc{}.
An instance $(G,R)$ is vital if it has a unique
$R$-cycle~$S$ and $V(S)=V(G)$.
Thus, any rerouting that produces another $R$-cycle,
in particular one avoiding a vertex, contradicts vitality.

The main idea is that a deep intrusion of~$S$ into a
terminal-free region must be protected against such
reroutings by further intrusions.
We show that these protections force repeated branching:
depth grows linearly, while the number of disjoint
substructures grows exponentially.
This yields the following logarithmic bound.

\begin{restatable}{theorem}{vitalsol}
    \label{thm:MainVitalSolution}
    There exists a function
    $\beta \colon \mathbb{N}^2 \to \mathbb{N}$ such that
    for every vital instance $(G,R)$ of \Tc{}, it holds that~$\mathsf{tw}(G)
        \leq \beta(\mathsf{bidim}(G,R),|G|)
        \in
        \mathbf{poly}(\mathsf{bidim}(G,R))\cdot\log |G|$.
\end{restatable}

\begin{figure}[t]
 \centering
 \begin{tikzpicture}

 \pgfdeclarelayer{background}
		\pgfdeclarelayer{foreground}
			
		\pgfsetlayers{background,main,foreground}

 \begin{pgfonlayer}{background}
 \pgftext{\includegraphics[width=11cm]{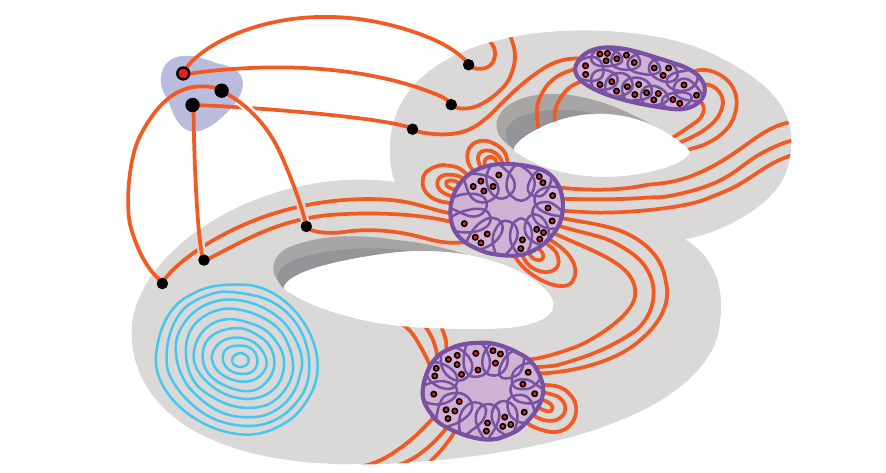}} at (C.center);
 \end{pgfonlayer}
			
 \begin{pgfonlayer}{main}
 \node (C) [v:ghost] {};
 
 \end{pgfonlayer}
 
 \begin{pgfonlayer}{foreground}
    \node (A) [v:ghost] at (-3.75,2) {$A$};

    \node (CCC) [v:ghost] at (-3.5,-2.4) {$\mathcal{C}$};
 \end{pgfonlayer}

 \end{tikzpicture}
 \caption{A sketch of an almost embedding with a small apex set $A$ and set of \textcolor{BrightTurquoise!80!black}{blue} concentric cycles~$\mathcal{C}$ embedded in some disc on the surface.
 The \textcolor{LavenderFloral}{purple} discs are the vortices and the \textcolor{Persimmon}{orange} paths are part of an $R$-cycle.
 Notice that the vertices of $R$ are confined entirely to the vortices and the apex set.}
 \label{fig:AlmostEmbedding}
\end{figure}

To explain the geometric setting of this argument, we require the full power of Robertson and Seymour's graph minor structure theory.
Towards this goal, we briefly sketch some fundamental concepts while noting that these are mostly for intuition and do not necessarily reflect the full technical detail of the definitions used.
Informally, an \emph{$a$-almost embedding} of a graph $G$ in a
surface~$\Sigma$ represents $G$ after deleting an
\emph{apex set}~$A$ of at most~$a$ vertices.
This representation is an embedding of $G-A$ in $\Sigma$ with the exception of a bounded number of regions called \emph{vortices}, where non-planar behaviour is allowed but controlled by a bounded-width path-decomposition respecting the vortex boundary.
The number of vortices is the \emph{breadth}, and their
complexity is measured by the \emph{depth}.
See \zcref{fig:AlmostEmbedding} for an illustration.

The Graph Minor Structure Theorem of Robertson and Seymour \cite{RobertsonS2003GraphMinorsXVI,GorskySW2025Polynomial} asserts that every $H$-minor-free graph can be built from $a(|H|)$-almost-embeddable graphs (into surfaces of Euler-genus at most $g(|H|)$ and with breadth at most $b(|H|)$ and depth at most $d(|H|)$) by gluing along cliques.

This notion of almost-embeddings is usually used to describe the \textsl{global} structure of $H$-minor-free graphs.
We do require a local variant more closely related to theorems of Robertson and Seymour \cite{RobertsonS2003GraphMinorsXVI,RobertsonS1999GraphMinorsXVII,GorskySW2025Polynomial}.
The representation of an almost embedding in this local version is much more involved and we use a formalisation known as \textsl{surface renditions} \cite{RobertsonS1999GraphMinorsXVII,KawarabayashiTW2021Quickly,GorskySW2025Polynomial}.
See \zcref{subsec_grpahMinors} for the formal definitions.
The notion of almost embeddings, however, is fully sufficient as an intuitive concept for the purpose of this introduction.
For annotated graphs of bounded bidimensionality, the
local structure theorem additionally confines all red vertices
to the vortices and the apex set
\cite{ProtopapasThilikosWiederrecht2025ColorfulMinors,
GorskyPW2026Quickly}.
By the clique-minor argument above, a vital instance
already excludes $K_{\mathbf{poly}(b)}$ as a minor.
The linear grid bound and the local structure theorem
therefore provide a large family~$\mathcal{C}$ of
concentric cycles represented in a vortex-free
disc~$\Delta\subseteq\Sigma$.
Crucially, obtaining $r$ such cycles requires treewidth
only $\mathbf{poly}(b)\cdot r$.
This preserves the linear dependence on the number of
cycles that our logarithmic threshold requires.

The apices require particular care.
Along~$S$, at most~$2|A|$ vertices outside~$A$ are adjacent to
the apex set.
By preparing~$2|A|+1$ disjoint candidate discs, we therefore
find one whose crop contains no such vertex.
All red vertices are outside this crop as well.
Inside the selected disc, vitality allows us to draw the
remaining pieces attached through at most three vertices
without crossings: the solution traverses all internal
vertices of each piece along a single path, and any edge
shortcutting that path would give another~$R$-cycle.

We next choose the concentric cycles appropriately, reserve
the outermost cycle as the boundary of the disc, and split~$S$
at its contacts with this boundary.
We retain only the resulting paths that meet one of the remaining cycles.
Standard rerouting arguments control how the resulting paths
meet the retained cycles.
Each path reaches a deepest cycle and returns to the boundary,
with its two portions meeting the surrounding cycles in order.
Moreover, unless~$C_i$ is the outermost retained cycle, a path
turning at~$C_i$ forces another path on its outward side that
turns at the next cycle~$C_{i+1}$.
These reroutings preserve the endpoint pairs of the paths,
so they can be joined to the unchanged parts of~$S$.
Vitality therefore forces the original paths already to have
the required behaviour.

\emph{Forcing branching.}
Our counting argument concerns all these restricted paths.
Each path cuts off a region between itself and the disc
boundary.
These regions are nested or have disjoint interiors.
Furthermore, nested paths must turn at different cycles.

\begin{figure}[t]
    \centering
    \includegraphics[width=0.75\linewidth]{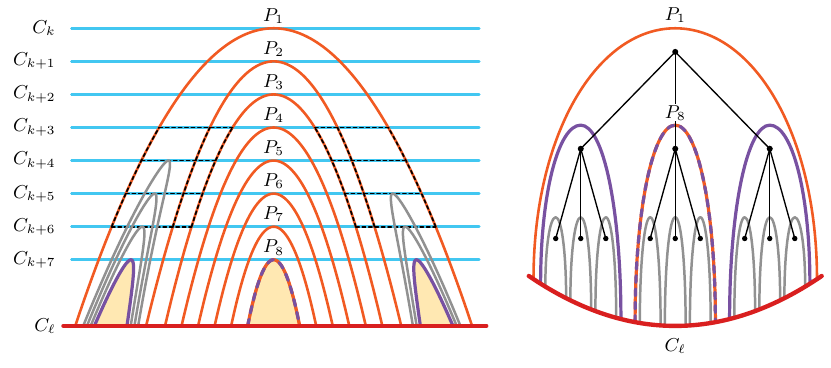}
    \caption{Illustrations of the branching forced by the vitality of the instance, the rerouting lemma, and the assumption of some paths intersecting somewhere deep inside the nested cycles from $\mathcal{C}$.
    On the left, we illustrate how the rerouting lemma forces a branching in three directions.
    The paths~$P_1,\dots,P_8$ are coloured \textcolor{Persimmon}{orange}.
    The two side branches and~$P_8$ are coloured \textcolor{LavenderFloral}{purple} and the \textcolor{DarkBananaYellow}{light yellow} areas they enclose (together with parts of the outermost \textcolor{red}{red} cycle~$C_{\ell} \in \mathcal{C}$) indicate regions with pairwise disjoint interiors.
    The dashed meshes indicate the areas that could potentially allow for an application of the rerouting lemma.
    On the right, we indicate how the branching continues when we continue through the surrounding cycles.}
    \label{fig:branch}
\end{figure}

Consider eight nested paths turning at consecutive
cycles~$C_k,\ldots,C_{k+7}$.
The arms on each side of the first three paths, together with
portions of four subsequent cycles, form a~$(4\times3)$-mesh.
If the solution met any of these meshes only in its three vertical
paths, our local rerouting lemma would contradict vitality.
Thus, an additional restricted path must meet each mesh.
The nesting order forces each of these paths to cut off a region
whose interior is disjoint from the region cut off by the third path.
Comparing their contacts with the outermost cycle of the meshes
shows that these two regions also have disjoint interiors.
Applying the preceding next-cycle property within both regions
gives two paths turning at~$C_{k+7}$.
Together with the eighth path of the original stack, these give
three branches in regions with disjoint interiors.
See \cref{fig:branch} for an illustration.

Every seven layers therefore give a branching step.
Repeating the argument independently inside the three regions produces~$3^q$ paths with distinct vertices after~$q$ steps.
Since the graph has only~$n$ vertices, this bounds the number
of concentric cycles by~$\mathbf{O}(\log n)$.
Obtaining~$r$ cycles required treewidth only~$\poly(b)\cdot r$, so sufficiently large treewidth would supply more cycles than this bound permits.

\SWin{add a comparison paragraph to the Thilikos-Wiederrecht examples here}

\begin{figure}[t]
 \centering
 \begin{tikzpicture}

 \pgfdeclarelayer{background}
		\pgfdeclarelayer{foreground}
			
		\pgfsetlayers{background,main,foreground}

 \begin{pgfonlayer}{background}
 \pgftext{\includegraphics[width=16cm]{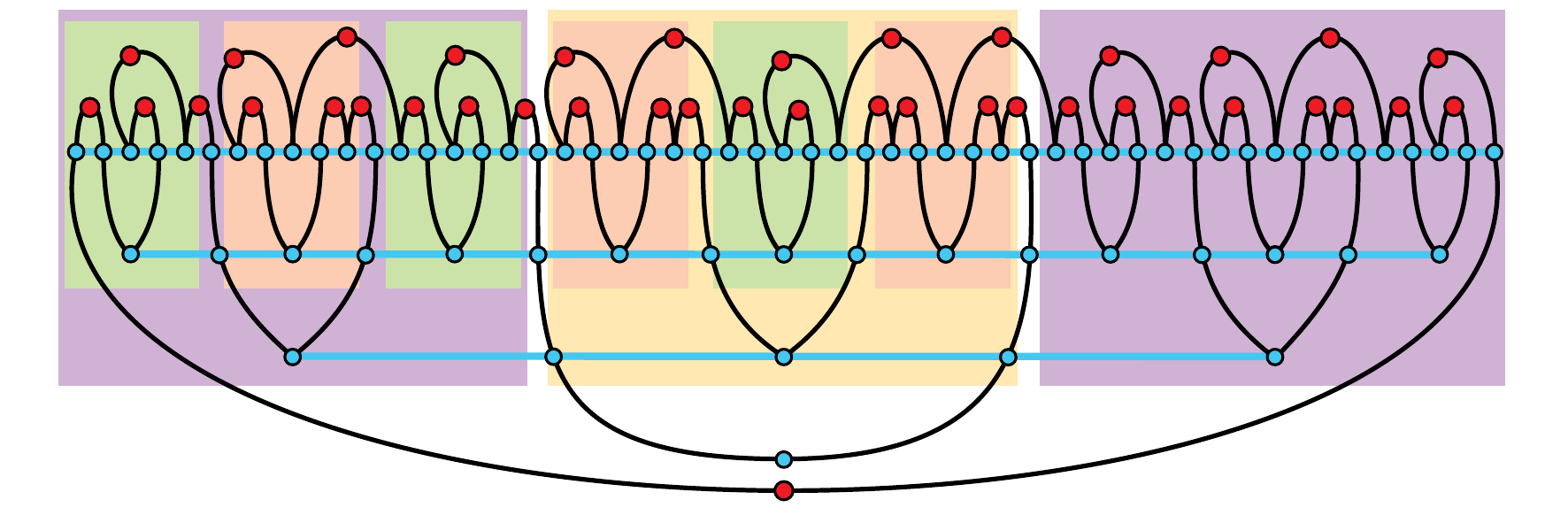}} at (C.center);
 \end{pgfonlayer}
			
 \begin{pgfonlayer}{main}
 \node (C) [v:ghost] {};
 
 \end{pgfonlayer}
 
 \begin{pgfonlayer}{foreground}
 \end{pgfonlayer}

 \end{tikzpicture}
 \caption{An illustration of the recursive construction of the counterexample family $\mathcal{H}$ to the irrelevant vertex property of \Tc{} in red-minors of the $2$-annotated $(k \times k)$-grids.
 The annotated graph displayed is $(H_4,R_4)$.
 The full annotated graph is a vital instance of \Tc{} recursively created from two types of gadgets (the \textcolor{LavenderFloral}{purple} and \textcolor{DarkBananaYellow}{yellow} pieces and on the level below the \textcolor{AppleGreen}{green} and \textcolor{Persimmon}{orange} pieces respectively). Notably, the treewidth of these graphs grows linearly in the number of layers while their vertex-number grows exponentially in the number of layers.}
 \label{fig:counterexamples}
\end{figure}

\paragraph{The Thilikos-Wiederrecht counterexamples.}
The class $\mathcal{D}$ defined above is precisely the class of annotated graphs $(G,R)$ for which there exists some $k \in \mathbb{N}$ such that $(G,R)$ is a red-minor of the $2$-annotated $(k \times k)$-grid.
This is precisely how Thilikos and Wiederrecht prove the lower bound for their boundary on the applicability of the irrelevant vertex technique for \Tc{} \cite{ThilikosW2026Spanning}.
Within this class they identify a family of examples $\mathcal{H}$ with the properties that each $(H_i,R_i)$ has treewidth $\Omega(\log |H_i|)$ while it is also a vital instance for \Tc{}.
The example in \zcref{fig:counterexamples} illustrates precisely the branching behaviour predicted by our analysis as shown in \zcref{fig:branch}.

\paragraph{A weak insulation lemma.}
The bound on vital instances does not yet identify an
irrelevant vertex in an arbitrary instance.
To bridge this gap, we prove a variant of the
\emph{annulus combing lemma} of Golovach, Stamoulis,
and Thilikos~\cite{GolovachST2023Combing}.
Informally speaking, combing reroutes a solution through a
terminal-free annulus so that, within a central
subannulus and away from its two boundary circles,
it coincides with a union of selected radial paths
called \emph{rails}.
This exposes the connections through which the
solution interacts with the region inside the annulus.
The standard route from combing to irrelevant vertices faces
two related obstacles in our setting:
\begin{description}
    \item[Problem 1: A logarithmic interface.]
    The bound on vital instances initially gives a
    combed solution using
    $\mathbf{poly}(b)\log n$
    rails.
    Thus, the number of connections to be preserved
    inside the annulus may still depend on~$n$.

    \item[Problem 2: Preparing the inner region.]
    Existing irrelevant vertex arguments require a
    sufficiently large wall with additional properties.
    Specifically, \emph{suspension} and \emph{homogeneity} are required.
    Homogeneity ensures that the bounded routing
    configurations available locally, including
    connections involving apices, recur throughout
    the wall.
    Homogenisation can require a polynomial increase in the initial wall order~\cite{GorskySW2026Price}, and finding a suspended wall by the available algorithm has a super-single-exponential dependence on that order
    \cite{SauST2024More}.
    Applying these tools at logarithmic scale would
    exceed our computational budget and require super-linear dependency on $\log |G|$ for our structural bounds.
\end{description}

Our local rerouting lemma resolves the first obstacle.
Once a solution is combed, any three consecutive
occupied rails, together with four suitable portions
of the surrounding cycles, form a mesh after contracting
the relevant contact paths.
Our local rerouting lemma applies to this mesh.
Re-routing reduces the number of crossings by two.
Moreover, the additional portions that turn back towards the
same boundary can be excluded by trimming a constant
number of layers.
Repeating this operation leaves at most two occupied
rails.
Since the initial number of crossings is
$\mathbf{O}_b(\log n)$ and each operation consumes
only a constant number of layers, the total amount
of insulation required remains logarithmic.

This constant-size interface allows us to separate
the two structural scales.
Using the linear grid theorem
\cite{GorskySTW2026Catching} and a variant of the
Flat Wall Theorem whose required wall order is linear
in the desired output order
\cite{GorskySW2025Polynomial}, we find a flat wall~$W$
of order $2^{\mathbf{poly}(b)}\log n$ after deleting
an apex set~$A$ of size $\mathbf{poly}(b)$.
Its outer layers provide the logarithmic insulation,
while a central region of size depending only on~$b$
provides room for suspension and homogenisation.
Once insulated, this region interacts with a solution
through at most two rails and the incidences of the
apices.
After recording these connections, the required rerouting can
be encoded by a rooted minor of detail at most three with~$\mathbf{poly}(b)$ roots.
We can therefore apply the more expensive preparation
tools entirely at a scale depending only on~$b$ with one caveat: the annulus must
avoid the incidences used by the apices in the
solution, which the algorithm does not know.
We address this by constructing~$2|A|+1$ nested
candidate annuli with pairwise disjoint domains,
all surrounding the same central region.
A cycle uses at most two incidences at each apex,
so for every solution at least one candidate annulus
avoids all these incidences and permits combing.
The successful annulus may depend on the solution,
but the central region is fixed.
Inside this region, we find a sufficiently large
suspended and homogeneous wall and select a vertex
that is irrelevant for all the bounded routing
configurations arising from the possible solutions
\cite{CavallaroGKTW2026Optimal}.
Whichever candidate annulus supplies the insulation,
the corresponding solution can consequently be
rerouted to avoid this same vertex.
The vertex is thus irrelevant for the original instance of
\Tc{} and can be found without knowing
a solution.
This completes the passage from the structural
bound on vital instances to the algorithmic rule
in \zcref{thm:MainIrrelevantVertex}.

\paragraph{From logarithmic treewidth to polynomial time.}
The final ingredient is the following single-exponential (deterministic) algorithm for \Tc{} parameterized by treewidth.

\begin{restatable}{theorem}{treewidthdp}
    \label{thm:t-cycle-treewidth}
    Let~$k \geq 1$ be an integer.
    There exists an algorithm that takes as input
    an annotated graph $(G,R)$ where
    $\mathsf{tw}(G) \leq k$ and solves
    \textsc{Terminal Cyclability} in time
    $\mathbf{O}(1783^k \cdot |G|)$.

    Moreover, if a tree-decomposition of width at
    most $k$ for $G$ is supplied as part of the input,
    the algorithm runs in time
    $\mathbf{O}(36^k \cdot k \cdot |G|)$.
\end{restatable}

The algorithm uses dynamic programming over a
tree-decomposition.
If implemented in a na\"ive way, the dynamic program has~$k^{\mathbf{O}(k)}$ entries.
We instead keep representative families using the framework of Bodlaender, Cygan, Kratsch, and Nederlof~\cite{BodlaenderCyganKratschNederlof2015}.
This reduces the dependence on the treewidth~$k$ to single exponential and this improvement is essential to the final running-time bound.
Repeatedly applying \zcref{thm:MainIrrelevantVertex}
either deletes a non-empty irrelevant set or supplies
a tree-decomposition of width~$k\in 2^{\mathbf{poly}(b)}\log n$.
The single-exponential dynamic program then runs in~$n^{2^{\mathbf{poly}(b)}}$ time, giving the claimed XP-algorithm.
By comparison, the direct $k^{\mathbf{O}(k)}$ implementation would yield
only a quasi-polynomial bound for fixed~$b$.

We mention that we may assume to be dealing with a $K_{\mathbf{poly}(\mathsf{bidim}(G,R))}$-minor-free graph due to our preprocessing and that many problems are known to admit single-exponential parameterized algorithms when parameterized by treewidth on $H$-minor-free graphs.
However, we are not aware of an existing proof that \Tc{} admits such an algorithm.

\section{Preliminaries}
\label{sec:prelim}
We introduce basic concepts as well as more advanced ideas from the toolkit of graph minors.

\subsection{Basics}
We open with a short array of definitions for commonly used concepts and notation.
By $\mathbb{N}$ we denote the set of non-negative integers.
Given any two integers $a,b\in\mathbb{N}$, we write $[a,b]$ for the set $\{z\in\mathbb{N} \mid a\leq z\leq b\}.$
Notice that the set $[a,b]$ is empty whenever $a>b.$
For any positive integer $c$ we set $[c]\coloneqq [1,c].$
We denote the multiset union by~$\uplus$.
For a graph~$G$, we denote the set of vertices by~$V(G)$ and the set of edges by~$E(G)$.
We use the shorthands~$|G| = |V(G)|$ and~$|\!|G|\!| = |V(G)| + |E(G)|$.

For a set~$V$, we denote by~$\binom{V}{k}$ the set of all subsets of~$V$ of size exactly~$k$ and by~$2^V$ the power set of $V$, that is, the set of all subsets of~$V$.
For a set~$U$, let~$\pema(U)$ denote the set of perfect matchings on~$U$, that is, the set of all subsets of~$\binom{U}{2}$ in which every element of~$U$ appears exactly once.
For a graph~$G$ and a collection~$\mathcal{H}$ of subgraphs of $G$, we denote by $V(\mathcal{H})$ the set $\bigcup_{H \in \mathcal{H}}V(H)$ and by~$E(\mathcal{H})$ the set $\bigcup_{H \in \mathcal{H}}E(H)$.

\paragraph{Treewidth.}
A \emph{tree-decomposition} for a graph $G$ is a pair $(T,\beta)$ such that $T$ is a tree, ${\beta\colon V(T)\to 2^{V(G)}}$ assigns to each node of $T$ a subset of the vertices of $G$ known as a \emph{bag}, $\bigcup_{t\in V(T)}G[\beta(t)] = G$, and for each $v\in V(G)$, the set $\{ t\in V(T)  \mid  v\in\beta(t) \}$ is connected.
The \emph{width} of $(T,\beta)$ is defined as $\max_{t\in V(T)} |\beta(t)|-1$.
The \emph{treewidth} of a graph $G$, denoted by $\mathsf{tw}(G)$, is the smallest integer $k$ such that $G$ has a tree-decomposition of width at most~$k$.

We use rooted \emph{nice tree-decompositions} whose nodes have one of the following types.
A \emph{leaf node}~$x$ has a bag~$\beta(x)=\{v\}$ and introduces~$v$.
An \emph{introduce-vertex node}~$x$ has one child~$y$ and~$\beta(x)=\beta(y)\cup\{v\}$ for some~$v\notin \beta(y)$.
An \emph{introduce-edge node}~$x$ has one child~$y$, satisfies~$\beta(x)=\beta(y)$, and introduces one edge with both endpoints in~$\beta(x)$.
A \emph{forget-vertex node}~$x$ has one child~$y$ and~$\beta(x)=\beta(y) \setminus\{v\}$ for some~$v\in \beta(y)$.
Finally, a \emph{join node}~$x$ has two children~$y,z$ with~$\beta(x)=\beta(y)=\beta(z)$.
We require every edge to be introduced exactly once.

Unlike the usual definition, we allow the root bag to contain two vertices.
The standard refinement of tree-decompositions shows that, given a tree-decomposition of width~$k \geq 1$ and a set~$X$ contained in one of its bags, one can construct in~$|\!|G|\!| \cdot \poly(k)$ time a nice tree decomposition of the same width, with root bag~$X$ and~$\mathbf{O}(k|G|)$ nodes.
See, for example, Bodlaender, Cygan, Kratsch, and Nederlof~\cite{BodlaenderCyganKratschNederlof2015}.

\paragraph{Red-minors and bidimensionality.}
Let $H$ and $G$ be graphs.
A \emph{minor model} of $H$ in $G$ is a family of connected subgraphs $\{ G_v\}_{v \in V(H)} \cup \{ G_e\}_{e \in E(H)}$ of $G$ such that
\begin{enumerate}
    \item $V(G_v) \cap V(G_u) = \emptyset$ for all $u \neq v \in V(H)$,
    \item $|V(G_u) \cap V(G_{uv})| = 1$ for all $uv \in E(H)$ and the intersection vertex is an endpoint of~$G_{uv}$,
    \item $V(G_u) \cap V(G_{e}) = \emptyset$ if $u \notin e$ for all $u \in V(H)$ and $e \in E(H)$,
    \item $G_e$ is a path in $G$ for every $e \in E(H)$, and
    \item $G_{e}$ and $G_{e'}$ are internally vertex-disjoint for all $e \neq e' \in E(H)$.
\end{enumerate}
In case $G_e$ is isomorphic to $K_2$ we do not require $G_e$ to be part of the model explicitly.
This also means that $\{ G_v\}_{v \in V(H)}$ may be a minor model of $H$ in $G$ if for all $uv \in E(H)$, there is an edge in~$G$ between $V(G_u)$ and $V(G_v)$.
Most of the time, we will be working with this simpler version of a model, but sometimes it will be convenient to provide explicit models for edges.
In particular, all clique-minor models considered below are represented
solely by their branch sets, with an edge between every two distinct
branch sets.

An annotated graph $(H,B)$ is a \emph{red-minor} of an annotated graph $(G,R)$ if there exists a minor-model~$\{ G_v\}_{v\in V(H)}$ of~$H$ in~$G$ such that $R \cap V(G_v) \neq \emptyset$ for all $v\in B$.
The \emph{bidimensionality} of an annotated graph $(G,R)$, denoted by $\mathsf{bidim}(G,R)$, is the largest integer $k$ such that there exists~$(\Gamma_k,V(\Gamma_k))$ as a red-minor in $(G,R)$ where $\Gamma_k$ denotes the $(k \times k)$-grid.
If~$R = \emptyset$, then we set~$\mathsf{bidim}(G,R) = 0$.
We say that a subgraph of~$G$ is blank if it does not contain any vertices in~$R$.

\paragraph{Paths, linkages, and vitality.}
A \emph{linkage} $\mathcal{L}$ in a graph $G$ is a set of pairwise vertex-disjoint paths.
Given the set $T \subseteq V(G)$ of the endpoints of the paths in $\mathcal{L}$, we say that $\mathcal{L}$ is a~\emph{$T$-linkage}.
Based on~$T$, we define the \emph{pattern of $\mathcal{L}$} as $\pi(\mathcal{L}) \coloneqq \{\{s,t\}  \mid  \text{ there is }L\in \LLL \text{ with endpoints } s \text{ and } t\}$.
In some cases, we are already given a set $\mathcal{T} = \{ (s_i,t_i)  \mid  i\in[k]\}$ of terminal pairs.
In such a situation, we say that a linkage $\mathcal{L}$ is a \emph{$\mathcal{T}$-linkage} if $|\mathcal{L}| = k$ and for every $i\in[k]$ there is a path $L \in \mathcal{L}$ with endpoints $s_i$ and $t_i$.
In \textsc{Disjoint Paths}, one is given as input a graph~$G$ and a set~$T$ of terminal pairs and the question is whether a~$T$-linkage exists.
Given a $T$-linkage~$\mathcal{L}$ in a graph $G$ we say that~$\mathcal{L}$ is \emph{vital} if $V(\mathcal{L}) = V(G)$ and $\mathcal{L}$ is the unique $T$-linkage in $G$ with its pattern.

We say that a path $P$ in $G$ is \emph{internally disjoint} from a set $X \subseteq V(G)$ if $V(P) \cap X$ does not contain any vertex of $P$ that is not an endpoint vertex.
Given a graph $G$ and two subsets $A, B \subseteq V(G)$, an \emph{$A$-path} in $G$ is a path with both endpoints in $A$ and internally disjoint from $A$, and an \emph{$A$-$B$-path} is a path with one endpoint in $A$, the other in $B$, and internally disjoint from $A \cup B$.
An \emph{$A$-$B$-linkage} in $G$ is a linkage consisting of $A$-$B$-paths.
If $H$ is a subgraph of $G$, an \emph{$H$-path} is a $V(H)$-path of length at least one with no edge in $E(H)$.

We will make heavy use of the following minor-testing algorithm.

\begin{proposition}[Cavallaro, Gorsky, Kreutzer, Thilikos, and Wiederrecht \cite{CavallaroGKTW2026Optimal}]\label{prop:FindMinor}
There exists an algorithm that takes as input two graphs $H$ and $G$ and either correctly decides that $G$ does not contain $H$ as a minor or outputs a minor model of $H$ in $G$ in time $2^{ 2^{\mathbf{poly}(|H|)} } \cdot |G|^2$.
\end{proposition}

\paragraph{Surfaces.}
By a \emph{surface} we mean a compact $2$-dimensional manifold $\Sigma$ with or without boundary.
Let $X \subseteq \Sigma$.
The boundary and interior of $X$ will be denoted $\mathsf{bd}(X)$ and $\mathsf{int}(X)$, respectively.
The topological closure of $X$ is denoted by $\mathsf{cl}(X)$.

Let $\gamma \sth [0,1]_{\mathbb{R}}\to \Sigma$ be continuous and injective on $[0,1[_{\mathbb{R}}$, such that either $\gamma(1) = \gamma(0)$ or~$\gamma(1) \notin \gamma([0,1[_{\mathbb{R}})$.
We say that the image $\gamma([0,1]_{\mathbb{R}})$ of $\gamma$ is a \emph{curve} in $\Sigma$ and $\gamma(0),\gamma(1)$ are its \emph{endpoints}.
A curve is \emph{closed} if its endpoints agree.

\subsection{Tools more specific to graph minor structure theory}\label{subsec_grpahMinors}
We continue by briefly introducing some key concepts for handling the general structure of $H$-minor-free graphs.
The definitions we introduce here originate from the work of Robertson and Seymour~\cite{RobertsonS1999GraphMinorsXVII}.
They were later refined by Kawarabayashi, Thomas, and Wollan \cite{KawarabayashiTW2021Quickly} and then adopted and reiterated by several authors (see for example \cite{ThilikosW2024Excluding, PaulPTW2024Obstructionsa, PaulPTW2025Local, GorskySW2025Polynomial}).

\paragraph{Brambles.}
Let $G$ be a graph.
A \emph{bramble} in $G$ is a collection of connected subgraphs $\mathcal{B}$ of $G$ such that for all $B_1,B_2 \in \mathcal{B}$ either $V(B_1) \cap V(B_2) \neq \emptyset$ or there exists an edge with one end in $B_1$ and the other in $B_2$.
A \emph{hitting set} for a bramble $\mathcal{B}$ is a set $S \subseteq V(G)$ such that $S \cap V(B) \neq \emptyset$ for all~$B \in \mathcal{B}$.
The \emph{order} of a bramble $\mathcal{B}$ is the minimum size of a hitting set for $\mathcal{B}$.

\begin{proposition}[Reed \cite{Reed1997Tree}]\label{prop_brambles}
Let $k \geq 0$ be an integer and $G$ be a graph.
If $G$ has a bramble of order $k+1$, then $\mathsf{tw}(G) \geq k$.
\end{proposition}

\paragraph{Separations and tangles.}
Let $G$ be a graph and $k$ be a positive integer.
A separation of $G$ is a pair $(A,B)$ such that $A,B \subseteq V(G)$, $A \cup B = V(G)$, and there is no edge with one end in $A \setminus B$ and the other in $B \setminus A$.
The \emph{order} of a separation $(A,B)$ of $G$ is defined as $|A \cap B|$.
We denote by~$\mathcal{S}_k(G)$ the collection of all separations $(A,B)$ of order less than $k$ in $G$.
An \emph{orientation} of $\mathcal{S}_k(G)$ is a set $\mathcal{O}$ such that for all $(A,B)\in\mathcal{S}_k(G)$ exactly one of $(A,B)$ and $(B,A)$ belongs to $\mathcal{O}$. 
A \emph{tangle} of order $k$ in $G$ is an orientation $\mathcal{T}$ of $\mathcal{S}_k(G)$ such that for all $(A_1,B_1),(A_2,B_2),(A_3,B_3)\in\mathcal{T}$, it holds that $G[A_1]\cup G[A_2]\cup G[A_3]\neq G$.
If $\mathcal{T}$ is a tangle and $(A,B)\in\mathcal{T}$ we call $A$ the \emph{small side} and $B$ the \emph{big side} of $(A,B)$.
Let $G$ be a graph and $\mathcal{T}$ and $\mathcal{D}$ be tangles of $G$.
We say that $\mathcal{D}$ is a \emph{truncation} of $\mathcal{T}$ if $\mathcal{D}\subseteq\mathcal{T}$.

Let $G$ and $H$ be graphs as well as $\mathcal{T}$ be a tangle in $G$.
We say that a minor-model $\mu$ of $H$ in~$G$ is \emph{controlled} by $\mathcal{T}$ if there does not exist a separation $(A,B)\in\mathcal{T}$ of order less than $|V(H)|$ and a vertex~$x \in V(H)$ such that $V(\mu(x))\subseteq A\setminus B$.
In particular, let $t \geq 2$ and $\mu$ be a minor model of~$K_t$, then for every separation $(A,B)$ of order less than $t$ there is some $v \in V(K_t)$ such that $\mu(v)$ is contained in either $A \setminus B$ or $B \setminus A$.
Moreover, for $X \neq Y \in \{ A,B \}$, if $V(\mu(v)) \subseteq X \setminus Y$, then there is no $u \in V(K_t)$ with $V(\mu(u)) \subseteq Y \setminus X$.
Indeed,
\begin{align*}
\mathcal{T}_{\mu} \coloneqq \big\{ (A,B) \in \mathcal{S}_{\lfloor \nicefrac{2}{3} \cdot t \rfloor}(G) \mid  \text{ there is } v \in V(K_t) \text{ with } V(\mu(v)) \subseteq B\setminus A \big\}
\end{align*}
is known to be a tangle and usually referred to as the \emph{tangle induced by $\mu$}~\cite{RobertsonS1991GraphMinorsX,BruhnUlmer2022Packing}.

If~$\mathcal{T}$ has order at least $1$, then we denote by the \emph{$\mathcal{T}$-big component} of~$G$ the unique component~$J$ of~$G$ such that $V(J) \subseteq B$ for all $(A,B) \in \mathcal{T}$ where~$A \cap B = \emptyset$.
We call the \((\mathcal T-A)\)-big component of \(G-A\) the \(\mathcal T\)-big component of \(G-A\).

Let $\mathcal{T}$ be a tangle of order $k \geq 2$ in a graph $G$ and let $A \subseteq V(G)$ be a set with $|A| < k-1$.
We denote by
\begin{align*}
    \mathcal{T} - A \coloneqq \big\{  (X,Y) \in \mathcal{S}_{k - |A|}(G-A)  \mid (X \cup A , Y \cup A) \in \mathcal{T}  \big\}
\end{align*}
the tangle of $G-A$ \emph{inherited} from $\mathcal{T}$.

\paragraph{Meshes.}
Let $n,m$ be integers with $n,m\geq 2$.
A \emph{$(n\times m)$-mesh} is a graph $M$ which is the union of paths $M=P_1\cup\ldots\cup P_n\cup Q_1\cup \ldots \cup Q_m$ where
    \begin{itemize}
        \item $P_1,\ldots,P_n$ are pairwise vertex-disjoint, and $Q_1,\ldots,Q_m$ are pairwise vertex-disjoint.
        \item for every $i\in [n]$ and $j\in [m]$, the intersection $P_i\cap Q_j$ induces a path,
        \item each $P_i$ is a  $V(Q_1)$-$V(Q_m)$-path intersecting the paths $Q_1,\ldots Q_m$ in the given order, and each $Q_j$ is a $V(P_1)$-$V(P_n)$-path intersecting the paths $P_1,\ldots, P_n$ in the given order.
    \end{itemize}
We say that the paths $P_1,\ldots,P_n$ are the \emph{horizontal paths}, and the paths $Q_1,\ldots,Q_m$ are the \emph{vertical paths}.
The union $P_{1} \cup P_{n} \cup Q_{1} \cup Q_{m}$ is a cycle called the \emph{perimeter} of $M$.
The unique cycle in the union $P_{i} \cup P_{i+1} \cup Q_{j} \cup Q_{j+1}$, where $i \in [n - 1]$ and $j \in [m - 1]$, is called a \emph{brick} of $M$.
A mesh $M'$ is a \emph{submesh} of a mesh $M$ if every horizontal (vertical) path of $M'$ is a subpath of a horizontal (vertical) path of~$M$, respectively.
We write \emph{$n$-mesh} as a shorthand for an $(n \times n)$-mesh.

Let $r \in \mathbb{N}$ with $r\geq 3$, let $G$ be a graph, and $M$ be an $r$-mesh in $G$.
Let $\mathcal{T}_M$ be the orientation of~$\mathcal{S}_r(G)$ such that for every $(A,B)\in\mathcal{T}_M$, the set $B\setminus A$ contains the vertex set of both a horizontal and a vertical path of $M$.
Then, $\mathcal{T}_M$ is the tangle \emph{induced} by $M$.
If $\mathcal{T}$ is a tangle in $G$, we say that $\mathcal{T}$ \emph{controls} the mesh $M$ if $\mathcal{T}_M$ is a truncation of $\mathcal{T}$.
A minor model is controlled by \(M\) if it is controlled by \(\mathcal T_M\).

The significance of meshes lies in the fact that every mesh has large treewidth and, reversely, every graph of large treewidth must contain a big mesh.
This theorem is originally due to Robertson and Seymour \cite{RobertsonS1986GraphMinorsV}.
The variant of this theorem we state here guarantees a linear relation between the size of the grid minor and the treewidth in $H$-minor-free graphs. 

Often a special case of meshes known as ``walls'' is used in the literature.
Let $n,m \in \mathbb{N}$ be two positive integers.
The \emph{$(n \times m)$-grid} is the graph $G$ with the vertex set $V(G) = [n] \times [m]$ and the edges
\begin{align*}
E(G) = \big\{ \{ (i, j) , (\ell , k) \} \mid    & i, \ell \in [n], \ j,k \in [m], \text{ and } \ \\
 & ( |i - \ell| = 1  \text{ and } j = k ) \text{ or } ( |j - k| = 1 \text{ and } i = \ell ) \big\} .
\end{align*}
An \emph{$n$-grid} is an $(n \times n)$-grid.

The \emph{elementary $(n \times m)$-wall} is in turn derived from the $(n \times 2m)$-grid by deleting all edges in the following set
\begin{align*}
\big\{ \{ (i, j) , (i+1 , j) \} \mid i \in [n - 1], \ j \in [2m], \text{ and } {i \not\equiv j \mod{2} } \big\}
\end{align*}
and removing all vertices of degree at most 1 in the resulting graph.
An \emph{$(n \times m)$-wall} is a subdivision of the elementary $(n \times m)$-wall.
An \emph{$n$-wall} is an $(n \times n)$-wall.

It is relatively easy to see that every $(n \times m)$-wall is also an $(n \times m)$-mesh for each~$n,m \geq 2$.
Conversely, every~$(2n \times 2m)$-mesh contains a $(n \times m)$-wall as a subgraph.

\begin{proposition}[Gorsky, Stamoulis, Thilikos, and Wiederrecht \cite{GorskySTW2026Catching}]\label{prop:LinearGrid}
There exists a function~$f_{\ref{prop:LinearGrid}}\colon \mathbb{N}^3 \rightarrow \mathbb{N}$ and an algorithm that, given an integer~$k \geq 2$, a $t$-vertex graph~$H$ of Euler genus $g_H$, and an $H$-minor-free graph $G$, outputs in $2^{\poly(t)} \cdot |G|^{\mathbf{O}(1)}$ time either a $(k \times k)$-mesh in~$G$ or a tree-decomposition of $G$ of width~$f_{\ref{prop:LinearGrid}}(t,g_H,k) \in \mathbf{O}(g_H \cdot k + (g_H + 1) \cdot \poly(t))$.
\end{proposition}

\paragraph{Paintings in surfaces.}
A \emph{painting} in a surface $\Sigma$ is a pair $\Gamma = (U,N)$, where $N \subseteq U \subseteq \Sigma$, $N$ is finite, $U$ is closed, $U \setminus N$ has a finite number of arcwise-connected components, called \emph{cells} of $\Gamma$, and for every cell $c$, the closure of $c$, denoted by $\overline{c}$, is a closed disc where $N_\Gamma(c) \coloneqq \overline{c} \cap N \subseteq \mathsf{bd}(\overline{c})$.
If $|N_\Gamma(c)| \geq 4$, the cell $c$ is called a \emph{vortex}.
We further let $N(\Gamma) \coloneqq N$, let $U(\Gamma) \coloneqq U$, and let $C(\Gamma)$ be the set of all cells of $\Gamma$.

Any given painting $\Gamma = (U,N)$ defines a hypergraph with $N$ as its vertices and one hyperedge \(N_\Gamma(c)\) for each cell \(c\in C(\Gamma)\).
Accordingly, we call $N$ the \emph{nodes} of $\Gamma$.

\paragraph{$\Sigma$-renditions.}
Let $G$ be a graph and $\Sigma$ be a surface.
A \emph{$\Sigma$-rendition} of $G$ is a triple $\rho = (\Gamma, \sigma, \pi)$, where
\begin{itemize}
    \item $\Gamma$ is a painting in $\Sigma$,
    \item for each cell $c \in C(\Gamma)$, $\sigma(c)$ is a subgraph of $G$, and
    \item $\pi \colon N(\Gamma) \to V(G)$ is an injection,
\end{itemize}
such that
\begin{description}
    \item[R1] $G = \bigcup_{c \in C(\Gamma)}\sigma(c)$,
    \item[R2] for all distinct $c,c' \in C(\Gamma)$, the graphs $\sigma(c)$ and $\sigma(c')$ are edge-disjoint,
    \item[R3] $\pi(N_\Gamma(c)) \subseteq V(\sigma(c))$ for every cell $c \in C(\Gamma)$, and
    \item[R4] for every cell $c \in C(\Gamma)$, we have $V(\sigma(c) \cap \bigcup_{c' \in C(\Gamma) \setminus \{ c \}} (\sigma(c'))) \subseteq \pi(N_\Gamma(c))$.
\end{description}
We write $N(\rho)$ for the set $N(\Gamma)$, let $N_\rho(c) = N_\Gamma(c)$ for all $c \in C(\Gamma)$, and similarly, we lift the set of cells from $C(\Gamma)$ to $C(\rho)$.
If it is clear from the context which $\rho$ is meant, we will sometimes simply write $N(c)$ instead of $N_\rho(c)$, and if the $\Sigma$-rendition $\rho$ for $G$ is understood from the context, we usually identify the sets $\pi(N(\rho))$ and $N(\rho)$ along $\pi$ for ease of notation.

\paragraph{Blank renditions.}
Let $\rho$ be a $\Sigma$-rendition of an annotated graph $(G,R)$.
If $$\pi(N(\rho)) \cup \bigcup \{ V(\sigma(c)) \mid c \in C(\rho) \text{ and } c \text{ is not a vortex}\}$$ is disjoint from $R$, we call $\rho$ a \emph{blank rendition (of $(G,R)$)}.

\paragraph{Societies.}
Let $\Omega$ be a cyclic ordering of the elements of some set which we denote by $V(\Omega)$.
A \emph{society} is a pair $(G,\Omega)$, where $G$ is a graph and $\Omega$ is a cyclic ordering with $V(\Omega)\subseteq V(G)$.
For a given set $S \subseteq V(\Omega)$ a vertex $s \in S$ is an \emph{endpoint} of $S$ if there exists a vertex $t \in V(\Omega) \setminus S$ that immediately precedes or succeeds $s$ in $\Omega$.
We call $S$ a \emph{segment} of $\Omega$ if $S$ is a set of consecutive vertices in the cyclic ordering~$\Omega$.

Let $(G,\Omega)$ be a society and let $\Sigma$ be a surface with one boundary component $B$ homeomorphic to the unit circle.
A \emph{rendition} of $(G,\Omega)$ in $\Sigma$ is a $\Sigma$-rendition $\rho$ of $G$ such that the image under $\pi_{\rho}$ of~$N(\rho) \cap B$ is $V(\Omega)$ and $\Omega$ is one of the two cyclic orderings of $V(\Omega)$ defined by the way the points of~$\pi_{\rho}^{-1}(V(\Omega))$ are arranged in the boundary $B$.

\paragraph{Traces of paths and cycles.}\SW{add an illustration explaining traces of paths and cycles in a rendition}
Let $\rho$ be a $\Sigma$-rendition of a graph $G$.
For every cell $c \in C(\rho)$ with $|N_\rho(c)| = 2$, we select one of the components of $\mathsf{bd}(c) - N_\rho(c)$.
This selection will be called a \emph{tie-breaker in $\rho$}, and we assume that every rendition comes equipped with a tie-breaker.

Let $G$ be a graph and $\rho$ be a $\Sigma$-rendition of $G$.
Let $Q$ be a cycle or path in $G$ that uses no edge of~$\sigma(c)$ for every vortex $c \in C(\rho)$.
We say that $Q$ is \emph{grounded} if either $Q$ is a path of length at least one with both endpoints in~$N(\rho)$, or $Q$ is a cycle that uses edges of $\sigma(c_{1})$ and $\sigma(c_{2})$ for two distinct cells $c_{1}, c_{2} \in C(\rho).$
If~$Q$ is grounded we define the \emph{trace} of $Q$ as follows.
Let~$P_1,\dots,P_k$ be distinct nontrivial maximal subpaths of~$Q$ such that $P_i$ is a subgraph of $\sigma(c)$ for some cell $c$.
Fix~$i \in [k]$.
The maximality of $P_i$ implies that its endpoints are $\pi(n_1)$ and $\pi(n_2)$ for distinct nodes $n_1,n_2 \in N(\rho)$.
If $|N_\rho(c)| = 2$, let $L_i$ be the closure of the component of $\mathsf{bd}(c) - \{ n_1,n_2 \}$ selected by the tie-breaker, and if $|N_\rho(c)| = 3$, let $L_i$ be the closure of the component of $\mathsf{bd}(c) - \{ n_1,n_2 \}$ that is disjoint from $N_\rho(c)$.
We define $L_i'$ by pushing $L_i$ slightly so that it is disjoint from all cells in $C(\rho)$, while maintaining that the resulting curves intersect only at a common endpoint.
The \emph{trace} of $Q$ is defined to be the curve $\gamma$ obtained from $\bigcup_{i\in[k]} L_i'$ by slightly pushing it in order to ensure that $\gamma$ intersects the painting of $\rho$ in nodes only.
If $Q$ is a cycle, its trace is therefore the homeomorphic image of the unit circle, and otherwise, it is an arc in $\Sigma$ with both endpoints in $N(\rho)$.

\paragraph{Aligned discs and grounded subgraphs.}
Let $G$ be a graph and let $\rho = (\Gamma, \sigma, \pi)$ be a $\Sigma$-rendition of $G$. 
We say that a 2-connected subgraph $H$ of $G$ is \emph{grounded (in $\rho$)} if every cycle in $H$ is grounded and no vertex of $H$ is drawn by $\Gamma$ in a vortex of $\rho$.
A disc in $\Sigma$ is called \emph{$\rho$-aligned} if its boundary only intersects $\Gamma$ in nodes.
If $H$ is planar, we say that it is \emph{flat in $\rho$} if there exists a $\rho$-aligned disc~$\Delta \subseteq \Sigma$ which contains all cells $c \in C(\rho)$ with $E(\sigma(c)) \cap E(H) \neq \emptyset$ and $\Delta$ does not contain any vortices of $\rho$.

For any $\rho$-aligned disc $\Delta$, we call the subgraph of $G$ that is drawn by $\Gamma$ onto $\Delta$ the \emph{crop of $G$ by $\Delta$ (in $\rho$)}.
Furthermore, the \emph{restriction $\rho'$ of $\rho$ by $\Delta$} is defined as the $\Delta$-rendition that consists of the restriction of~$\Gamma$, $\sigma$, and $\pi$ to $\Delta$.

This allows us to define a society associated to $\Delta$ as follows.
Let $V(\Omega_{\Delta})$ be the set of all vertices whose corresponding nodes are drawn in the boundary of~$\Delta$ and let $\Omega_{\Delta}$ be the cyclic ordering of~$V(\Omega_{\Delta})$ obtained by traversing along the boundary of $\Delta$ in the anticlockwise direction.
Now, let~$G_{\Delta}$ be the crop of $G$ by $\Delta$.
We call the society $(G_{\Delta}, \Omega_{\Delta})$ the \emph{$\Delta$-society (in $\rho$)}.
If $\rho$ is clear from the context, we do not mention it.
We also call the restriction of $\rho$ by $\Delta$, the \emph{restriction of $\rho$ to~$(G_{\Delta}, \Omega_{\Delta})$.}

Let $\rho$ be a $\Sigma$-rendition of a graph~$G$, where~$\Sigma$
is connected. If~$C$ is a grounded cycle whose trace bounds a
closed disc~$\Delta'\subseteq\Sigma$, we call~$\Delta'$ a
\emph{$C$-disc} in~$\rho$.
If~$\Sigma$ is the sphere, then there are two discs bounded by the trace of~$C$.
If the trace is contained in a
specified disc, we choose its $C$-disc inside that disc; otherwise, we choose one of the two discs arbitrarily.

\paragraph{Flat Meshes.}
Let $n \ge 2$ be an integer.
Let $G$ be a graph, and let $M \subseteq G$ be an $n$-mesh.
We say that $M$ is a \emph{flat mesh} in $G$ if there exists a sphere-rendition $\rho$ of $G$ with a unique vortex $c_0$ such that $M$ is grounded and flat in $\rho$, and the closed disc $\Delta_M$ bounded by the trace of the perimeter of $M$
whose interior avoids $c_0$ contains every node corresponding to a vertex of $M$.
For this rendition, the perimeter-disc of $M$ is always chosen to be $\Delta_M$.
We say that $\rho$ \emph{witnesses} the flatness of $M$.

\paragraph{Inner and outer graphs of cycles and aligned curves.}
Let $G$ be a graph with a $\Sigma$
-rendition $\rho$ and let $\gamma$ be a (closed) curve in $\Sigma$.
We say that $\gamma$ is \emph{$\rho$-aligned} if it intersects $\rho$ in nodes only.
That is, $\gamma$ must be disjoint from all cells of $\rho$.\SW{add an illustration of a rendition with some cycle, its inner and outer graph, and some aligned curves}
Further, let $C$ be a grounded cycle or a $\rho$-aligned closed curve whose trace -- or~$C$ itself, if it is a closed curve -- bounds a disc $\Delta_C$ and the $\Delta_C$-society $(G', \Omega').$
If~$C$ is a grounded cycle, then we call $G' \cup C$ the \emph{inner graph of $C$ (in $\rho$)} and call $G'$ itself the \emph{proper inner graph of $C$ (in $\rho$).}
Let $B = \pi(N(\rho) \cap \mathsf{bd}(\Delta_C)).$
We define the \emph{proper outer graph of $C$ (in $\rho$)} as $G'' \coloneqq G[B \cup (V(G) \setminus V(G'))]$ and call $G'' \cup C$ the \emph{outer graph of $C$ (in $\rho$).} 
If~$C$ is a~$\rho$-aligned closed curve, then we call~$G'$ the inner graph of~$C$ and~$G''$ the outer graph of~$C$.

\paragraph{Transactions and nests.}
Let $(G, \Omega)$ be a society. 
A \emph{transaction} in $(G, \Omega)$ is an $A$-$B$-linkage for disjoint segments $A, B$ of $\Omega$ consisting of $V(\Omega)$-paths. 

Let~$\rho$ be a $\Sigma$-rendition of a graph~$G$.
Further, let $\mathcal{C} = \langle C_1,\ldots,C_s\rangle$ be a sequence of pairwise disjoint cycles in $G,$ such that each of them is grounded in $\rho$ and we have $\Delta_1 \subsetneq \ldots \subsetneq \Delta_s \subseteq \Sigma,$ where $\Delta_i$ is the $C_i$-disc for each $i \in [s]$.
We say that the cycles in $\mathcal{C}$ are \emph{concentric (in $\rho$)} and call $C_1$ the \emph{inner} cycle and $C_s$ the \emph{outer cycle} of $\mathcal{C}$ in $\rho$.\SW{add illustration for cylindrical renditions and nests with radial linkages} 

Let now~$\rho$ be a rendition of a society~$(G,\Omega)$.
A \emph{nest $\mathcal{C}$ (in $\rho$)} is a set of concentric cycles with the inner cycle $C$ such that every vortex of $\rho$ is found in the $C$-disc.
The order of a nest is its number of cycles.

\paragraph{Depth of vortices.}
Let $G$ be a graph and $\rho$ be a $\Sigma$-rendition of $G$ with a vortex cell $c_0.$
Notice that $c_0$ defines a society $(\sigma(c_0), \Omega_{c_0})$, where $V(\Omega_{c_0})$ is the set of vertices of $G$ corresponding to~$N_\rho(c_0).$
The ordering $\Omega_{c_{0}}$ is obtained by traversing along the boundary of the closure of $c_0$ in anti-clockwise direction.
We call $(\sigma(c_0),\Omega_{c_0})$ as obtained above the \emph{vortex society} of $c_0.$

We define the \emph{depth} of a society $(G, \Omega)$ as the maximum cardinality of a transaction in $(G, \Omega)$.
The \emph{depth} of the vortex $c_{0}$ is thereby defined as the depth of its vortex society.

Given a $\Sigma$-rendition $\rho$, we define the \emph{breadth of $\rho$} as the number of vortex cells of $\rho$ and the \emph{depth of $\rho$} as the maximum depth of its vortex societies, or zero if there are no vortices.

\subsection{The structure of annotated graphs of small bidimensionality}\label{subsec_redMinors}

We use the following consequence of the structure theorem of Gorsky, Protopapas, and Wiederrecht~\cite{GorskyPW2026Quickly} that establishes polynomial bounds for the structure of annotated graphs of small bidimensionality.
We slightly modify the original statement and use a refinement of its first outcome described by Cavallaro, Gorsky, Kreutzer, Thilikos, and Wiederrecht~\cite{CavallaroGKTW2026Optimal}.

\begin{proposition}[Gorsky, Protopapas, and Wiederrecht \cite{GorskyPW2026Quickly}]\label{prop:localstructure}
There exist functions $\mathsf{apex}_{\ref{prop:localstructure}},\mathsf{depth}_{\ref{prop:localstructure}} \colon \mathbb{N}^3 \to \mathbb{N}$ and $\mathsf{mesh}_{\ref{prop:localstructure}} \colon \mathbb{N}^4 \to \mathbb{N}$ such that for all integers $k,t,r \geq 1$, and $w \geq 3$, every annotated graph $(G,R)$ with $\mathsf{bidim}(G,R) \leq r$, and every $\mathsf{mesh}_{\ref{prop:localstructure}}(r, k, t, w)$-mesh $M \subseteq G$ one of the following holds.
\begin{enumerate}

 \item $(G,R)$ has a blank $K_{\lfloor \nicefrac{5t}{2} \rfloor}$-minor model $\psi$ and a separation $(X,Y)$ of order at most $t-1$ such that the $\mathcal{T}_\psi$-big component of $G - (X \cap Y)$ is blank,

 \item $(G, R)$ has a red $K_t$-minor model controlled by $M$, or
 \item there exists a set $A \subseteq V(G)$ with $|A| \leq \mathsf{apex}_{\ref{prop:localstructure}}(k, t, r)$ and a surface $\Sigma$ of Euler genus less than $9t^2$ such that $(G - A, R \setminus A)$ has a blank $\Sigma$-rendition $\rho$ with breadth at most~$\nicefrac{3}{2}(t-1)(3t-4) + (3r+2)(3r+1)-3$, depth at most $\mathsf{depth}_{\ref{prop:localstructure}}(k, t, r)$, and there exists a vortex-free, $\rho$-aligned disc $\Delta \subseteq \Sigma$ such that the restriction of~$\rho$ to~$\Delta$ contains a~$w$-wall that is grounded and flat in~$\rho$.
 Moreover, if we denote by $\mathcal{V}$ the vortices of $\rho$, then for every $v \in \mathcal{V}$ there exists a $\rho$-aligned disc $\Delta_{v}$ such that
 \begin{itemize}
    \item $\Delta_v$ contains~$v$ and is disjoint from $\Delta$ and from $\Delta_{v'}$ for all $v' \in \mathcal{V}\setminus\{ v\}$, and 
    \item there exists a nest of order $k$ in the restriction of $\rho$ to $\Delta_v$.
 \end{itemize}
\end{enumerate}
Moreover, it holds that $\mathsf{apex}_{\ref{prop:localstructure}}(k,t,r), \mathsf{depth}_{\ref{prop:localstructure}}(k,t,r) \in \poly(k+t+r)$ and $\mathsf{mesh}_{\ref{prop:localstructure}}(r,k,t,w) \in \poly(k+t+r) + \mathbf{O}(t^2w)$.

There also exists an algorithm that, given $t$, $k$, $r$, $w$, an annotated graph~$(G,R)$, and a mesh $M$ as above as input, finds one of these outcomes in time $\poly(t+k+r+w)|\!| G |\!|^3$.\SW{add an illustration}
\end{proposition}

\section{Preparing the toolkit}\label{sec:Preparations}
We require some additional tools that can be imported almost entirely from the literature.
Those include a subroutine finding an irrelevant vertex whenever our instance $(G,R)$ contains a large clique minor relative to its bidimensionality, and some fundamental operations to deal with paths in a surface relative to a large set of concentric cycles embedded in some disc.
The purpose of this section is the preparation of these tools for later use.

\subsection{What to do with a clique}\label{subsec:clique}
While we are not restricting the number of vertices in $R$, the fact that we bound the bidimensionality of our instances $(G,R)$ allows us to easily find irrelevant vertices in large enough clique minors.
This observation has two distinct purposes in our proof:
First, it allows us to show that vital instances do not contain large clique minors -- where the size of the excluded minor is conditioned only on the bidimensionality of $(G,R)$.
Second, within our final algorithm it allows us to perform a preprocessing step in which we first detect all large clique minors in the graph and subsequently remove their irrelevant vertices.
Once this procedure terminates we know that we are now in the regime of $K_{\mathbf{O}(b^2)}$-minor-free graphs which allows us to call upon \zcref{prop:LinearGrid} to ensure that the graph either has treewidth in $f(b) \cdot \log n$, or contains a grid minor of order at least $g(b) \cdot \log n$ where $f$ and $g$ are functions we will determine later and $b$ denotes the bidimensionality of the instance $(G,R)$.
Key to this irrelevancy is the following theorem of Robertson and Seymour.

Let $G$ be a graph and $Z \subseteq V(G)$ be a set of vertices.
The \emph{weak $k$-folio relative to $Z$}, denoted by $k\text{-}\mathsf{WFolio}_G(Z)$, is the set of all patterns~$\pi(\mathcal L)$ of linkages~$\mathcal L$ in~$G$ with at most~$k$ paths and all endpoints in~$Z$. Singleton paths are allowed.
A vertex $v \in V(G) \setminus Z$ is said to be \emph{irrelevant} for the weak $k$-folio relative to $Z$ if and only if $k\text{-}\mathsf{WFolio}_{G-v}(Z) = k\text{-}\mathsf{WFolio}_{G}(Z)$.
In  particular, it is worthwhile pointing out that the weak $k$-folio relative to $Z$ defined above is essentially the special case where the detail $d = 0$ in the original statement.

\begin{proposition}[Robertson and Seymour \cite{RobertsonSeymour1995DisjointPaths}]\label{thm_cliqueirrelevantvertex}
 Let $G$ be a graph, let $k \in \mathbb{N}$ be a non-negative integer, and let $Z \subseteq V(G)$ with $|Z| \leq k$.
 Further let $\theta \coloneqq \lfloor \nicefrac{5}{2} \cdot k \rfloor + 1$, let $\varphi$ be a minor model of $K_{\theta}$ in $G$, and let $(A,B)$ be a separation in $G$ such that
 \begin{enumerate}
 \item \label{item:cliqueright} $A \cap V(\varphi(v)) = \emptyset$ for some $v \in V(K_{\theta})$,
 \item \label{item:terminalleft} $Z \subseteq A$,
 \item \label{item:smallsep} subject to \zcref{item:cliqueright} and \zcref{item:terminalleft}, $(A,B)$ is of minimum order, and
 \item subject to \zcref{item:cliqueright}, \zcref{item:terminalleft}, and \zcref{item:smallsep}, $A$ is maximal.
 \end{enumerate}
 Then for any $v \in B \setminus A$, $v$ is irrelevant for the weak $k$-folio of $G$ relative to $Z$.

 In particular, given $G$, $Z$, and $\varphi$ as above, such a vertex $v$ can be found in $\poly(k)|\!|G|\!|$-time.
\end{proposition}

We will also require the following result from \cite{ProtopapasThilikosWiederrecht2025ColorfulMinors} which refines some of the tools from \cite{RobertsonSeymour1995DisjointPaths}.

\begin{proposition}[Protopapas, Thilikos, and Wiederrecht \cite{ProtopapasThilikosWiederrecht2025ColorfulMinors}]\label{thm:colorfulclique}
 Let $t \geq 2$ and~$k \geq 1$ be integers with $k \geq \lfloor \nicefrac{3t}{2} \rfloor + t$.
 Let $(G,R)$ be an annotated graph such that $G$ contains a minor model $\varphi$ of $K_k$.
 Then one of the following is true:
 \begin{enumerate}
 \item There exists a red-minor model $\psi$ of a red $K_t$ in $(G,R)$ such that $\mathcal{T}_\psi$ is a truncation of $\mathcal{T}_\varphi$, or

 \item There exists a set $Z \subseteq V(G)$ of size at most $t - 1$ such that the $\mathcal{T}_\varphi$-big component of $G - Z$ is blank.
 \end{enumerate}
 Furthermore, there exists an algorithm that takes as input $t$, $(G,R)$, and $\varphi$ as above and finds one of the two outcomes in time $\poly(t)|\!|G|\!|$.
\end{proposition}

With this, we are ready to state the main lemma of this subsection.

\begin{lemma}\label{lemma:WahtToDoWithClique}
Let $b,t \geq 1$ be positive integers where $t \geq  6 \cdot (b+1)^2  + 1 $, let $(G,R)$ be an  instance of \textsc{Terminal Cyclability}.
If there exists a minor model $\varphi$ of $K_t$ in $G$, then either $\mathsf{bidim}(G,R) > b$ or there exists an irrelevant vertex $v \in V(G) \setminus R$.

Moreover, there exists an algorithm that takes as input $(G,R)$, $b$, and $\varphi$ as above and either correctly concludes that $\mathsf{bidim}(G,R) > b$ or finds the irrelevant vertex $v$ in time $\poly( b ) \cdot |\!|G|\!|$.
\end{lemma}

\begin{proof}
We first show that we may assume without loss of generality that~$|R| \geq 2$.
If~$R=\emptyset$, any vertex is irrelevant, since deleting one vertex leaves a clique minor of order at least three.
If~\(R=\{r\}\), then we can find a cycle containing~\(r\), or determine that none exists, in linear time.
In the latter case, any non-red vertex is irrelevant.
Otherwise, if the cycle is not spanning, any vertex outside it is irrelevant.
If it is spanning, the clique minor guarantees a chord, and one of the two cycles obtained using that chord contains~$r$ and omits a vertex.
Hence, we may assume that~$|R| \geq 2$.

We apply \zcref{thm:colorfulclique} to $(G,R)$, looking for a red $(b+1)^2$-clique minor.
In case we find this clique minor, it is clear that $(G,R)$ also contains a red $((b+1) \times (b+1))$-grid as a red-minor, thereby implying that $\mathsf{bidim}(G,R) \geq b+1$.
Hence, we may assume that we find the second outcome of \zcref{thm:colorfulclique}:
A set $Z \subseteq V(G)$ of size at most $(b+1)^2 - 1$ such that the $\mathcal{T}_{\varphi}$-big component of~$G-Z$ is blank.
In other words, there exists a separation $(A,B)$ with~$A \cap B = Z$ and of order at most $(b+1)^2 - 1$ such that $(B,A) \in \mathcal{T}_{\varphi}$ and $R \subseteq B$.
Moreover, this also means that there exists a minor model $\varphi'$ of $K_{\lfloor 5 \cdot (b+1)^2 \rfloor + 1} = K_{\lfloor \nicefrac{5}{2} \cdot 2(b+1)^2 \rfloor + 1}$ in $G[A \setminus Z]$ since~$Z = A\cap B$ can intersect at most $(b+1)^2-1$ branch sets of $\varphi$ and~$(B,A) \in \mathcal{T}_{\varphi}$.

Now notice that any solution $S$ intersecting~$A\setminus Z$
contains at least two distinct vertices of $Z$, since~$A\setminus Z$ is blank and $|R|\geq2$.
Splitting $S$ at its vertices in $Z$, let $\mathcal{L}$ consist
of the resulting paths with non-empty interiors contained in
$A\setminus Z$.
We encode these paths by their terminal pairs in $Z$.
To make this formally correct, we need to first observe that for any solution $S$ of $(G,R)$, each vertex of $Z$ contained in $S$ has degree $2$.
Hence, the corresponding instance of \textsc{Disjoint Paths} in~$G[A]$ uses any vertex of~$Z$ for at most $2$ terminals.
Thus, we obtain $G'$ from $G[A]-E(G[Z])$ by replacing every
vertex of $Z$ by two non-adjacent copies with the same neighbours
in $A\setminus Z$.
This is also the reason why we stressed that $\varphi'$ is a minor model of a clique with $\lfloor \nicefrac{5}{2} \cdot 2(b+1)^2 \rfloor + 1$ vertices.
Let $Z'$ be the resulting set and notice that $A\setminus Z = V(G') \setminus Z'$.
We include every unused vertex of~$Z'$ as a singleton path
in the linkage pattern, ensuring that replacement paths
remain internally disjoint from $Z'$.

Suppose there is some vertex $v \in A \setminus Z$ such that $v$ is irrelevant for the weak $|Z'|$-folio of $G'$ relative to $Z'$.
Then we may reroute the lifted linkage of any solution $S$
intersecting $A\setminus Z$ in $G'-v$ while preserving its pattern.
Discarding the singleton paths, identifying the copies, and
restoring the unchanged part of $S$ in $G[B]$ yields a simple
$R$-cycle: The replacement paths have pairwise disjoint interiors
in $A\setminus Z$, two distinct original endpoints each, and at least two
edges each, since $Z'$ is independent.
Solutions disjoint from $A\setminus Z$ already avoid $v$.
It follows therefore that $v$ is irrelevant for $(G,R)$.

We can find a separation $(A',B')$ of $G'$ satisfying the requirements of \zcref{thm_cliqueirrelevantvertex} in~$\mathbf{poly}(b)|\!|G|\!|$ time~\cite{KawarabayashiKobayashiReed2012Quadratic}.
By applying \zcref{thm_cliqueirrelevantvertex} to $G'$, $(A',B')$,
and $Z'$ with $k=2(b+1)^2$, we find a vertex
$v\in B'\setminus A'\subseteq A\setminus Z$ that is irrelevant
for the weak $2(b+1)^2$-folio relative to $Z'$.
Since~$|Z'|\leq2(b+1)^2$, this vertex is also irrelevant for the
weak $|Z'|$-folio and hence for $(G,R)$.
\end{proof}

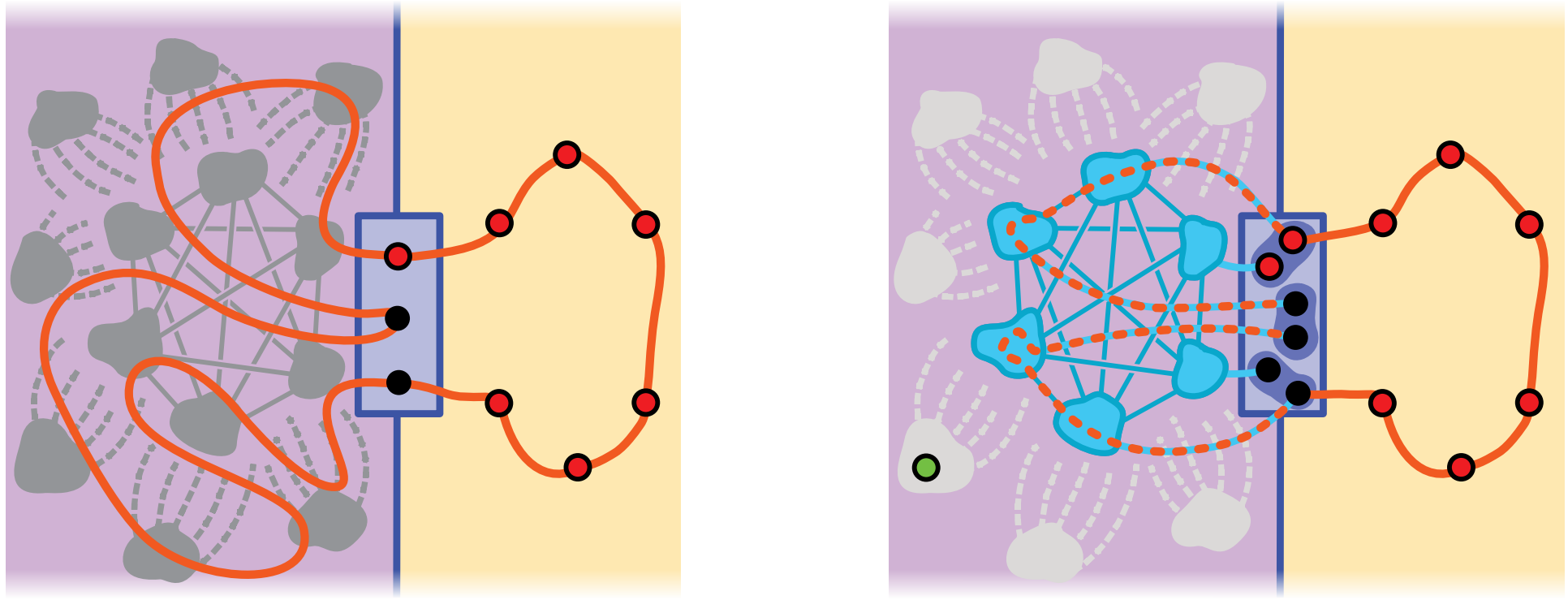
\begin{figure}[t]
 \centering
 \begin{tikzpicture}

 \pgfdeclarelayer{background}
		\pgfdeclarelayer{foreground}
			
		\pgfsetlayers{background,main,foreground}
			
 \begin{pgfonlayer}{main}
 \node (C) [v:ghost] {};

 \node(L) [v:ghost] at (-3.6,0) {
 \begin{tikzpicture}

 \pgfdeclarelayer{background}
		 \pgfdeclarelayer{foreground}
			
		 \pgfsetlayers{background,main,foreground}

 \begin{pgfonlayer}{background}
 \pgftext{\includegraphics[width=5.5cm]{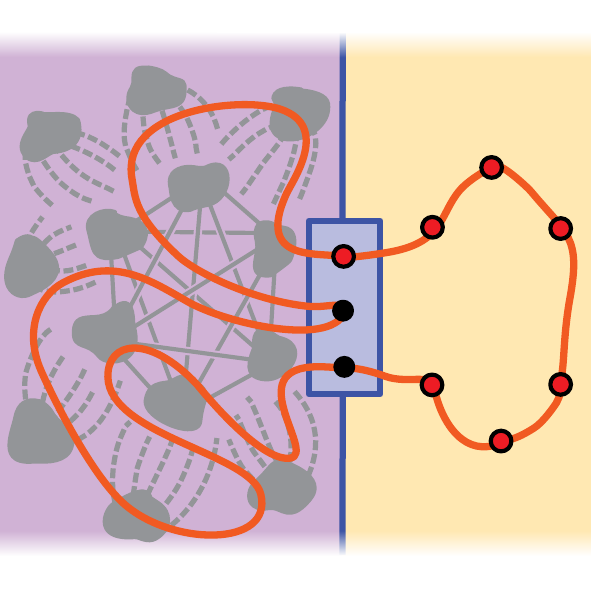}} at (C.center);
 \end{pgfonlayer}
			
 \begin{pgfonlayer}{main}
 \node (C) [v:ghost] {};
 
 \end{pgfonlayer}
 
 \end{tikzpicture}
 };

 \node(R) [v:ghost] at (3.6,0) {
 \begin{tikzpicture}

 \pgfdeclarelayer{background}
		 \pgfdeclarelayer{foreground}
			
		 \pgfsetlayers{background,main,foreground}

 \begin{pgfonlayer}{background}
 \pgftext{\includegraphics[width=5.5cm]{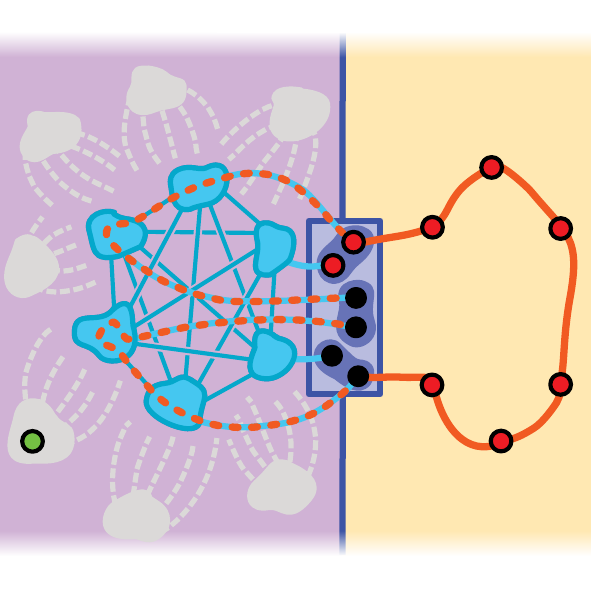}} at (C.center);
 \end{pgfonlayer}
			
 \begin{pgfonlayer}{main}
 \node (C) [v:ghost] {};
 
 \end{pgfonlayer}
 \end{tikzpicture}
 };
 \end{pgfonlayer}

 \end{tikzpicture}
 \caption{An illustration of the inner workings of the proof of \zcref{lemma:WahtToDoWithClique}.
 On the left, the situation after the application of \zcref{thm:colorfulclique} with a small order separation between the red vertices and the clique-minor (indicated in \textcolor{DimGray}{grey}). The \textcolor{Persimmon}{orange} curve illustrates the possible behaviour of an $R$-cycle $C$.
 On the right we see the corresponding situation in $G'$ with the duplicated vertices $Z'$ in the separator.
 The dashed \textcolor{Persimmon}{orange} paths indicate how the clique-minor may be used for routing the left side of $C$ through the size-constrained clique minor. The \textcolor{AppleGreen}{green} vertex is irrelevant.}
 \label{fig:RWhatToDoWithAClique}
\end{figure}%

\subsection{The wells are running dry}\label{subsec:LastDrop}

In this subsection we recall some fundamental results about linkages intersecting a partial embedding of a graph in a disc which contains a family $\mathcal{C}$ of concentric cycles.
It has been observed several times that such linkages -- and the cycles themselves -- can always be deformed such that the paths interact with the cycles in a ``convex'' manner.
This property is crucial for several parts in our analysis and we make use of the terminology and results of Cavallaro, Gorsky, Kreutzer, Thilikos, and Wiederrecht \cite{CavallaroGKTW2026Optimal}.

Let $s$ be a positive integer.
An \emph{$s$-prewell} $\mathcal{W} = (G, \Omega, \rho, \mathcal{C}, \mathcal{P})$ consists of a society $(G,\Omega)$ with a vortex-free rendition $\rho$ in a disc $\Delta$, a family of pairwise internally vertex-disjoint nontrivial~$V(\Omega)$-paths~$\mathcal P$, and a family of concentric cycles $\mathcal{C} = \langle C_1, \dots, C_s \rangle$ such that
\begin{itemize}
    \item for every $P \in \mathcal{P}$ there is $C \in \mathcal{C}$ such that $V(P) \cap V(C) \neq \emptyset$, and
    \item $G = \bigcup \mathcal{C} \cup \bigcup \mathcal{P}$.
\end{itemize}
We choose the traces of the paths in~$\mathcal P$ to meet~$\mathsf{bd}(\Delta)$ only at their endpoints.
If $\mathcal{W}$ also satisfies the following condition, it is called an \emph{$s$-well}:
\begin{itemize}
    \item there does not exist any path $P \in \mathcal{P}$ containing a grounded $C_1$-path $P' \subseteq P$ whose trace, except for its endpoints, is contained in the interior of the $C_1$-disc in~$\rho$.
\end{itemize}

Indeed, we are interested in a slightly stronger property as follows.

Let $\mathcal{C} = \langle C_1, \ldots, C_s\rangle$ be a sequence of concentric cycles in a $\Delta$-rendition $\rho$ of a graph $G$ in a disc $\Delta$ and let $\Delta_i$ be the $C_i$-disc in $\rho$ for each $i \in [s],$ with $\Delta_0 = \emptyset$.
We say that $\mathcal{C}$ is \emph{tight (in $\rho$)} if for all $i \in [s]$ there does not exist a grounded $C_i$-path whose trace is found in $\Delta_i \setminus \Delta_{i-1}$. 
Given a vortex $c$ with $c \subseteq \Delta_1,$ we say that $\mathcal{C}$ is \emph{tight around $c$ (in $\rho$)} if there does not exist a grounded $C_1$-path whose trace is found in $\Delta_1$.
In general, we call a set of concentric cycles $\mathcal{C}'$ tight in a rendition $\rho'$ (around a vortex) if we can find a disc such that the restriction of $\rho'$ to $\Delta'$ allows us to call $\mathcal{C}'$ tight.
A (pre)well $\mathcal{W} = (G', \Omega, \rho', \mathcal{C}', \mathcal{P})$ is \emph{tight} if $\mathcal{C}'$ is tight in $\rho'$.

The existence of tight concentric cycles is relatively easy to establish.
In the literature many algorithmic proofs going from a family of concentric cycles to a tight one have been given, see for example the discussion of cosy nests in \cite{GorskySW2025Polynomial}.
Thus we state the following result without proof.

\begin{proposition}\label{prop:makecyclestight}
 Let $\mathcal{C} = \langle C_1, \ldots, C_s\rangle$ be a set of concentric cycles in a $\Delta$-rendition $\rho$ of a graph $G$ in a disc $\Delta$ with $\Delta_s$ being the $C_s$-disc in $\rho$.
 Then there exists a tight concentric set of cycles $\mathcal{C}' = \langle C_1', \ldots, C_s'\rangle$ such that the trace of each cycle in $\mathcal{C}'$ lies in $\Delta_s$.
\end{proposition}

It is easy to see that every tight prewell is in fact a well.

\begin{observation}\label{obs:tight_prewell_is_well}
    Let $s$ be a positive integer and let $\mathcal{W} = (G, \Omega, \rho, \mathcal{C}, \mathcal{P})$ be an $s$-prewell.
    If $\mathcal{C}$ is tight, then $\mathcal{W}$ is an $s$-well. 
\end{observation}

In an $s$-well $\mathcal{W} = (G, \Omega, \rho, \mathcal{C}, \mathcal{P})$, we may take the trace $\gamma$ in $\rho$ of any path $P \in \mathcal{P}$ and observe that $\Delta-\gamma$ contains a unique component whose intersection with the trace of $C_1$ is empty.
The \emph{interior of \(P\)}, denoted by \(\Delta^P\), is the closure of this unique component of \(\Delta-\gamma\).\SW{Illustrate a well and the discs defined by the paths}

We say that an $s$-well $\mathcal{W} = (G, \Omega, \rho, \mathcal{C}, \mathcal{P})$ is \emph{drained} if 
\begin{itemize}
 \item for each $P\in\mathcal P$ and every $i\in[s-1]$, if~$P\cap C_i$ is non-empty, then there exists $Q\in\mathcal P\setminus\{P\}$ whose trace is contained in $\Delta^P$ and for which $Q\cap C_{i+1}$ is non-empty;

 \item for all $i \in [s],$ no $C_i$-path $P'$ contained in a member of~$\mathcal{P}$ that is disjoint from $V(\mathcal{C} \setminus \{ C_i \})$ is contained in a cell $c \in C(\rho)$ such that $\sigma(c)$ contains an edge of $C_i$.
\end{itemize}

In drained wells the vortex-free rendition tells us that any path that moves deeply into a drained well implies the existence of many other paths that also go deep.
This is captured in the following observation which we will apply implicitly whenever we work with drained wells.

\begin{observation}\label{obs:fullbucketsindrainedwells}
 Let $\mathcal{W} = (G, \Omega, \rho, \mathcal{C}, \mathcal{P})$ be a drained $s$-well and let $P \in \mathcal{P}$ be a path such that~${C_i \cap P}$ is non-empty for some $i \in [s]$.
 Then there exist $(s - i) + 1$ distinct paths~$P_s, P_{s-1}, \ldots, P_{i+1}, P_i = P$ such that~$P_j \cap C_j$ is non-empty for all~$j \in [i,s]$ and we have~$\Delta^{P_{j+1}}\subsetneq\Delta^{P_j}$ for every
$j\in[i,s-1]$.
\end{observation}

Tightness allows us to further strengthen the properties of drained wells.
We say that a well~${\mathcal{W} = (G, \Omega, \rho, \mathcal{C} = \langle C_1, \ldots, C_s\rangle, \mathcal{P})}$ is \emph{dry} if $\mathcal{W}$ is drained and for all $P \in \mathcal{P}$ we have that
\begin{enumerate}
 \item the graph $C_1 \cap P$ is either empty or a single path,

 \item there exists a unique $i \in [s]$ such that $C_i \cap P$ is a single path -- we call this path the \emph{apex} of $P$,

 \item for all $j \in [i-1]$ the graph $C_j \cap P$ is empty, and

 \item for all $j \in [i+1,s]$ the graph $C_j \cap P$ consists of exactly two paths.
\end{enumerate}

\begin{proposition}[Cavallaro, Gorsky, Kreutzer, Thilikos, and Wiederrecht \cite{CavallaroGKTW2026Optimal}]\label{prop:DryWell}
Let~$s$ be a positive integer and let~$W=(G,\Omega,\rho,\mathcal C,\mathcal P)$ be a tight~$s$-well with~$|\mathcal P|\geq2$.
If the cells of~$\rho$ represent individual edges
or isolated vertices and the traces of all cycles
in~$\mathcal C$ lie in~$\mathsf{int}(\Delta)$, then there exists a tight, dry~$s$-well~$\mathcal{W}' = (G', \Omega, \rho', \mathcal{C}, \mathcal{P}')$ such that $G' \subseteq G,$ $\rho'$ is the rendition of $G'$ induced by $\rho,$ $|\mathcal{P}| = |\mathcal{P}'|,$ and for each $P \in \mathcal{P}$ there exists a $P' \in \mathcal{P}'$ with the same endpoints.
\end{proposition}

\subsection{Building villages}\label{subsec:BuildingVillages}
With all the preparations in place, we may now move on to the first proper refinement step for our vital instances.
In the future we will study counterexamples $(G,R)$ to \zcref{thm:MainVitalSolution}.
Due to \zcref{lemma:WahtToDoWithClique} we know that such counterexamples must have large treewidth but exclude a clique minor whose size is dependent only on the bidimensionality of the instance.
This means that we may always assume that there exists a large wall that allows for the application of \zcref{prop:localstructure}.
This gives rise to some surface $\Sigma$ where $(G,R)$ has a rendition under the removal of a small set of apices and with a small number of vortices, all of which have bounded depth and to where $R$ is confined.
The goal of this subsection is to establish that we may assume some additional properties satisfied by such an instance.

\paragraph{Villages.}
Let $b \geq 1$ and $w \geq 3$ be integers and $(G,R)$ be a vital instance of \textsc{Terminal Cyclability} with solution $S$ such that $\mathsf{bidim}(G,R) \leq b$.
Suppose further that there exists
\begin{enumerate}
    \item a set $A \subseteq V(G)$ with $|A| \leq \mathsf{apex}_{\ref{prop:localstructure}}(3,  6 \cdot (b+1)^2  + 1, b)$ and
    \item a surface $\Sigma$ of Euler genus less than $9(  6 \cdot (b+1)^2 + 1 )^2$ such that 
    \item $(G - A, R \setminus A)$ has a blank $\Sigma$-rendition $\rho$ with
    \begin{itemize}
        \item breadth at most $\nicefrac{3}{2}( 6 \cdot (b+1)^2 )(18 \cdot (b+1)^2 - 1) + (3b + 2)(3b + 1)-3$,
        \item depth at most $\mathsf{depth}_{\ref{prop:localstructure}}(3,  6 \cdot (b+1)^2  + 1, b)$, and
        \item there exists a vortex-free, $\rho$-aligned disc $\Delta \subseteq \Sigma$ such that the restriction of $\rho$ to $\Delta$ contains a sequence $\mathcal{C} = \langle C_1,C_2,\dots,C_w \rangle$ of $w$ concentric cycles where
        \item no vertex of $N_S(A)$ is contained in the crop of $G$ by $\Delta$.
        \item Moreover, if we denote by $\mathcal{V}$ the vortices of $\rho$, then for every $v \in \mathcal{V}$ there exists a $\rho$-aligned disc $\Delta_{v}$ such that $\Delta_v$ contains $v$ and is disjoint from $\Delta$ and from $\Delta_{v'}$ for all $v' \in \mathcal{V}\setminus\{ v\}$.
    \end{itemize}
\end{enumerate}
We refer to $(G,R,S,A,\rho,\Sigma,\Delta,\mathcal{C})$ as a \emph{$(w,b)$-village} of $(G,R)$.

\begin{theorem}\label{thm:BuildAVillage}
There exists a function $f_{\ref{thm:BuildAVillage}} \colon \mathbb{N}^2 \to \mathbb{N}$ such that for all integers $b\geq 1$ and $w\geq 3$, if~$(G,R)$ is a vital instance of \textsc{Terminal Cyclability} with $\mathsf{tw}(G) \geq f_{\ref{thm:BuildAVillage}}(b,w)$ and $\mathsf{bidim}(G,R) = b$, then there exists a $(w,b)$-village for $(G,R)$.
Moreover, $f_{\ref{thm:BuildAVillage}}(b,w) \in \mathbf{poly}(b) \cdot w$.
\end{theorem}

\begin{proof}
Let us set up the function $f_{\ref{thm:BuildAVillage}}$ before we begin with the proof.
\begin{align*}
    t(b) & \coloneqq  6 \cdot (b+1)^2  + 1\\
    f_{\ref{thm:BuildAVillage}}(b,w) & \coloneqq f_{\ref{prop:LinearGrid}} \big( t(b) , t(b)^2 , \mathsf{mesh}_{\ref{prop:localstructure}}( b , 3 , t(b) , (2 \cdot \mathsf{apex}_{\ref{prop:localstructure}}( 3 , t(b) , b)  + 1 ) \cdot (2w+2) ) \big) +1
\end{align*}
We fix $S$ to be the unique solution of $(G,R)$.

By \zcref{lemma:WahtToDoWithClique} we know that $G$ cannot contain a $K_{t(b)}$-minor as otherwise the existence of an irrelevant vertex would contradict the vitality of $(G,R)$.

Hence, with $\mathsf{tw}(G) > f_{\ref{prop:LinearGrid}} \big( t(b) , t(b)^2 , \mathsf{mesh}_{\ref{prop:localstructure}}( b , 3 , t(b) , (2 \cdot \mathsf{apex}_{\ref{prop:localstructure}}( 3 , t(b) , b)  + 1 ) \cdot (2w+2) ) \big)$, \zcref{prop:LinearGrid} supplies us with a mesh $M_0$ of order $\mathsf{mesh}_{\ref{prop:localstructure}}( b , 3 , t(b) , (2 \cdot \mathsf{apex}_{\ref{prop:localstructure}}( 3 , t(b) , b)  + 1 ) \cdot (2w+2) )$ in $G$.
Notice that $f_{\ref{prop:LinearGrid}} \big( t(b) , t(b)^2 , \mathsf{mesh}_{\ref{prop:localstructure}}( b , 3 , t(b) , (2 \cdot \mathsf{apex}_{\ref{prop:localstructure}}( 3 , t(b) , b)  + 1 ) \cdot (2w+2) ) \big) \in \mathbf{poly}(b) \cdot \mathsf{mesh}_{\ref{prop:localstructure}}( b , 3 , t(b) , (2 \cdot \mathsf{apex}_{\ref{prop:localstructure}}( 3 , t(b) , b)  + 1 ) \cdot (2w+2) ) $.
Moreover, we have that $\mathsf{mesh}_{\ref{prop:localstructure}}( b , 3 , t(b) , (2 \cdot \mathsf{apex}_{\ref{prop:localstructure}}( 3 , t(b) , b)  + 1 ) \cdot (2w+2) ) \in \mathbf{poly}(b) \cdot w$ and thus the desired bound on $f_{\ref{thm:BuildAVillage}}$ holds.

We now apply \zcref{prop:localstructure} to $M_0$.
Since $G$ does not contain a $K_{t(b)}$-minor, the first two outcomes of the theorem are impossible and so we must be in the third outcome.
Hence, we obtain a set $A \subseteq V(G)$ with $|A| \leq \mathsf{apex}_{\ref{prop:localstructure}}( 3 , t(b) , b)$ and a surface $\Sigma$ of Euler genus less than $9t(b)^2$ such that $(G-A,R\setminus A)$ has a blank $\Sigma$-rendition $\rho$ with breadth at most $\nicefrac{3}{2}(t(b) - 1)(3t(b) - 4)+(3b+2)(3b+1)-3$, depth at most $\mathsf{depth}_{\ref{prop:localstructure}}( 3 , t(b) , b )$, and there exists a vortex-free, $\rho$-aligned disc $\Delta' \subseteq \Sigma$ such that the restriction of $\rho$ to $\Delta'$ contains a flat $ ( (2\mathsf{apex}_{\ref{prop:localstructure}}( 3 , t(b) , b) + 1) \cdot (2w+2))$-mesh $M_1$ that is grounded and flat in~$\rho$.

For any vertex $a \in A$, we have~$|N_S(a)|=2$.
Hence, $|N_S(A)| \leq 2|A|$.
We may now find $2|A| + 1$ submeshes of $M_1$, each of order $2w+2$ and with pairwise disjoint compasses, that is, disjoint inner graphs of their perimeter.
By the pigeonhole principle this means that there is one of those $2w+2$-meshes, say $M_2$ whose compass is disjoint from $N_S(A)$.
By discarding the perimeter of~$M_2$ and then selecting $\mathcal{C}$ to be the sequence of $w$ cycles obtained by iteratively removing the perimeter from $M_2$, we have found our desired village in $(G,R)$.
\end{proof}

\subsection{Teaching villagers how to cycle}\label{subsec:VillagersCycle}

Having built villages in the previous subsection, we next control the
interaction of the solution with the concentric cycles inside their protected
disc. For this purpose, it suffices to consider the restrictions of the solution
then use \zcref{prop:DryWell} to obtain a dry well.

\paragraph{Water is getting scarce.}
Let~$\mathfrak{V}=(G,R,S,A,\rho,\Sigma,\Delta,\mathcal C)$ be a
$(w,b)$-village, where~$\mathcal C=\langle C_1,\ldots,C_w\rangle$.
We say that~$\mathfrak{V}$ is \emph{thirsty} if the following holds.
All crops in this definition are taken through the rendition~$\rho$ of~$G-A$.
Inside~$\Delta$, the rendition is refined into cells representing individual
edges or isolated vertices, so that the crop is explicitly embedded in~$\Delta$.
The cycles of~$\mathcal C$ are tight and their traces lie in~$\mathsf{int}(\Delta)$.
Let~$B$ be the set of graph vertices represented on~$\mathsf{bd}(\Delta)$, and let \[\mathcal P \coloneqq \{P\subseteq S \mid P\text{ is a nontrivial }B\text{-path
 contained in the crop by }\Delta \text{ and } V(P)\cap V(\mathcal C)\neq\emptyset\}.\]
Let~$W \coloneqq \bigcup\mathcal P\cup\bigcup\mathcal C$, and let~$\Omega$ be the cyclic order of~$V(W)\cap B$ inherited from~$\mathsf{bd}(\Delta)$. We require that~$\mathfrak{W}=(W,\Omega,\rho_\Delta,\mathcal C,\mathcal P)$ is a dry~$w$-well, where~$\rho_\Delta$ is the induced rendition.
We refer to~$\mathfrak{W}$ as the \emph{well} of~$\mathcal V$ and to the
members of~$\mathcal P$ as its \emph{legs}.

\begin{theorem}\label{thm:dry-village-preparation}
There exists a function~$f_{\ref{thm:dry-village-preparation}}\colon\mathbb N^2\to\mathbb N$ such that
for all integers~$b\geq1$ and~$w\geq3$, if~$(G,R)$ is a vital instance
of~\Tc{} with~$\mathsf{tw}(G)\geq f_{\ref{thm:dry-village-preparation}}(b,w)$ and~$\mathsf{bidim}(G,R)=b$, then there exists a thirsty~$(w,b)$-village for~$(G,R)$. Moreover, $f_{\ref{thm:dry-village-preparation}}(b,w)\in\poly(b)\cdot w$.
\end{theorem}

\begin{proof}
Let~$f_{\ref{thm:dry-village-preparation}}(b,w) \coloneqq f_{\ref{thm:BuildAVillage}}(b,w+1)$. By \zcref{thm:BuildAVillage}, there exists a~$(w+1,b)$-village~$(G,R,S,A,\rho,\Sigma,\Delta_0, \langle C^0_1,\ldots,C^0_{w+1}\rangle)$ for~$(G,R)$.
We first refine the ordinary cells inside~$\Delta_0$ without changing~$G$.
Consider such a cell~$c$ with an internal vertex. Every connected component of~$S\cap\sigma(c)$ containing an internal vertex is a path whose two distinct endpoints belong to~$N_\rho(c)$.
Since~$|N_\rho(c)|\leq 3$, there is exactly one such component~$P$, and it contains every internal vertex of~$\sigma(c)$.
Every remaining edge has an endpoint at the possible third boundary node outside~$P$.
Consequently, $\sigma(c)$ can be embedded in~$\mathsf{cl}(c)$ while fixing its boundary nodes.
If~$c$ has no internal vertex, its graph has at most three vertices and can be embedded directly.

We refine these drawings into cells representing individual edges or isolated vertices.
At most one of the pairwise disjoint grounded cycles~$C^0_i$ uses edges of a given cell.
We choose the drawing so that its passage through that cell is on the same side of any remaining boundary node as its original trace.
The cycles therefore remain concentric.
The refinement changes neither the graph nor its vortices, and the crop by~$\Delta_0$ remains disjoint from~$R\cup N_S(A)$.
We then apply \zcref{prop:makecyclestight} in the refined rendition to obtain tight concentric cycles~$C_1,\ldots,C_{w+1}$ inside~$\Delta_0$.
Let~$\Delta \coloneqq \Delta_{C_{w+1}}$ and~$\mathcal{C} \coloneqq \langle C_1,\ldots,C_w\rangle$.
Note that the retained cycles lie strictly inside~$\Delta$.

Define~$B$,~$\mathcal P$, $W$, and~$\Omega$ as above.
Splitting~$S$ at all its contacts with~$B$ makes the members of~$\mathcal P$ pairwise internally vertex-disjoint~$V(\Omega)$-paths. Every vertex of a retained cycle lies on one of these paths: $S$ spans~$G$, the retained cycles are strictly inside~$\Delta$, and
no edge of~$S$ leaves the crop through an apex.
Following~$S$ from such a vertex in either direction reaches~$B$, since~$S$ contains a red vertex outside~$\Delta$.
In particular, $V(W)=V\bigl(\bigcup\mathcal P\bigr)$.
Thus, $\mathfrak{W}=(W,\Omega,\rho_\Delta,\mathcal C,\mathcal P)$ is a
tight prewell and hence a well.

Notice that vitality implies~$|\mathcal P| \geq 2$.
Furthermore, distinct members of~$\mathcal P$ have distinct endpoint pairs.
\zcref{prop:DryWell} now provides a dry well whose paths have the same endpoint pairs as~$\mathcal P$.
We replace the paths of~$\mathcal P$ in~$S$ by these paths.
Their interiors avoid~$V(\Omega)$ and are pairwise vertex-disjoint.
Moreover, $W$ meets the unchanged part of~$S$ only at the prescribed boundary endpoints.
Hence, this replacement preserves the cyclic order of the unchanged
pieces of~$S$ and gives a simple~$R$-cycle, also when two selected paths share a boundary endpoint.
Vitality forces this cycle to be~$S$, and therefore each selected path remains unchanged.
Thus,~$\mathfrak{W}$ is already dry.
This concludes the proof.
\end{proof}

\subsection{The local rerouting lemma}

To conclude this section, we prove our local rerouting lemma.
Recall that the goal is to generalise the rerouting lemma of Gahlawat, Rathod, and Zehavi~\cite{GRZ25} to graphs that need not be planar.
In particular, we show that a solution cycle whose intersection
with a terminal-free~$(4\times3)$-mesh consists precisely of
its three vertical paths can be rerouted inside the mesh.
This local rerouting will later allow us to reduce the number of times a solution crosses a railed annulus.

\begin{lemma}\label{lemma:Reroute4x3Mesh}
     Let~$(G,T)$ be an annotated graph and let~$C$ be a $T$-cycle in $G$. Moreover, let~$M$ be a~$(4 \times 3)$-mesh with paths~$P_1,P_2,P_3,P_4,Q_1,Q_2,$ and~$Q_3$ such that~$C \cap M = Q_1 \cup Q_2 \cup Q_3$ and~$V(M) \cap T = \emptyset$.
    Then, there exists a $T$-cycle $C'$ such that~$V(C') \cap V(Q_2) \cap V(P_2) = \emptyset$ and~$V(C') \cap (V(G) \setminus V(M)) = V(C) \cap (V(G) \setminus V(M))$.
\end{lemma}

\begin{proof}
    We start with some notation.
    We consider the paths~$Q_i$ directed in such a way that the vertices of~$V(P_1) \cap V(Q_i)$ are first on~$Q_i$ and the vertices of~$V(P_4) \cap V(Q_i)$ are last, which is possible by definition of meshes.
    Consider \cref{fig:mesh} for an example.
    Therein, we consider the paths~$Q_i$ to be directed from top to bottom.
    \begin{figure}[t!]
        \centering
        \begin{tikzpicture}
            \foreach \i in {1,4}{
                \node[circle,draw,inner sep=3pt] (a\i1) at(1,-\i){};
                \node[circle,draw,inner sep=3pt] (a\i2) at(2,-\i){};
                \node[circle,draw,inner sep=3pt] (a\i3) at(3,-\i){};
            }
            
            \foreach \i in {2,3}{
                \node[circle,draw,inner sep=3pt] (a\i1) at(1,-\i){};
                \node[circle,draw,inner sep=3pt] (a\i2) at(1.75,-\i+.25){};
                \node[circle,draw,inner sep=3pt] (b\i2) at(2.25,-\i-.25){};
                \node[circle,draw,inner sep=3pt] (a\i3) at(3,-\i){};
                \draw[DarkMagenta,very thick] (a\i2) to (b\i2);
            }
            
            \node at(1,-.65) {\small $a_{1,1}$};
            \node at(3,-.65) {\small $a_{1,3}$};
            \node at(1.5,-1.4) {\small $a_{2,2}$};
            \node at(2.25,-1.85) {\small $b_{2,2}$};
            \node at(1,-4.35) {\small $a_{4,1}$};
            \node at(3,-4.35) {\small $a_{4,3}$};
            
            \node at(.5,-1) {$P_1$};
            \node at(.5,-2) {$P_2$};
            \node at(.5,-3) {$P_3$};
            \node at(.5,-4) {$P_4$};
            \node at(1,0) {$Q_1$};
            \node at(2,0) {$Q_2$};
            \node at(3,0) {$Q_3$};
            
            \foreach \i in {1,4}{
                \draw[blue,very thick] (a\i1) to (a\i2);
                \draw[blue,very thick] (a\i2) to (a\i3);
            }
            
            \foreach \i in {2,3}{
                \draw[blue,very thick] (a\i1) to (a\i2);
                \draw[blue,very thick] (b\i2) to (a\i3);
            }

            \foreach \j in {1,3}{
                \draw[red,very thick] (a1\j) to (a2\j);
                \draw[red,very thick] (a2\j) to (a3\j);
                \draw[red,very thick] (a3\j) to (a4\j);
            }
            \draw[red,very thick] (a12) to (a22);
            \draw[red,very thick] (b22) to (a32);
            \draw[red,very thick] (b32) to (a42);
    
            \begin{pgfonlayer}{background}
                \draw[path] (2,-.5) -- (a12.center) -- (a22.center) -- (b22.center) -- (a32.center) -- (b32.center) -- (a42.center) -- (2,-4.5);
                \foreach \j in {1,3}{
                    \draw[path] (\j,-.5) -- (\j,-4.5);
                }
            \end{pgfonlayer}
        \end{tikzpicture}
        \caption{An example of a~$4 \times 3$-mesh. The paths~$P_i$ are coloured blue, the paths~$Q_j$ are coloured red, and edges that belong to both a path~$P_i$ and a path~$Q_j$ are coloured purple. The three paths~$Q_1, Q_2$, and~$Q_3$ that are part of the solution~$C$ are highlighted.}
        \label{fig:mesh}
    \end{figure}
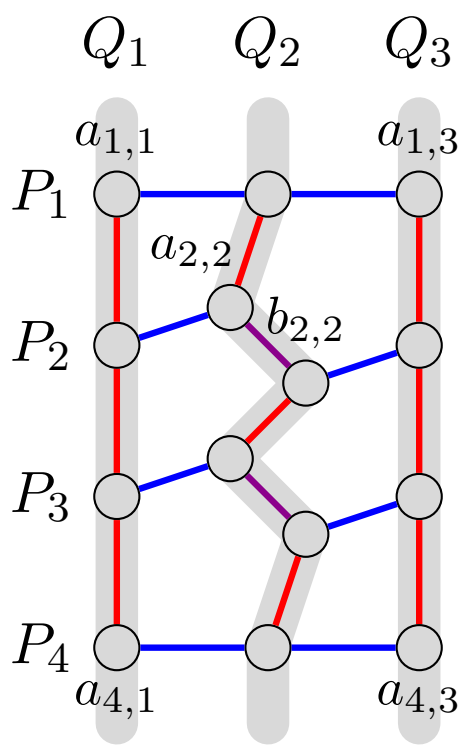
    For each~$i \in [4]$ and~$j \in [3]$, let~$a_{i,j}$ be the first vertex of~$Q_j$ in~$V(P_i)$ and let~$b_{i,j}$ be the last vertex of~$Q_j$ in~$V(P_i)$.
    In the nontrivial intersections shown in \cref{fig:mesh}, the $a$-vertices are the top-left endpoints of purple edges, and the $b$-vertices are the corresponding bottom-right endpoints.
    
    We construct the~$T$-cycle~$C'$ in two steps.
    First, we remove the subpath of~$Q_2$ between~$b_{1,2}$ and~$b_{4,2}$ and the subpath of~$Q_3$ between~$a_{1,3}$ and~$a_{4,3}$.
    Note that since~$C$ is a simple cycle and these two paths are vertex disjoint, removing them from~$C$ (but retaining their endpoints) yields two vertex-disjoint paths.
    We now consider two cases for the second step.
    First, we try to add the subpath of~$P_1$ between~$b_{1,2}$ and~$a_{1,3}$ and the subpath of~$P_4$ between~$b_{4,2}$ and~$a_{4,3}$.
    Let~$C_1$ be the resulting graph.
    See \cref{fig:meshcase1} (left side) for an illustration.
        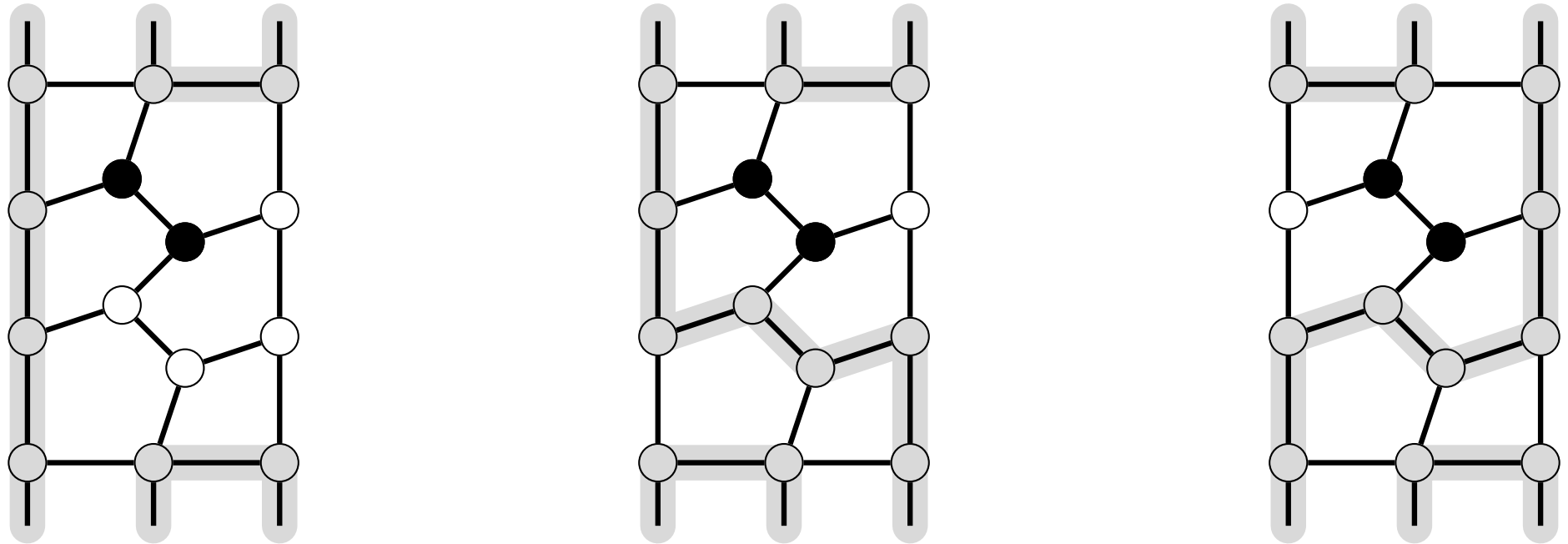
\begin{figure}[tp]
        \centering
        \begin{tikzpicture}
            \def\x{5}
            \def\y{10}
            \foreach \i in {1,4}{
                \node[circle,draw,inner sep=3pt] (a\i1) at(1,-\i){};
                \node[circle,draw,inner sep=3pt] (a\i2) at(2,-\i){};
                \node[circle,draw,inner sep=3pt] (a\i3) at(3,-\i){};
            }

            \foreach \i in {2,3}{
                \node[circle,draw,inner sep=3pt] (a\i1) at(1,-\i){};
                \node[circle,draw,inner sep=3pt] (a\i2) at(1.75,-\i+.25){};
                \node[circle,draw,inner sep=3pt] (b\i2) at(2.25,-\i-.25){};
                \node[circle,draw,inner sep=3pt] (a\i3) at(3,-\i){};
                \draw[very thick] (a\i2) to (b\i2);   
            }
            
            \foreach \i in {1,4}{
                \draw[black,very thick] (a\i1) to (a\i2);
                \draw[black,very thick] (a\i2) to (a\i3);
            }

            \foreach \i in {2,3}{
                \draw[black,very thick] (a\i1) to (a\i2);
                \draw[black,very thick] (b\i2) to (a\i3);
            }
            
            \foreach \j in {1,3}{
                \draw[black,very thick] (\j,-.5) to (a1\j);
                \draw[black,very thick] (a1\j) to (a2\j);
                \draw[black,very thick] (a2\j) to (a3\j);
                \draw[black,very thick] (a3\j) to (a4\j);
                \draw[black,very thick] (a4\j) to (\j,-4.5);
            }
            \draw[black,very thick] (2,-.5) to (a12);
            \draw[black,very thick] (a12) to (a22);
            \draw[black,very thick] (b22) to (a32);
            \draw[black,very thick] (b32) to (a42);
            \draw[black,very thick] (a42) to (2,-4.5);
            \node[circle,draw,fill=black,inner sep=3pt] at(1.75,-1.75) {};
            \node[circle,draw,fill=black,inner sep=3pt] at(2.25,-2.25) {};
            
            \begin{pgfonlayer}{background}
                \draw[path] (1,-.5) -- (1,-4.5);
                \draw[path] (2,-.5) -- (a12.center) -- (a13.center) -- (3,-.5);
                \draw[path] (2,-4.5) -- (a42.center) -- (a43.center) -- (3,-4.5);
            \end{pgfonlayer}

            \foreach \i in {1,4}{
                \node[circle,draw,inner sep=3pt] (a\i1) at(\x+1,-\i){};
                \node[circle,draw,inner sep=3pt] (a\i2) at(\x+2,-\i){};
                \node[circle,draw,inner sep=3pt] (a\i3) at(\x+3,-\i){};
            }

            \foreach \i in {2,3}{
                \node[circle,draw,inner sep=3pt] (a\i1) at(\x+1,-\i){};
                \node[circle,draw,inner sep=3pt] (a\i2) at(\x+1.75,-\i+.25){};
                \node[circle,draw,inner sep=3pt] (b\i2) at(\x+2.25,-\i-.25){};
                \node[circle,draw,inner sep=3pt] (a\i3) at(\x+3,-\i){};
                \draw[very thick] (a\i2) to (b\i2);
            }
            
            \foreach \i in {1,4}{
                \draw[black,very thick] (a\i1) to (a\i2);
                \draw[black,very thick] (a\i2) to (a\i3);
            }

            \foreach \i in {2,3}{
                \draw[black,very thick] (a\i1) to (a\i2);
                \draw[black,very thick] (b\i2) to (a\i3);
            }
            
            \foreach \j in {1,3}{
                \draw[black,very thick] (\x+\j,-.5) to (a1\j);
                \draw[black,very thick] (a1\j) to (a2\j);
                \draw[black,very thick] (a2\j) to (a3\j);
                \draw[black,very thick] (a3\j) to (a4\j);
                \draw[black,very thick] (a4\j) to (\x+\j,-4.5);
            }
            \draw[black,very thick] (\x+2,-.5) to (a12);
            \draw[black,very thick] (a12) to (a22);
            \draw[black,very thick] (b22) to (a32);
            \draw[black,very thick] (b32) to (a42);
            \draw[black,very thick] (a42) to (\x+2,-4.5);
            \node[circle,draw,fill=black,inner sep=3pt] at(\x+1.75,-1.75) {};
            \node[circle,draw,fill=black,inner sep=3pt] at(\x+2.25,-2.25) {};
            
            \begin{pgfonlayer}{background}
                \draw[path] (\x+1,-.5) -- (a31.center) -- (a32.center) -- (b32.center) -- (a33.center) -- (\x+3,-4.5);
                \draw[path] (\x+2,-.5) -- (a12.center) -- (a13.center) -- (\x+3,-.5);
                \draw[path] (\x+2,-4.5) -- (a42.center) -- (a41.center) -- (\x+1,-4.5);
            \end{pgfonlayer}
            
            \foreach \i in {1,4}{
                \node[circle,draw,inner sep=3pt] (a\i1) at(\y+1,-\i){};
                \node[circle,draw,inner sep=3pt] (a\i2) at(\y+2,-\i){};
                \node[circle,draw,inner sep=3pt] (a\i3) at(\y+3,-\i){};
            }

            \foreach \i in {2,3}{
                \node[circle,draw,inner sep=3pt] (a\i1) at(\y+1,-\i){};
                \node[circle,draw,inner sep=3pt] (a\i2) at(\y+1.75,-\i+.25){};
                \node[circle,draw,inner sep=3pt] (b\i2) at(\y+2.25,-\i-.25){};
                \node[circle,draw,inner sep=3pt] (a\i3) at(\y+3,-\i){};
                \draw[very thick] (a\i2) to (b\i2);
            }
            
            \foreach \i in {1,4}{
                \draw[black,very thick] (a\i1) to (a\i2);
                \draw[black,very thick] (a\i2) to (a\i3);
            }

            \foreach \i in {2,3}{
                \draw[black,very thick] (a\i1) to (a\i2);
                \draw[black,very thick] (b\i2) to (a\i3);
            }
            
            \foreach \j in {1,3}{
                \draw[black,very thick] (\y+\j,-.5) to (a1\j);
                \draw[black,very thick] (a1\j) to (a2\j);
                \draw[black,very thick] (a2\j) to (a3\j);
                \draw[black,very thick] (a3\j) to (a4\j);
                \draw[black,very thick] (a4\j) to (\y+\j,-4.5);
            }
            \draw[black,very thick] (\y+2,-.5) to (a12);
            \draw[black,very thick] (a12) to (a22);
            \draw[black,very thick] (b22) to (a32);
            \draw[black,very thick] (b32) to (a42);
            \draw[black,very thick] (a42) to (\y+2,-4.5);
            \node[circle,draw,fill=black,inner sep=3pt] at(\y+1.75,-1.75) {};
            \node[circle,draw,fill=black,inner sep=3pt] at(\y+2.25,-2.25) {};
            
            \begin{pgfonlayer}{background}
                \draw[path] (\y+3,-.5) -- (a33.center) -- (b32.center) -- (a32.center) -- (a31.center) -- (\y+1,-4.5);
                \draw[path] (\y+1,-.5) -- (a11.center) -- (a12.center) -- (\y+2,-.5);
                \draw[path] (\y+2,-4.5) -- (a42.center) -- (a43.center) -- (\y+3,-4.5);
            \end{pgfonlayer}
        \end{tikzpicture}
        \caption{The three solution candidates~$C_1$ (left side), $C_2$ (middle) and~$C_3$ (right side). The vertices of~$V(P_2) \cap V(Q_2)$ are coloured black and are not part of any solution candidate.}
        \label{fig:meshcase1}
    \end{figure}
    Note that each vertex has degree 2 in~$C_1$ and no vertex in~$V(P_2) \cap V(Q_2)$ belongs to~$C_1$.
    Hence, if~$C_1$ is connected, then we can return~$C' = C_1$.
    So assume that~$C_1$ is not connected.
    As shown above, before adding the two subpaths of~$P_1$ and~$P_4$, we had two connected components and since we assume~$C_1$ to not be connected, it must still consist of two connected components.
    Let~$R_1$ be the connected component of~$C_1$ that contains the added subpath of~$P_1$ and let~$R_2$ be the connected component of~$C_1$ that contains the added subpath of~$P_4$.
    We now make a final case distinction whether~$Q_1$ belongs to~$R_1$ or~$R_2$ and in both cases create the~$T$-cycle~$C'$.
    
    If~$Q_1$ belongs to~$R_1$, then we remove the subpath of~$Q_1$ between~$a_{3,1}$ and~$a_{4,1}$ and the subpath of~$P_4$ between~$b_{4,2}$ and~$a_{4,3}$ and replace them by the subpath of~$P_3$ between~$a_{3,1}$ and~$a_{3,3}$, the subpath of~$Q_3$ between~$a_{3,3}$ and~$a_{4,3}$, and the subpath of~$P_4$ between~$a_{4,1}$ and~$b_{4,2}$.
    We call the resulting graph~$C_2$ and return~$C'=C_2$.
    See \cref{fig:meshcase1} (middle) for an illustration.
    Note that since~$C_1$ is the disjoint union of two cycles, it holds that the part of~$C_1$ that is not contained in~$M$ is a set of disjoint paths, one of which has endpoints~$b_{4,2}$ and~$a_{4,3}$.
    Moreover, by construction the paths of~$C$ outside of~$M$ are not modified and therefore~$V(C2) \cap (V(G) \setminus V(M)) = V(C) \cap (V(G) \setminus V(M))$.
    Since~$M$ does not contain any vertices of~$T$, it also holds that~$C_2$ contains all vertices of~$T$.
    It remains to show that~$C_2$ is connected and that it does not contain any vertices of~$V(P_2) \cap V(Q_2)$.
    Note that the latter holds trivially by construction.
    To show the former, let~$L_1$ be the $a_{3,1}$-$a_{4,1}$-path remaining in~$R_1$ after removing the specified subpath of~$Q_1$, and let~$L_2$ be the $b_{4,2}$-$a_{4,3}$-path remaining in~$R_2$ after removing the specified subpath of~$P_4$.
    Starting at~$a_{3,1}$, we can traverse~$L_1$ to~$a_{4,1}$, then follow~$P_4$ to~$b_{4,2}$, then traverse~$L_2$ to~$a_{4,3}$, before following~$Q_3$ to~$a_{3,3}$, and finally following~$P_3$ back to~$a_{3,1}$.
    Since the prescribed walk (which is also a simple cycle) only uses edges of~$C_2$ and contains all vertices of~$C_2$, it holds that~$C_2$ is connected.

    It remains to consider the case where~$Q_1$ belongs to~$R_2$.
    This case is almost completely symmetric, the only difference being that a symmetric replacement would use vertex~$a_{2,2}$.
    However, this can easily be circumvented.
    Instead, we remove the subpath of~$Q_1$ between~$a_{1,1}$ and~$a_{3,1}$ and the subpath of~$P_1$ between~$a_{1,2}$ and~$a_{1,3}$.
    We replace them by the subpath of~$Q_3$ between~$a_{1,3}$ and~$a_{3,3}$, the subpath of~$P_3$ between~$a_{3,1}$ and~$a_{3,3}$, and the subpath of~$P_1$ between~$a_{1,1}$ and~$a_{1,2}$.
    Let~$C_3$ be the resulting graph; see \cref{fig:meshcase1} (right side) for an illustration.
    Note that no vertex of~$V(P_2) \cap V(Q_2)$ is contained in~$C_3$.
    The proof for connectivity and that it contains all vertices of~$T$ is analogous to the previous case and hence omitted.
    This concludes the proof.
\end{proof}

\section{Branching inside a well}
\label{sec:branch}

\todo[inline]{MB: start of new part}
In this section, we prove that the well of a thirsty village in a vital
instance has logarithmic depth.
Our local rerouting lemma forces branching among nested legs, and every
seven additional layers give three independent branches.
Afterwards, we show that this bounds the treewidth of vital instances.

Let~$\mathfrak{V}=(G,R,S,A,\rho,\Sigma,\Delta,\mathcal C)$ be a thirsty
$(w,b)$-village, where~$\mathcal C=\langle C_1,\ldots,C_w\rangle$,
and let~${\mathfrak{W}=(W,\Omega,\rho_\Delta,\mathcal C,\mathcal P)}$
be its well.

We first perform a simple preprocessing step.
Let~$H$ be obtained from~$S\cup\bigcup\mathcal C$ by contracting every
nontrivial component of~$S\cap C_i$ for each~$i\in[w]$, while retaining
parallel edges.
Denote the images of~$S$ and~$C_i$ by~$T$ and~$D_i$, respectively, and
let~$\mathcal Q$ be the family of images of the paths in~$\mathcal P$.
Moreover, let~$\widehat\Delta$ be the disc obtained from~$\Delta$ by the
same contractions.
All contracted paths are blank and lie strictly inside~$\Delta$.
Hence, the vertices of~$R$ and the endpoints of the paths in~$\mathcal P$ remain unchanged.
Each~$C_i$ has at least two edges outside~$S$, since a unique such
edge would shortcut a blank subpath of~$S$ and therefore contradict vitality.
Thus, the~$D_i$ remain pairwise disjoint embedded circles and~$T$ remains a spanning cycle.
Note that~$E(T)\cap E(D_i)=\emptyset$ for all~$i\in[w]$.

In the following, we call the members of~$\mathcal Q$ legs.
For~$P\in\mathcal Q$, let~$a(P)$ be the index of its apex cycle,
and let~$\widehat\Delta^P$ denote its ordinary interior, that is,
the closure of the component of~$\widehat\Delta$ minus the trace
of~$P$ that does not meet the trace of~$D_1$.
The contractions preserve the nesting of the ordinary interiors
and the successor property supplied by drainedness.
Each leg now meets its apex cycle in one vertex and every later
cycle in exactly two vertices.
We next show a laminar structure of the interiors of the legs in~$\mathcal{Q}$ and that each leg whose apex is not on~$D_w$ has a nested
leg with apex on the next cycle.

\begin{lemma}\label{lem:dry-well-caps}
For any two distinct legs~$P,Q\in\mathcal Q$, their ordinary interiors
have disjoint open interiors, or one is properly contained in the other.
Moreover, $\widehat\Delta^Q\subsetneq\widehat\Delta^P$ implies
$a(Q)>a(P)$.
If~$a(P)<w$, then there exists a leg~$Q\in\mathcal Q$ with
$a(Q)=a(P)+1$ and~$\widehat\Delta^Q\subsetneq\widehat\Delta^P$.
\end{lemma}

\begin{proof}
The legs are internally disjoint proper arcs in~$\widehat\Delta$ and all ordinary interiors are chosen on the side not meeting~$D_1$.
Hence they are laminar.

Now, let~$i$ be the index such that~$a(P)=i$.
The cycles preceding~$D_i$ are disjoint from~$P$ and lie on its
$D_1$-side.
The cycle~$D_i$ meets~$P$ in one vertex, and the remainder
of~$D_i$ is connected and lies on the same side.
Thus~$\mathsf{int}(\widehat\Delta^P)$ meets none of~$D_1,\ldots,D_i$.

If~$\widehat\Delta^Q\subsetneq\widehat\Delta^P$, the leg~$Q$ cannot meet these cycles in~$\mathsf{int}(\widehat\Delta^P)$, nor can it meet them on~$P$, because the retained cycles lie strictly inside~$\widehat\Delta$ and different legs are internally vertex-disjoint.
Thus, we have that~$a(Q)>i$.
Finally, the inherited successor property supplies a different leg inside~$\widehat\Delta^P$ meeting~$D_{i+1}$.
Its ordinary interior is properly contained in~$\widehat\Delta^P$, so its apex index is larger than~$i$ and at most~$i+1$.
\end{proof}

The next goal is to formalise the branching step.

\begin{lemma}\label{lem:dry-well-branching}
Let~$P\in\mathcal Q$ be a leg with apex on~$D_k$, where~$k+7\leq w$.
Then, there exist three legs in~$\mathcal Q$ with apex on~$D_{k+7}$ whose ordinary interiors are contained in~$\widehat\Delta^P$ and have pairwise disjoint open interiors.
\end{lemma}

\begin{proof}
Let~$P_1,P_2,\ldots,P_8\in\mathcal Q$ be legs such that~$P_1=P$, $a(P_i)=k+i-1$, and~$\widehat\Delta^{P_8}\subsetneq\ldots\subsetneq \widehat\Delta^{P_1}$.
Note that such legs exist by \zcref{lem:dry-well-caps}.
Consider the cycles~$D_{k+3},\ldots,D_{k+6}$.
Each such cycle meets each of~$P_1,P_2,P_3$ in exactly two vertices.
On the part of any one of these cycles inside~$\widehat\Delta^{P_1}$, their contacts occur in the nested order~$x_1,x_2,x_3,y_3,y_2,y_1$, where~$x_i$ and~$y_i$ are its contacts with~$P_i$.
Choose the~$x$-side consistently along the three legs.
For each~$j\in\{3,4,5,6\}$, let~$A_j$ be the corresponding subpath of~$D_{k+j}$ from~$P_1$ to~$P_3$, passing through~$P_2$.
For each~$i \in \{1,2,3\}$, let~$X_i$ be the subpath of this arm of~$P_i$ between $D_{k+3}$ and~$D_{k+6}$.

The paths~$A_3,A_4,A_5,A_6,X_1,X_2,X_3$ form a~$(4\times3)$-mesh~$M$.
If~$T\cap M=X_1\cup X_2\cup X_3$, then \zcref{lemma:Reroute4x3Mesh} gives an~$R$-cycle omitting a vertex, contradicting the vitality of~$(H,R)$.
Hence, some vertex of a horizontal path outside the three vertical paths belongs to another leg~$U_1\in\mathcal Q$.
Here, we use the fact that~$T$ spans~$H$ and that every vertex on a retained cycle belongs to a leg.
The other arms of~$P_1,P_2,P_3$ do not meet the selected horizontal paths, so~$U_1$ is different from these legs.

This contact lies in~$\mathsf{int}(\widehat\Delta^{P_1})$ and outside~$\widehat\Delta^{P_3}$.
The interior of~$U_1$ cannot cross~$P_1$ or touch~$\bd(\widehat\Delta)$,
so~$\widehat\Delta^{U_1}\subsetneq\widehat\Delta^{P_1}$.
Due to the laminar structure, either~$\widehat\Delta^{U_1}$ and~$\widehat\Delta^{P_3}$ have disjoint open interiors or~$\widehat\Delta^{P_3}\subsetneq\widehat\Delta^{U_1}$.
In the latter case, \cref{lem:dry-well-caps} implies~$k<a(U_1)<k+2$ and therefore~$a(U_1)=k+1$.
However, $\widehat\Delta^{U_1}$ and~$\widehat\Delta^{P_2}$ both contain~$\widehat\Delta^{P_3}$ and are therefore comparable, contradicting~$U_1\neq P_2$.
Thus, $\widehat\Delta^{U_1}$ and~$\widehat\Delta^{P_3}$ have disjoint
open interiors.

Since~$U_1$ meets one of~$D_{k+3},\ldots,D_{k+6}$,
we have~$a(U_1)\leq k+6$.
By a symmetric argument, there is also a leg~$U_2$ that has disjoint open interiors with~$P_3$ on the $y$-side.
Note that we have~$\widehat\Delta^{U_i}\subsetneq\widehat\Delta^{P_1}$,
$\mathsf{int}(\widehat\Delta^{U_i}) \cap \mathsf{int}(\widehat\Delta^{P_3})=\emptyset$, and~$a(U_i)\leq k+6$ for each~$i \in \{1,2\}$.

Let~$t=k+6$ and let~$A_1$ and~$A_2$ be the subpaths
of~$D_t$ from~$x_1$ to~$x_3$ and from~$y_3$ to~$y_1$,
respectively, using the contact vertices on~$D_t$.
For each~$i \in \{1,2\}$, it holds that~$U_i$ meets~$A_i$.
Now assume towards a contradiction that~$\widehat\Delta^{U_1}$ and~$\widehat\Delta^{U_2}$ are equal or one contains the other, and let~$B$ be the larger one.
By the one or two contacts of its defining leg with~$D_t$, the intersection of~$B$ with the trace of~$D_t$ is a connected arc or a single vertex.
It contains a point of each of~$A_1$ and~$A_2$.
Since~$B\subseteq\widehat\Delta^{P_1}$, it therefore contains the intervening subpath of~$D_t$ inside~$\widehat\Delta^{P_3}$.
An interior point of an edge of this subpath belongs to~$\mathsf{int}(B)\cap \mathsf{int}(\widehat\Delta^{P_3})$:
the retained cycles lie strictly inside~$\widehat\Delta$ and their edges do not belong to~$T$, a contradiction.
Thus~$U_1$ and~$U_2$ are distinct and their ordinary interiors have disjoint open interiors.

Repeatedly applying~\cref{lem:dry-well-caps} inside each of~$\widehat\Delta^{U_1}$ and~$\widehat\Delta^{U_2}$ gives
legs~$U'_1$ and~$U'_2$ with apices on~$D_{k+7}$.
Together with~$P_8$, these legs have the required properties, concluding the proof.
\end{proof}

We next bound the size of thirsty villages.

\begin{theorem}\label{thm:dry-well-height}
Let~$\mathfrak{V}=(G,R,S,A,\rho,\Sigma,\Delta,\mathcal C)$ be a thirsty
$(w,b)$-village for a vital instance~$(G,R)$.
Then, $w\leq7\lfloor\log_3|G|\rfloor+7$.
\end{theorem}

\begin{proof}
First, we construct~$H$, $T$, $D_i$ for each~$i\in[w]$, and~$\mathcal Q$ as in the beginning of the section.
Since~$T$ spans~$H$ and every vertex of~$D_1$ belongs to a leg, some leg has its apex on~$D_1$.
Repeatedly applying~\cref{lem:dry-well-branching} shows that whenever~$7q+1\leq w$, there are~$3^q$ legs with apex on~$D_{7q+1}$ whose ordinary interiors have pairwise disjoint open interiors.
These legs have distinct apex vertices, since the retained cycles
lie strictly inside~$\widehat\Delta$ and the legs are internally
vertex-disjoint.
Thus, $3^q\leq|H|\leq |G|$.
If~$w\geq7\lfloor\log_3 |G|\rfloor+8$, then consider~$q=\lfloor\log_3 |G|\rfloor+1$.
It holds that~$7q+1 \leq w$ and~$3^q>|G|$, a contradiction.
This concludes the proof.
\end{proof}

\todo[inline]{MB: End of new part.}

To conclude this section, we now combine the preparation of a thirsty village with the logarithmic bound on the depth of its well in order to finalise the proof of \zcref{thm:MainVitalSolution}.

\vitalsol*

\begin{proof}
We assume~$f_{\ref{thm:dry-village-preparation}}$ to be nondecreasing in both arguments.
This can be achieved by replacing any entry~$f_{\ref{thm:dry-village-preparation}}(b,w)$ which is smaller than~$f_{\ref{thm:dry-village-preparation}}(b-1,w)$ or~$f_{\ref{thm:dry-village-preparation}}(b,w-1)$ by~${\max(f_{\ref{thm:dry-village-preparation}}(b-1,w),f_{\ref{thm:dry-village-preparation}}(b,w-1))}$.
For~$b,n\in\mathbb{N}$, let~$h(n) \coloneqq 7\lfloor\log_3\max\{3,n\}\rfloor+8$ and~$\beta(b,n) \coloneqq \max\{2,f_{\ref{thm:dry-village-preparation}}(\max\{1,b\},h(n))\}$.
For~$n\geq3$, this gives~$\beta(b,n)\in\poly(b)\cdot\log n$.

Let~$(G,R)$ be vital and let~$S$ be its unique solution. Then~$|G|\geq3$.
If~$R=\emptyset$, any edge outside~$S$ would be a chord and yield another
cycle. Hence~$G=S$ and~$\mathsf{tw}(G)=2$. We may therefore assume
that~$\mathsf{bidim}(G,R)\geq1$.

Suppose towards a contradiction that~$\mathsf{tw}(G)\geq\beta(\mathsf{bidim}(G,R),|G|)$.
By \cref{thm:dry-village-preparation}, there exists a thirsty
$(h(|G|),\mathsf{bidim}(G,R))$-village for~$(G,R)$. However,
\cref{thm:dry-well-height} gives~$h(|G|) \leq 7\lfloor\log_3|G|\rfloor+7$, a contradiction. This concludes the proof.
\end{proof}

\section{Combing a spanning cycle}\label{sec:Combing}
All of the previous sections have been dedicated to proving \zcref{thm:MainVitalSolution} which provides a bound on the treewidth of vital instances of \textsc{Terminal Cyclability}.
However, in order to establish an irrelevant vertex rule we need to lift this statement to non-vital instances and make it algorithmic in the sense that we need a way to efficiently detect an irrelevant vertex in case the treewidth exceeds a certain bound.
To do this we employ a strategy pioneered by Golovach, Stamoulis, and Thilikos~\cite{GolovachST2023Combing} known as ``annulus combing''.
The core idea is that whenever a graph contains a large collection of ``concentric'' cycles which are almost embedded in a sphere rendition, with one vortex on either side of the ``annulus'' created by these cycles, then for any solution only a bounded number of subpaths are able to pass from one side of the annulus to the other.
The original annulus combing lemma was formulated for the \textsc{Disjoint Paths} problem and it was later improved by Cavallaro, Gorsky, Kreutzer, Thilikos, and Wiederrecht \cite{CavallaroGKTW2026Optimal}.
While this improvement already assumes bounded bidimensionality for the terminals, it is not able to handle the spanning requirement we impose.
Recently, Thilikos and Wiederrecht introduced a first variant of the lemma for a version of \textsc{Disjoint Paths} where the paths are required to contain a specified set of red vertices \cite{ThilikosW2026Spanning}.
However, this variant still does not fit our needs because it only works for annotated graphs $(G,R)$ where a parameter strictly larger than bidimensionality is bounded.
For this reason, we have to open the black box of this proof once again in order to establish the necessary properties for our case.

We start by introducing the necessary terminology.

\paragraph{Cylindrical meshes.}
To formalise fully what a ``concentric family of cycles'' is meant to be in the context of annulus combing, we make use of so-called cylindrical meshes.

Let $m,n \geq 2$ be positive integers, let $M$ be a graph, and let $C_1, \ldots, C_m$ be cycles and $P_1, \ldots, P_n$ be paths in $M$ such that the following holds for all $i \in [m]$ and $j \in [n]$:
\begin{itemize}
 \item $C_1, \ldots, C_m$ are pairwise vertex-disjoint, $P_1, \ldots, P_n$ are pairwise vertex-disjoint, and $M = C_1 \cup \ldots \cup C_m \cup P_1 \cup \ldots \cup P_n$.

 \item $C_i \cap P_j$ is a path, and if $i \in \{ 1, m \}$ or $j \in \{ 1, n \}$, then $C_i \cap P_j$ has exactly one vertex,

 \item when traversing $C_i$ starting from an endpoint of $P_1\cap C_i$, then either the paths $P_1, \ldots, P_n$ are encountered in the order listed or the next $P_j$ one encounters is $P_n$ and from here the paths are encountered in the order $P_n,\dots,P_1$, and

 \item $P_j$ has one end in $C_1$ and the other in $C_m$, and when traversing $P_j$ starting from its endpoint on $C_1$, the cycles $C_1, \ldots, C_m$ are encountered in the order listed.
\end{itemize}
If the above conditions hold for $M$, we call $M$ an \emph{$(n \times m)$-cylindrical mesh}.
The cycles $C_1, \ldots, C_m$ are called the \emph{concentric cycles}, or \emph{cycles}, of $M$ and the paths $P_1, \ldots, P_n$ are called the \emph{radial paths}, or \emph{rails}, of $M$.
We also call $(n \times n)$-cylindrical meshes \emph{$n$-cylindrical meshes}.

\paragraph{The compass of a (cylindrical) mesh.}
Let $M$ be a flat mesh in a graph $G$ where $\rho$ is a sphere rendition of $G$ witnessing the flatness of $M$.
The \emph{compass} of $M$ is the inner graph of its perimeter.

Let now $G$ be a graph and $\rho$ be a $\Sigma$-rendition of $G$ for some surface $\Sigma$.
Let~$M$ be a cylindrical mesh grounded in~$\rho$ whose outermost and innermost cycle traces bound a specified annulus~$\circledcirc$.
The compass of~$M$, with respect to this annulus, is the union of all graphs~$\sigma(c)$ over the cells~$c$ contained in~$\circledcirc$.

\paragraph{A partially embedded railed annulus.}
Let $w,r \geq 2$ be two integers.
A \emph{$(w,r)$-railed annulus} is a graph $A_{w,r}$ such that there exist pairwise vertex-disjoint cycles $C_1,\dots, C_w$, called the \emph{circles}, and pairwise vertex-disjoint paths $R_1,\dots R_r$, called the \emph{rails}, such that
\begin{enumerate}
 \item $A_{w,r} = \bigcup_{i\in [w]} C_i \cup \bigcup_{i \in [r]} R_i$ is a planar graph,
 \item $R_i$ has one endpoint on $C_1$, the other on $C_w$ for all $i\in[r]$, and is otherwise disjoint from $C_1\cup C_w$,
 \item for all $i\in[r]$, $j\in[w]$, $R_i \cap C_j$ is a path, and
 \item for each $i\in[r]$, $R_i$ visits the cycles $C_1, \dots,C_w$ in the order listed when traversing along $R_i$ starting from its endpoint on $C_1$.
\end{enumerate}
We call $C_1$ and $C_w$ the \emph{boundary circles} of $A_{w,r}$.
We say that a graph $A$ is a \emph{railed annulus} if there exist $r$ and $w$ such that $A$ is a $(w,r)$-railed annulus.

Now let $\Sigma$ be a surface and $G$ be a graph, $A\subseteq V(G)$ some set of vertices and $\rho$ be a $\Sigma$-rendition of $G-A$.
We say that a railed annulus $H \subseteq G$ is \emph{separating} in $\rho$ if
\begin{enumerate}
    \item $H$ is grounded in $\rho$,
    \item if $H$ has at least two circles and at least two rails, then there exists an annulus $\circledcirc \subseteq \Sigma$ such that the traces of the boundary circles of $H$ coincide with the two boundary components of $\circledcirc$, and the trace of every grounded subpath of a rail of~$H$ is contained in~$\circledcirc$,
    \item $\circledcirc$ is disjoint from the closures of the vortices of $\rho$, and
    \item for every circle $C$ of $H$, the trace of $C$ is a separating curve in $\Sigma$.\SW{add an illustration}
\end{enumerate}
We call the annulus $\circledcirc$ above the \emph{domain} of $H$ in $\rho$.
The \emph{crop of $G-A$ by $\circledcirc$} is the union of all graphs $\sigma(c)$ whose cell $c$ of $\rho$ is contained in $\circledcirc$.
Finally, if $(G,R)$ is an annotated graph, $A \subseteq V(G)$, $\rho$ is a $\Sigma$-rendition of $G-A$, and $H \subseteq G-A$ is a separating railed annulus with domain $\circledcirc$, we say that $H$ is \emph{blank} if the crop of $G-A$ by $\circledcirc$ does not contain any vertices from $R$.
Moreover, if $S$ is an $R$-cycle in $G$, we say that $H$ is \emph{$S$-calm} if it is blank and the crop of $G-A$ by $\circledcirc$ is disjoint from $N_S(A \cap V(S))$.

Let $p\geq 0$ and $q\geq 2$ be integers.
Let $A = A_{2p+q,r}$ be a railed annulus.
The \emph{$p$-trimming} of $A$ is the railed annulus $A_{q,r}$ obtained from $A$ by removing the first and last $p$ circles and restricting the rails to their minimal $C_{p+1}$-$C_{p+q}$-subpaths.

\paragraph{Combed $R$-cycles.}
Let $(G,R)$ be an instance of \textsc{Terminal Cyclability}, let $A \subseteq V(G)$ be some set, $\Sigma$ be a surface and $\rho$ be a $\Sigma$-rendition of $G-A$.
Moreover, let $H \subseteq G-A$ be a separating railed annulus with domain $\circledcirc$ in $\rho$.
Let $G_{\circledcirc}$ be the crop of $G-A$ by $\circledcirc$.
An~$R$-cycle~$S$ is \emph{combed in~$H$} if, within the crop
of~$G-A$ by~$\circledcirc$ after deleting the vertices of the two
boundary circles, it coincides with a union of rails of~$H$.
We say that~$S$ uses at most~$q$ rails if at most~$q$ rails
suffice.
\SW{add illustration}

We will also need the following tool from the work of Cavallaro, Gorsky, Kreutzer, Thilikos, and Wiederrecht \cite{CavallaroGKTW2026Optimal}.

Let $\Sigma$ be a surface. 
A \emph{topological linkage} $\mathcal{T}$ in $\Sigma$ with pattern $\pi$ is a set of pairwise disjoint simple curves such that $\mathcal{T}$ contains for each pair $\{a,b\} \in \pi$ a curve with endpoints $a$ and $b$.
We denote the \emph{pattern} of $\mathcal{T}$ by $\pi(\mathcal{T})$.

A pattern $\pi$ is \emph{realisable} (in $G$) if there is a linkage $\mathcal{L}$ in $G$ with $\pi(\mathcal{L}) = \pi$ and it is \emph{topologically feasible} (in $\Sigma$) if there is a topological linkage $\mathcal{T}$ in $\Sigma$ with $\pi(\mathcal{T}) = \pi$.

\begin{proposition}[Cavallaro, Gorsky, Kreutzer, Thilikos, and Wiederrecht \cite{CavallaroGKTW2026Optimal}]\label{prop:two-sided-cylinder}
 Let $G$ be a graph with a vortex-free rendition $\rho$ into a disc $\Delta$ and let $\mathcal{C} \coloneqq \{ C_1,
 \dots, C_t\}$ with $t \geq 2k$ a sequence of disjoint grounded cycles in $G$ such that the intersection of $\mathsf{bd}(\Delta)$ and $\rho$ is the set of nodes of $C_1$ and for all $i < j \in[2,t]$, the trace of $C_i$ separates the trace of $C_1$ from the trace of $C_j$.
 Let $\mathcal{P}$ be a $V(C_t){-}V(C_1)$-linkage of order $2k$ in $G$ such that $C_i \cap P$ is a path for all $1 \leq i \leq t$ and all $P\in \mathcal{P}$.
 Finally, let $\pi$ be a pattern of order $k$ on $(V(C_t) \cup V(C_1)) \cap V(\mathcal{P})$.

 If $\pi$ is topologically feasible in the closed annulus between $C_1$ and $C_t$, then it is realisable in $G$.
\end{proposition}

We are now ready to state our annulus combing lemma for $R$-cycles.

\begin{theorem}\label{thm:PutTheCycleInTheAnus}
There exist functions $f_{\ref{thm:PutTheCycleInTheAnus}},g_{\ref{thm:PutTheCycleInTheAnus}},h_{\ref{thm:PutTheCycleInTheAnus}}\colon \mathbb{N}^2 \to \mathbb{N}$ such that for every instance $(G,R)$ of \textsc{Terminal Cyclability} with $\mathsf{bidim}(G,R) \leq b$, every set $A \subseteq V(G)$, every surface $\Sigma$, and every $\Sigma$-rendition $\rho$ of $G-A$, the following holds:

If $S$ is an $R$-cycle in $G$ and $H=A_{a,s}$ is an $S$-calm railed annulus with $a \geq f_{\ref{thm:PutTheCycleInTheAnus}}(b,|G|)$ and $s \geq g_{\ref{thm:PutTheCycleInTheAnus}}(b,|G|)$ in $G-A$ and $\rho$, then there exists an $R$-cycle $S^{\star}$ in $G$ such that
\begin{enumerate}
    \item $S^{\star}$ is combed in the $h_{\ref{thm:PutTheCycleInTheAnus}}(b,|G|)$-trimming $H^{\star}$ of $H$, and
    \item $S^{\star}$ uses at most $2$ rails of $H^{\star}$.
\end{enumerate}
Moreover, $f_{\ref{thm:PutTheCycleInTheAnus}}(b,n),g_{\ref{thm:PutTheCycleInTheAnus}}(b,n),h_{\ref{thm:PutTheCycleInTheAnus}}(b,n) \in \mathbf{poly}(b) \cdot \log n + 2^{\mathbf{poly}(b)}$.
\end{theorem}

A striking feature of \zcref{thm:PutTheCycleInTheAnus} is that the number of rails our solution $S$ gets combed onto is bounded by a constant that is independent of $\mathsf{bidim}(G,R)$.
As we will find out in the proof, this is due to \zcref{lemma:Reroute4x3Mesh}.
Indeed, the usual way to prove the various annulus combing lemmas is to create a vital instance consisting only of the solution and the circles of the annulus.
This allows us to apply the treewidth bound for vital instances in order to control the way the solution interacts with the circles.
Then the actual combing happens and the paths of the solution -- of which there now can only be few -- are routed onto the rails of a central annulus.
Once this is done, we implement a second refinement step.
By repeatedly applying \zcref{lemma:Reroute4x3Mesh} to those paths crossing through the central annulus and, in each step moving slightly more inwards, we reduce the number of crossing paths more and more.
Since the initial number of crossing paths was already bounded by the preceding arguments, if we reserve enough space inside the combed annulus, this procedure will finally come to a halt and produce the desired outcome.

Let us proceed with the proof of \zcref{thm:PutTheCycleInTheAnus}.

\begin{proof}[Proof of \zcref{thm:PutTheCycleInTheAnus}]
We begin by defining our functions by setting
\begin{align*}
    z(b,n) & \coloneqq 6 \cdot (\beta(b,n) + 1) + 10\\
    f_{\ref{thm:PutTheCycleInTheAnus}}(b,n) & \coloneqq 22 \cdot \beta(b,n) + 36,\\
    g_{\ref{thm:PutTheCycleInTheAnus}}(b,n) & \coloneqq 4 \cdot \beta(b,n) + 4\text{, and}\\
    h_{\ref{thm:PutTheCycleInTheAnus}}(b,n) & \coloneqq 11 \cdot \beta(b,n) + 13,
\end{align*}
where $\beta$ denotes the function from \zcref{thm:MainVitalSolution}.
Notice that we may assume $R \neq \emptyset$ as otherwise there is nothing to prove.

\todo[inline]{MB: Here is a new part}

Let~$\mathcal C=\langle C_1,\ldots,C_a\rangle$ be the circles of~$H$.
For the following preparation, let~$H_0$ be the $1$-trimming
of~$H$ and let~$\mathcal C_0 \coloneqq \langle C_2,\ldots,C_{a-1}\rangle$.
We choose an~$R$-cycle~$S_1$ among all~$R$-cycles of~$G$ for which~$H$ is calm such that~$S_1$ minimises~$|E(S_1)\setminus E(\mathcal C_0)|$.
Let~$I \coloneqq \{i\in\{2,\ldots,a-1\} \mid V(C_i)\cap V(S_1)\neq\emptyset\}$ and~$G_1 \coloneqq S_1\cup\bigcup_{i\in I}C_i$.
Note that~$G_1$ is the connected component containing~$S_1$ in~$S_1\cup\bigcup\mathcal C_0$.
If~$I=\emptyset$, then~$S_1$ already avoids the required central annulus.
We therefore assume that~$I\neq\emptyset$.
Every~$R$-cycle in~$G_1$ is still calm for~$H$, since all edges of~$G_1$ incident with~$A$ belong to~$S_1$.
Thus, the minimality of~$S_1$ applies to every such cycle.

We first refine the relevant part of the rendition into an
actual embedding.
The support cells of the circles in~$\mathcal C_0$ lie inside the original domain of~$H$ and are therefore blank and disjoint from~$N_{S_1}(A)$.
Consider such a cell~$c$.
The nontrivial part of~$S_1\cap\sigma(c)$, if present, is one path~$P$ between two boundary nodes.
Any further vertex of~$S_1$ in this cell is its possible third boundary node~$z$.
Moreover, at most one circle of~$\mathcal C_0$ uses edges of~$\sigma(c)$, and its nontrivial intersection with this cell is a path~$Q$.

Every subpath of~$Q$ with distinct endpoints on~$P$ and interior disjoint from~$P$ must contain~$z$.
Otherwise, replacing the corresponding blank subpath of~$P$ would decrease the number of edges outside~$\mathcal C_0$.
There is at most one such excursion through~$z$.
Hence, the union in the cell is a subdivision of a path with possible additional edges to the third boundary node and can be embedded while fixing the boundary nodes.
If~$P$ is absent, the required embedding is immediate.
We choose these embeddings consistently with the circle traces, preserving their nesting.

In this drawing, we split~$S_1$ at every contact with $C_2\cup C_{a-1}$.
Let~$\mathcal S$ consist of the nontrivial resulting paths whose interiors lie in the open annulus between these two circles.
We call a member of~$\mathcal S$ a \emph{mountain}, a \emph{valley}, or a \emph{river} according as its endpoints lie on~$C_2$, on~$C_{a-1}$, or on different boundary circles.
These paths are internally vertex-disjoint.

\todo[inline]{MB: End of new part}

\paragraph{Generating a vital instance.}
We now go through two phases of reductions for $G_1$, finally ending with an annotated graph $(G_3,R)$ for which we then show that it constitutes a vital instance of \textsc{Terminal Cyclability} with some additional properties.

\textbf{Step 1:} \textsl{Remove non-red vertices of degree $2$.}
We iteratively pick some vertex $v \in V(G_1) \setminus R$ of degree exactly $2$, then we pick one of its incident edges $e$ and contract that edge -- from here on we allow the formation of loops and the occurrence of parallel edges in the following situations.
It is possible that the intersection of some circle $C_i$ with $S_1$ is a single path $P$.
If $P$ has length at least one, this operation leaves $C_i$ with two edges, one representing $P$ and the other $C_i-P$.
If $P$ is only a singleton, $C_i$ shrinks into a loop at this singleton.
In order to maintain the correct degree counts in this setting of multi graphs, we define the degree of a vertex to be the number of incidences.
This means, we count each parallel edge separately and each loop adds two to the degree of its vertex as it adds two distinct incidences. 

In case the other endpoint of $e$ belonged to $R$, we replace that vertex with the contraction vertex, thereby maintaining the total number of vertices in $R$.
Let $(G_2,R)$ be the resulting annotated graph.
For every~$i\in I$, let~$C_i''$ be the image of~$C_i$ in~$G_2$.
Each of these circles survives.
No vertex of~$R$ belongs to these circles or to the working annulus.
Moreover, every vertex of~$G_2$ is either red or has degree at least three and at most four.
Let~$S_2$ denote the image of~$S_1$.

\textbf{Step 2:} \textsl{Remove vertices of degree $3$.}
Next, for all $i\in I$ and any edge $e \in E(S_2) \cap E(C_i'')$ we now contract $e$.
As before, we allow for the formation of loops in this step.

As a result, we obtain a new annotated graph~$(G_3,R)$ and the image~$S_3$ of~$S_2$.
For every~$i\in I$, let~$C_i'$ be the image of~$C_i''$.
These circles are pairwise vertex-disjoint and satisfy~$E(S_3)\cap E(C_i')=\emptyset$.
Every vertex of~$G_3$ is either red and has degree two, or belongs to some~$C_i'$ and has degree four.

We perform these operations in the drawing described above, retaining loops and parallel edges as distinct drawn edges.
For each~$P\in\mathcal S$, we denote by~$P'$ its image in~$G_3$.
These images remain proper arcs in the working annulus.
Indeed, identifying the endpoints of a mountain or valley would require a contracted common solution-boundary-circle path joining them.
Together with the original piece, this would be the whole cycle~$S_1$, contained in the blank working annulus, contradicting~$R\neq\emptyset$.

\begin{claim}\label{claim:StillBoundedBidim}
The new instance $(G_3,R)$ satisfies $\mathsf{bidim}(G_3,R) \leq b$.
\end{claim}

\begin{claimproof}
    The claim is easy to see since $(G_3,R)$ is indeed a red-minor of $(G,R)$ and thus $\mathsf{bidim}(G_3,R) \leq \mathsf{bidim}(G,R) \leq b$.
    This fact is not changed by allowing for parallel edges or loops since these cannot contribute to the existence of minor models of simple graphs.
\end{claimproof}

\todo[inline]{MB: start of new part (many things are similar to before, some larger changes; it became to many things to point out every one individually...)}

\begin{claim}\label{clm:combing-vital}
The new instance $(G_3,R)$ is vital with solution $S_3$.
Moreover, ${\mathsf{tw}(G_3)\leq \beta(b,|G|)}$.
\end{claim}

\begin{claimproof}
Suppose that some vertex~$v$ of~$G_3$ does not belong to~$S_3$.
Then~$v\notin R$, and hence~$v$ belongs to some~$C_i'$ and
has degree four.
Its two incidences outside~$C_i'$ belong to~$S_3$,
a contradiction.
Thus~$S_3$ spans~$G_3$.

For every edge~$e$ of~$S_3$, let the cost of~$e$ be the number of original edges outside~$\mathcal C_0$ represented by~$e$.
The cost of every circle edge is zero.
Every edge of~$S_3$ has positive cost, since all common solution-circle paths were contracted in Step~2.
Reversing the reductions lifts any~$R$-cycle of~$G_3$ to
an~$R$-cycle of~$G_1$ whose number of edges outside~$\mathcal C_0$ equals its cost.
For a contracted common path, we join the two selected attachments along that path, or use only their common endpoint if both attachments occur there.
Note that all contracted common paths are blank.

Now suppose that~$S'\neq S_3$ is another~$R$-cycle in~$G_3$.
It omits an edge of~$S_3$, and its positive-cost edges form a proper subset of those of~$S_3$.
Its lift therefore contradicts the minimality of~$S_1$.
This proves uniqueness.

If~$|G_3| \leq 2$, then~$\mathsf{tw}(G_3) \leq 1 \leq \beta(b,|G|)$.
Otherwise, the spanning cycle~$S_3$ has at least three vertices.
It cannot use an edge with a parallel twin, since exchanging
that edge for its twin would give another~$R$-cycle.
Let~$G'_3$ be obtained from~$G_3$ by deleting all
loops and retaining one representative of each class of parallel edges.
Then~$(G'_3,R)$ is a simple vital instance with the same treewidth as~$G_3$.
By \zcref{thm:MainVitalSolution} and~$|G'_3|\leq|G|$, we have~$\mathsf{tw}(G_3) = \mathsf{tw}(G'_3) \leq \beta(b,|G|)$.
\end{claimproof}

Our next goal is to show that mountains and valleys do not enter deeply into the working annulus.
We use the images~$P'$ for all~$P\in\mathcal S$ in~$G_3$ for the following bramble arguments.
For a mountain~$M$ in this drawing, let~$D_M$ be the closed disc bounded by~$M$ and the corresponding outer boundary arc on the side not containing the inner boundary.

\paragraph{The area is nice for cycling: No deep valleys
or high mountains.}

\begin{claim}\label{clm:combing-depth}
No mountain of~$S_1$ meets~$C_{2\beta(b,|G|)+5}$, and no valley
of~$S_1$ meets~$C_{a-2\beta(b,|G|)-4}$.
\end{claim}

\begin{proof}[Proof of claim]
As the arguments for valleys are symmetric, we only consider
mountains.
To this end, let~$d \coloneqq 2\beta(b,|G|)+5$ and assume towards a contradiction that a mountain meets~$C_d$.
Then~$C_2,\ldots,C_d$ all meet~$S_1$ and therefore survive in~$G_3$.
Let~$M_d$ be the image of this mountain.

We construct distinct mountains~$M_d,M_{d-1},\ldots,M_3$ such that~$M_j$ meets~$C_j'$ and
\begin{align*}
{D_{M_3}\subsetneq D_{M_4} \subsetneq\ldots\subsetneq D_{M_d}}.
\end{align*}
Suppose that~$M_{j+1}$ has already been chosen.
Since it reaches~$C_{j+1}'$ and has both endpoints on~$C_2'$, there is a subpath~$Q$ of~$C_j'$ with distinct endpoints on~$M_{j+1}$ and interior contained in~$\mathsf{int}(D_{M_{j+1}})$ and disjoint from~$M_{j+1}$.

If~$Q$ were a single edge, it would shortcut the blank subpath of~$M_{j+1}$ between its endpoints and give another~$R$-cycle, contradicting~\cref{clm:combing-vital}.
Hence, $Q$ has an internal vertex~$v$.
Since~$S_3$ spans~$G_3$, the vertex~$v$ belongs to another solution piece.
This piece cannot cross~$M_{j+1}$ and cannot reach the inner boundary, which is outside~$D_{M_{j+1}}$.
It is therefore a mountain~$M_j$ whose disc is properly contained in~$D_{M_{j+1}}$.
This completes the construction.

For each~$j\in[3,d-1]$, let~$B_j \coloneqq M_j\cup C_j'$ and let~$B_d \coloneqq M_d$.
Each of these graphs is connected, and any two intersect, since a mountain reaching~$C_k'$ meets every preceding circle.
Moreover, each interior vertex belongs to at most one selected mountain and one selected circle.
A boundary vertex belongs to at most two selected mountains and none of the selected circles.
Thus, $\mathcal{B} \coloneqq \{B_j \mid j\in[3,d] \}$ is a bramble with~$d-2=2\beta(b,|G|)+3$ members, each vertex belonging to at most two members.
Its order is therefore at least~$\beta(b,|G|)+2$.
By \zcref{prop_brambles}, this implies~$\mathsf{tw}(G_3)\geq \beta(b,|G|)+1$, contradicting \cref{clm:combing-vital}.
\end{proof}

\paragraph{Even the rivers are few.}
The only remaining members of $\mathcal{S}$ we have not yet taken care of are the rivers.
These must necessarily intersect all circles of~$\mathcal C_0$.
However, similar to before, there is a bramble hiding among the rivers and this time, if there are too many, the order of this bramble exceeds the bound imposed by \zcref{thm:MainVitalSolution}.

\begin{claim}\label{clm:combing-rivers}
The number~$r$ of rivers of~$S_1$ is at most~$2\beta(b,|G|)+2$.
\end{claim}

\begin{proof}[Proof of claim]
Suppose that there are~$2\beta(b,|G|)+3$ distinct rivers, and let
$J_1',\ldots,J_{2\beta(b,|G|)+3}'$ be their images in~$G_3$.
Every working circle meets every river and therefore survives.
For each~${i\in[2\beta(b,|G|)+3]}$, let~$B_i \coloneqq J_i'\cup C_{i+2}'$.
These graphs are connected and pairwise intersecting.

The selected circles lie strictly between the two working
boundary circles.
Hence, each interior vertex belongs to at most one selected river and one selected circle, while a boundary vertex belongs to at most two selected rivers and none of the selected circles.
Thus the graphs~$B_i$ form a bramble of order at least~$\beta(b,|G|)+2$,
contradicting \cref{clm:combing-vital}.
\end{proof}

We have now established the required properties of~$S_1$ in the working annulus~$H_0$.
Let~$H_1$ be the $(2\beta(b,|G|)+4)$-trimming of the original annulus~$H$.
Enumerate its circles as~$C_1^1,\ldots,C_{a_1}^1$, where~$a_1:=a-4\beta(b,|G|)-8$.
By \cref{clm:combing-depth}, $H_1$ is disjoint from all
mountains and valleys.
We continue to consider the rivers of~$S_1$ with respect to~$H_0$, since restricting them to~$H_1$ might create additional local mountains and valleys.

If there are no rivers, then~$S_1$ already satisfies the desired conclusion.
Otherwise, enumerate the rivers as~$J_1,\ldots,J_r$ in their cyclic order, where~$r\leq2\beta(b,|G|)+2$ by \cref{clm:combing-rivers}.
At a shared endpoint, use the order of the river incidences.
Similarly, enumerate the full rails of the original annulus~$H$ as~$R_1,\ldots,R_s$ in cyclic order and orient them from~$C_1$ towards~$C_a$.

Next, we set up a collection of $5$ vertex-disjoint railed annuli obtained from $H_1$ as follows:
\begin{itemize}
    \item $H_2$ is the railed annulus with $2\beta(b,|G|) + 2$ circles and $s$ rails obtained by taking the cycles $C^1_1,\dots,C^1_{2\beta(b,|G|)+2}$ and restricting the rails of $H_1$ to their $V(C^1_1)$-$V(C^1_{2\beta(b,|G|)+2})$-subpaths.
    \item $H_3$ is the railed annulus with $4\beta(b,|G|) + 4$ circles and $s$ rails obtained by taking the cycles $C^1_{2\beta(b,|G|)+3},\dots,C^1_{6\beta(b,|G|)+6}$ and restricting the rails of $H_1$ to their $V(C^1_{2\beta(b,|G|)+3})$-$V(C^1_{6\beta(b,|G|)+6})$-subpaths.
    \item $H^{\star}$ is the $(8\beta(b,|G|)+10)$-trimming of $H$.
    \item $H_4$ is the railed annulus with $4\beta(b,|G|) + 4$ circles and $s$ rails obtained by taking the cycles $C^1_{a_1 - 6\beta(b,|G|) - 5},\dots,C^1_{a_1 - 2\beta(b,|G|) - 2}$ and restricting the rails of $H_1$ to their $V(C^1_{a_1 - 6\beta(b,|G|) - 5})$-$V(C^1_{a_1 - 2\beta(b,|G|) - 2})$-subpaths.
    \item $H_5$ is the railed annulus with $2\beta(b,|G|) + 2$ circles and $s$ rails obtained by taking the cycles $C^1_{a_1 - 2\beta(b,|G|) - 1},\dots,C^1_{a_1}$ and restricting the rails of $H_1$ to their $V(C^1_{a_1 - 2\beta(b,|G|) - 1})$-$V(C^1_{a_1})$-subpaths.
\end{itemize}

For a circle~$D$ and a full rail~$R_j$, let
$\operatorname{in}_j(D)$ and~$\operatorname{out}_j(D)$ be
the first and last vertices of~$R_j\cap D$, respectively,
when traversing~$R_j$ towards~$C_a$.
Let
\begin{align*}
    X_2 &\coloneqq \{\operatorname{out}_j(C_{2\beta(b,|G|)+2}^1) \mid j\in[s]\},\\
    X_3 &\coloneqq \{\operatorname{out}_i(C_{6\beta(b,|G|)+6}^1) \mid i\in[r]\},\\
    X_4 &\coloneqq \{\operatorname{in}_i(C_{a_1-6\beta(b,|G|)-5}^1) \mid i\in[r]\}, \text{ and}\\
    X_5 &\coloneqq \{\operatorname{in}_j(C_{a_1-2\beta(b,|G|)-1}^1) \mid j\in[s]\}.
\end{align*}

\paragraph{Channelling rivers.}
For each~$i\in[r]$, let~$P_i^1$ be the initial subpath
of~$J_i$ from its endpoint on~$C_2$ to its first contact
with~$C_{2\beta(b,|G|)+2}^1$.
Similarly, let~$P_i^2$ be the subpath obtained by following
$J_i$ from its endpoint on~$C_{a-1}$ to its first contact
with~$C_{a_1-2\beta(b,|G|)-1}^1$.

We next construct auxiliary linkages connecting the river endpoints
on the two working boundaries to distinct rails through~$H_2$
and~$H_5$, respectively.
To this end, we start with the graphs~$H_2\cup\bigcup_{i\in[r]}P_i^1$ and~$H_5\cup\bigcup_{i\in[r]}P_i^2$.
In each graph, we split every shared working-boundary endpoint into one degree-one copy for each incident river prefix.
Let~$D_1,D_2$ be the resulting auxiliary graphs, and let~$x_i^1,x_i^6$ be the respective roots belonging to~$J_i$.
Let~$X_1 \coloneqq \{x_i^1 \mid i\in[r]\}$ and let~$X_6 \coloneqq \{x_i^6 \mid i\in[r]\}$.
The prefixes~$P_i^1$ and~$P_i^2$ in each auxiliary graph are now pairwise vertex-disjoint.

Note that~$D_1$ stays on the outer side of the inner circle of~$H_2$, and~$D_2$ stays on the inner side of the outer circle of~$H_5$.
This allows us to prove the following.

\begin{claim}\label{claim:ChannelOntoRails1}
There exists an $X_1$-$X_2$-linkage $\mathcal{J}_1$ in $D_1$ and an $X_5$-$X_6$-linkage $\mathcal{J}_2$ in $D_2$, each of size precisely $r$.
\end{claim}

\begin{claimproof}
We only prove the assertion for~$D_1$, as the other case is symmetric.
Assume towards a contradiction that no~$X_1$-$X_2$-linkage of order~$r$ exists.
By Menger's theorem, there is a set~$Z\subseteq V(D_1)$ of size at most~$r-1$ separating~$X_1$ from~$X_2$.

The~$r$ prefixes~$P_i^1$ are pairwise vertex-disjoint, so one of them avoids~$Z$.
Similarly, one of the~$2\beta(b,|G|)+2 \geq r$ circles of~$H_2$ avoids~$Z$.
Let this circle be~$C_j^1$.
For each full rail, consider its subpath from its first contact with the outer circle of~$H_2$ to its last contact with the inner circle of~$H_2$.
These~$s\geq r$ paths are pairwise vertex-disjoint and contained in~$H_2$, so one of them avoids~$Z$.
Let~$Q$ be this path.
The graph~$P_i^1\cup C_j^1\cup Q$ is connected, avoids~$Z$,
and contains a vertex of~$X_1$ and a vertex of~$X_2$.
This contradicts the choice of~$Z$.
\end{claimproof}

For each~$i\in[r]$, let~$\pi_1(i)$ be the index of the rail containing the~$X_2$-endpoint of the member of~$\mathcal J_1$ starting at~$x_i^1$.
Similarly, let~$\pi_2(i)$ be the index of the rail containing the~$X_5$-endpoint of the member of~$\mathcal J_2$ ending at~$x_i^6$.
Note that both maps are injective and preserve cyclic order.
Let~$X_3' \coloneqq \{\operatorname{in}_{\pi_1(i)}(C_{2\beta(b,|G|)+3}^1) \mid i\in[r]\}$ and~$X_4' \coloneqq \{\operatorname{out}_{\pi_2(i)}(C_{a_1-2\beta(b,|G|)-2}^1) \mid i\in[r]\}$.

\todo[inline]{MB: end of changes}

\begin{claim}\label{claim:ChannelOntoRails2}
There exists an $X_3'$-$X_3$-linkage $\mathcal{I}_1$ in $H_3$ and an $X_4$-$X_4'$-linkage $\mathcal{I}_2$ in $H_4$, each of size precisely $r$.

Moreover, for every $i \in[r]$, the endpoint in $X_3'$ of the path $P \in \mathcal{I}_1$ whose endpoint in $X_3$ lies on $R_i$ belongs to $R_{\pi_1(i)}$, and the endpoint in $X_4'$ of the path $P \in \mathcal{I}_2$ whose endpoint in $X_4$ lies on $R_i$ belongs to $R_{\pi_2(i)}$.
\end{claim}

\begin{claimproof}
We prove the claim only for~$H_3$ as the case for~$H_4$ is completely symmetric.
We first select~$2r$ rails whose indices include~$[r]\cup\pi_1([r])$.
If this set does not yet contain~$2r$ rails, then we add arbitrary rails until we have selected exactly~$2r$ rails.
Note that this is possible as~$s\geq4\beta(b,|G|)+4 \geq 2r$.

We then contract the contact path of every selected rail with each boundary circle of~$H_3$.
The selected rails now form a boundary-to-boundary linkage of order~$2r$, and each circle meets each of them in a path.
The boundary circles remain cycles, since each retains at least one distinct vertex from each of the~$s$ original rails.

The required pattern is topologically feasible because
$\pi_1$ preserves cyclic order.
Since~$H_3$ has~$4\beta(b,|G|)+4 \geq 2r$ circles, \zcref{prop:two-sided-cylinder} provides us with the required linkage in the contracted graph.

Finally, we lift the linkage through the contracted contact paths, joining the selected attachments along each such path.
At a terminal, we join the selected attachment to the prescribed original entry or exit vertex.
At most one linkage member uses each contracted vertex, so lifting preserves vertex-disjointness and gives the required linkage in~$H_3$.
This concludes the proof of the claim.
\end{claimproof}

We can now combine these linkages to force the rivers onto the rails~$R_1,\ldots,R_r$ in~$H^\star$.
To this end, fix any~$i\in[r]$ and start at~$x_i^1$.
We follow its path in~$\mathcal J_1$ to~$X_2$.
Afterwards, we follow $R_{\pi_1(i)}$ to~$X_3'$.
The corresponding path in~$\mathcal I_1$ brings us to the vertex of~$X_3$ on~$R_i$.
We next follow~$R_i$ to~$X_4$, and then follow the corresponding
path in~$\mathcal I_2$ to~$X_4'$.
From there, we follow~$R_{\pi_2(i)}$ to~$X_5$ and its path
in~$\mathcal J_2$ to~$x_i^6$.

Note that the interiors of the connecting rail segments avoid their endpoint routing annuli.
Together with the confinement of~$D_1,D_2$, the disjointness of the linkages, and the injectivity of~$\pi_1$ and~$\pi_2$, this shows that the concatenated paths are pairwise vertex-disjoint in the auxiliary graphs.

This allows us to identify the root copies with their original vertices
and replace each river by its corresponding new path.
At a shared root, this restores exactly the two original solution incidences.
All other intersections with the unchanged part of~$S_1$ are the prescribed river endpoints.
Thus, the replacements preserve the original cyclic connection pattern and give an~$R$-cycle~$S^\star$.
Within~$H^\star$, away from its boundary-circle vertices, this cycle follows precisely the rails~$R_1,\ldots,R_r$.
Hence~$S^\star$ is combed in~$H^\star$ using~$r\leq2\beta(b,|G|)+2$ rails.

\paragraph{Dropping the number of rivers to $2$.}
In the last step we have to reduce the number of rivers from at most $2 \beta(b,|G|) + 2$ to $2$.
We do this as follows.
First, $H^{\star}$ has at least $z(b,|G|) = 6 \cdot (\beta(b,|G|) + 1) + 10$ circles and~$S^\star$ is combed in~$H^\star$ using~$r$ rails.
If~$r\leq2$, then~$S^\star$ remains combed using at most two rails in the~$h_{\ref{thm:PutTheCycleInTheAnus}}(b,|G|)$-trimming of~$H$, and we are done.
Hence, we may assume that~$r\geq3$.
Let us denote the circles of $H^{\star}$ as $C^{\star}_1,\dots,C^{\star}_h$.
Let~$\mathcal X_1 \coloneqq \{C_2^\star,C_3^\star,C_{h-2}^\star,C_{h-1}^\star\}$.
These four circles lie strictly inside~$H^\star$.

We then choose any three consecutive used rails and, in the union of~$S^\star$ and the circles of~$\mathcal X_1$, contract their twelve contact paths with these circles.
In the resulting graph, their images and the corresponding circle subpaths form a~$(4\times3)$-mesh~$M$ whose intersection with the image of~$S^\star$ consists precisely of its three vertical paths.
Applying the construction in the proof of \zcref{lemma:Reroute4x3Mesh} to the image of~$S^\star$ and then lifting the resulting cycle by reversing the contact-path contractions yields the lifted cycle~$S_1^\star$.
Each contracted path is blank, and its selected attachments
can be joined along that path.
Recall the proof of \zcref{lemma:Reroute4x3Mesh} -- and consider \zcref{fig:meshcase1}.
It becomes apparent that this rerouting results, in particular, in a solution $S^{\star}_1$ which has two fewer rivers than $S^{\star}$ but introduces one additional valley $X$ and one additional mountain $Y$.
In each of the three constructions, the new cycle uses exactly one of the selected rails strictly between the second and third selected circles, and all changes between rails occur on the four selected circles.
Thus, the new mountain and valley and all changes between rails avoid the~$3$-trimming of~$H^\star$.
Moreover, the new solution $S^{\star}_1$ is combed for the $3$-trimming $H^{\star}_1$ of $H^{\star}$.

For~$i\in[\beta(b,|G|)+1]$, let~$\mathcal{X}_i \coloneqq \{C_{3i-1}^\star,C_{3i}^\star, C_{h-3i+1}^\star,C_{h-3i+2}^\star\}$ and~$\mathcal{Y}_i \coloneqq \{C_j^\star \mid j\in [3i+1,h-3i]\}$.
Moreover, let $H^{\star}_i$ denote the $3i$-trimming of $H^{\star}$.
Then $\mathcal{Y}_i$ is precisely the set of circles of $H^{\star}_i$.
Let~$\mathcal R_i^\star$ be the selected family of rails whose union agrees with~$S_i^\star$ in the crop after deleting the vertices of the two boundary circles of~$H_i^\star$.
We then iteratively reduce the number of rivers that intersect $\mathcal{Y}_i$ by two until at most two rivers are remaining.

Above we have described the start of the process where $i=1$.
Now suppose we have already completed $i-1$ steps for $i \geq 2$, yielding an $R$-cycle $S^{\star}_{i-1}$ that is combed in $H^{\star}_{i-1}$.
If~$|\mathcal R_{i-1}^\star| \leq 2$, then let~$i^\star \coloneqq i-1$ and stop the iteration.
Since $|\mathcal{R}^{\star}_{i-1}| \geq 3$, we may repeat the construction above again using three consecutive rails from $\mathcal{R}^{\star}_{i-1}$ and the circles from $\mathcal{X}_i$.
Another application of \zcref{lemma:Reroute4x3Mesh}, followed by the lifting, then yields a new $R$-cycle $S^{\star}_i$ which is still combed in $H^{\star}_i$ and~$|\mathcal{R}^{\star}_i| = |\mathcal{R}^{\star}_{i-1}| - 2$.

Since the number of rails of $H^{\star}_i$ used by $S^{\star}_i$ is reduced by $2$ in every iteration and we have at most~$2\beta(b,|G|) + 2$ such rails in $H^{\star}$, after at most~$\beta(b,|G|) + 1$ iterations the process must come to a halt for some $i^{\star} \leq \beta(b,|G|) + 1$.
At this point, $S_{i^\star}^\star$ is combed in~$H_{i^\star}^\star$ using at most two rails.
Indeed, since~$h \geq z(b,|G|) = 6 \cdot (\beta(b,|G|) + 1) + 10$ we know that $H^{\star}_{i^{\star}}$ has~$h-6i^\star \geq 10$ circles.
Moreover, $H_{i^\star}^\star$ is the ${(8\beta(b,|G|)+10+3i^\star)}$-trimming of~$H$, and~$8\beta(b,|G|)+10+3i^\star \leq11\beta(b,|G|)+13 = h_{\ref{thm:PutTheCycleInTheAnus}}(b,|G|)$.
In particular, $S^{\star}_{i^{\star}}$ is combed in any trimming of $H^{\star}_{i^{\star}}$ including the $h_{\ref{thm:PutTheCycleInTheAnus}}(b,|G|)$-trimming of $H$.
At this point we may return $S^{\star}_{i^{\star}}$ as our desired $R$-cycle.
\end{proof}

\section{Detecting an irrelevant vertex}\label{sec:FindIrrelevantVertex}
The first major subroutine we need to establish is an algorithm that finds an irrelevant vertex in~$(G,R)$ whenever the treewidth exceeds the required bound.
So far, we have shown that a solution can be combed through a sufficiently insulated annulus using at most two rails.
However, so far \zcref{thm:PutTheCycleInTheAnus} requires knowledge about the way the apices are used.
Moreover, we have not talked about how to integrate \zcref{lemma:WahtToDoWithClique}, how to find a partially embedded annulus in the first place and even how to safely ensure that a large enough wall survives for us to apply the irrelevant vertex argument.
The latter is particularly important because the required treewidth threshold must remain linear in~$\log |G|$.

The goal of this section is to eventually derive \zcref{thm:MainIrrelevantVertex}.

\subsection{Getting rid of clique minors}\label{subsec:CliqueReduction}
Let us first establish that we may always assume to be in the regime of $H$-minor-free graphs.
For this we simply implement a preprocessing algorithm that either correctly determines that the input instance has bidimensionality larger than $b$ or returns an equivalent instance excluding $K_{6(b + 1)^2 + 1}$ as a minor.

\begin{lemma}\label{lemma:RemoveCliqueMinors}
There exists an algorithm that takes as input an instance $(G,R)$ of \textsc{Terminal Cyclability} together with an integer $b \geq 1$ and either correctly determines that $\mathsf{bidim}(G,R) > b$ or returns an equivalent instance $(G',R)$ such that~$G'$ is an induced subgraph of~$G$ and~$G'$ does not contain a $K_{6(b + 1)^2+1}$-minor in time $2^{2^{\mathbf{poly}(b)}} \cdot |G|^3$.
\end{lemma}

\begin{proof}
The algorithm is relatively simple.
We first run the algorithm from \zcref{prop:FindMinor} for $K_{6( b + 1 )^2+1}$ and either correctly determine that $G$ does not contain it as a minor or we find a minor model $\varphi$ in $G$ in time $2^{ 2^{\mathbf{poly}(b)} }\cdot |G|^2$.
We then hand $\varphi$ over to \zcref{lemma:WahtToDoWithClique}.
With $6(b+1)^2 + 1$, we know that this will either correctly deduce that $\mathsf{bidim}(G,R) > b$, in which case we may stop, or it will find an irrelevant vertex $v \in V(G) \setminus R$ in time $\mathbf{poly}(b) \cdot |\!| G |\!|$.

From here on we simply feed $(G-v,R)$ back into the same algorithm until after at most $|G|$ repetitions, we either stop because $\mathsf{bidim}(G,R) > b$ or we have exhausted all $K_{6( b + 1 )^2+1}$-minors in $G$ and may return the remaining annotated graph.
\end{proof}

\subsection{Finding and preparing a flat mesh}\label{subsec:FlatMesh}
From now on we may assume that $G$ does not contain any large clique minors.
This is already great for us because now we have access to a linear grid theorem in the form of \zcref{prop:LinearGrid}.
However, we need to make several additional preparatory steps.
We require a series of technical definitions and auxiliary results before we can state the core lemma of this subsection.
In broad terms, however, we need to overcome the following difficulty:
In order to apply \zcref{thm:PutTheCycleInTheAnus} we need to know that we can be sure the large separating annulus is not being used by the apices for at least some solution.
Since we do not know the solution beforehand it is hard to choose such an annulus.
However, if we choose our initial flat mesh $M$ to be large enough, we can actually generate $2a + 1$ nested candidate annuli where $a$ is the number of apices.
Then, for \textsl{any} solution one of those annuli must be a valid choice for the application of \zcref{thm:PutTheCycleInTheAnus} and thus we know that the central part of $M$ is always insulated by \textsl{some} valid choice of a separating annulus where combing can be applied.
However, we do not have an algorithmic way to determine which of these annuli is the correct choice, so we need to prove that some fixed vertex in the compass of the central part of $M$ is irrelevant no matter which of the $2a+1$ annuli ends up being the correct choice.
This is typically done by ensuring that $M$ satisfies a number of additional properties known as ``homogeneity'' and ``being suspended'' -- both terms are defined below.
We first obtain a flat and suspended mesh using a version
of the Flat Wall Theorem whose required input order is
linear in the requested order.
Homogenisation, however, requires a mesh whose order is
polynomial in the desired output order.
Applying it to the entire logarithmic-sized mesh would
therefore lose the required linear dependence on $\log |G|$.
Instead, within the compass of the central submesh we find another flat wall $W$ satisfying additional connectivity assumptions as well as a robust homogeneity property that will allow us to identify an irrelevant vertex in the centre of $W$.

The goal of this subsection is to establish an intermediate theorem.
We prove that for any $K_{6(b+1)^2 + 1}$-minor-free instance $(G,R)$ of \textsc{Terminal Cyclability} we can either find that $\mathsf{bidim}(G,R) > b$, that the treewidth of $G$ is at most $\mathsf{poly}(b) \cdot \log |G| + 2^{\mathbf{poly}(b)}$, or find a flat mesh $M$ of order $\mathsf{poly}(b) \cdot \log |G| + 2^{\mathbf{poly}(b)}$ such that there exists a ``suspended'' and homogeneous flat mesh $M'$ of order $ 2^{\mathbf{poly}(b)}$ in the compass of the centre of $M$.
This lemma will be the backbone of our proof of \zcref{thm:MainIrrelevantVertex} in \zcref{subsec:NowWeDoIt}.

\subsubsection*{Flattening meshes and finding suspended ones.}
Flat meshes play a key role in the algorithmic theory of graph minors.
However, certain applications require additional conditions on the sphere-renditions witnessing this flatness.
In this subsection we introduce the necessary terminology and import the theorems required to flatten meshes and to realise the required properties.

We actually need two different types of statements, because we require a way to turn any mesh of logarithmic size into a flat one without polynomial overheads and, at the same time, we also need a theorem that produces a flat and suspended mesh.
Below we state the variant of the Flat Mesh Theorem due to Gorsky, Seweryn, and Wiederrecht \cite{GorskySW2025Polynomial} as our go-to flat mesh option.
The algorithm to find a suspended flat mesh is due to Sau, Stamoulis, and Thilikos \cite{SauST2024More}.

\begin{proposition}[Robertson and Seymour \cite{RobertsonSeymour1995DisjointPaths,GorskySW2025Polynomial}]\label{prop:FlatMesh}
    There exist functions $f_{\ref{prop:FlatMesh}} \colon \mathbb{N}^2 \to \mathbb{N}$ and $g_{\ref{prop:FlatMesh}} \colon \mathbb{N} \to \mathbb{N}$ such that for all integers $t \geq 5,r \geq 2$ such that the following holds for every integer $k \geq f_{\ref{prop:FlatMesh}}(t,r)$.

    For every graph $G$ containing a $(k \times k)$-mesh $M\subseteq G$ either
    \begin{itemize}
        \item $G$ contains a $K_t$-minor, or
        \item there exists an $(r \times r)$-submesh $M' \subseteq M$, and a set $A \subseteq V(G) \setminus V(M')$ with $|A| \leq g_{\ref{prop:FlatMesh}}(t)$ such that $M'$ is a flat mesh in $G-A$.
    \end{itemize}
    Furthermore, there exists an $\mathbf{O}(t^3 |\!| G |\!|)$-time algorithm that takes as input $G$ and $M$ and finds either a minor model of $K_t$ or $M'$ and $A$ as above together with a rendition of $G-A$ witnessing the flatness of $M'$.
    
    Moreover, it holds that $f_{\ref{prop:FlatMesh}}(t,r) \coloneqq 100t^3(r + 2t + 2)$ and $g_{\ref{prop:FlatMesh}} \coloneqq 16t^3$
\end{proposition}

With \zcref{prop:FlatMesh} we have a quick way to obtain a flat mesh by losing only a polynomial factor depending only on the size of the excluded minor in the size of the input mesh.
This is great because together with \zcref{prop:LinearGrid} this means that in $H$-minor-free graphs it is easy to preserve the logarithmic factor.
Moreover, the algorithm above is entirely polynomial in $r$.

The downside is, the flat mesh that is returned by \zcref{prop:FlatMesh} is not necessarily ``suspended''.
This may as well be a technical issue that can be overcome by a more careful analysis of the proof of \zcref{prop:FlatMesh}, but we believe this is beyond the scope of this paper and leave this question for another time.

\paragraph{Suspended meshes.}
Let $r \geq 4$ be an integer.
Let $G$ be a graph and $M \subseteq G$ be an $(r \times r)$-mesh in $G$.
We say that $M$ is a \emph{suspended $(r\times r)$-mesh} -- or simply a \emph{suspended mesh} -- in $G$ if there exists a sphere-rendition $\rho$ of $G$ such that $M$ is flat in $\rho$ and for every cell $c\in C(M)$ the following two properties hold:
\begin{enumerate}
 \item for every non-empty set $S \subseteq N(c)$ there exists a connected subgraph of $\sigma(c)$ containing all vertices from $S$ and avoiding all vertices from $N(c)\setminus S$ and
 \item there exist $|N(c)|$ vertex-disjoint paths from $N(c)$ to the perimeter of $M$.
\end{enumerate}
We say that $\rho$ \emph{witnesses} that $M$ is suspended in $G$.

Let $G$ be a graph and $W \subseteq G$ be a flat wall in $G$ witnessed by the sphere rendition $\rho$.
The \emph{treewidth of $W$ in $G$} is the smallest integer $w$ such that the treewidth of the compass of $W$ in $\rho$ is at most $w$.

\begin{proposition}[Sau, Stamoulis, and Thilikos \cite{SauST2024More}]\label{prop:FindGoodFlatWall}
There exist functions $f_{\ref{prop:FindGoodFlatWall}},g_{\ref{prop:FindGoodFlatWall}} \colon \mathbb{N} \to \mathbb{N}$ such that for every integer $t \geq 1$, all odd integers $r \geq 5$, and every graph $G$ one of the following holds:
\begin{enumerate}
 \item $G$ contains a $K_t$ minor,
 \item $G$ has a tree-decomposition of width at most $f_{\ref{prop:FindGoodFlatWall}}(t) \cdot r$, or
 \item there is a set $A \subseteq V(G)$, where $|A| \leq g_{\ref{prop:FindGoodFlatWall}(t)}$ together with an $r$-wall $M$ that is flat and suspended in $G-A$ as witnessed by the sphere-rendition $\rho$ and the treewidth of $M$ in $G$ is at most $f_{\ref{prop:FindGoodFlatWall}}(t) \cdot r$.
\end{enumerate}
Moreover, $f_{\ref{prop:FindGoodFlatWall}}(t) \in 2^{\mathbf{O}(t^2 \log t)}$, $g_{\ref{prop:FindGoodFlatWall}}(t) \in \mathbf{O}(t^{24})$, and there exists an algorithm that finds one of the three outcomes above in time $2^{2^{\mathbf{O}(t^2 \log t)} \cdot r \log r + \mathbf{O}(r^2)} \cdot |G| + 2^{ 2^{\mathbf{O}(t^2 \log t)} \cdot r^3 \log r }$.
\end{proposition}

To obtain suspension, we apply Theorem 8 of Sau, Stamoulis, and Thilikos~\cite{SauST2024More} with order~\(r+2\) and take the central \(r\)-subwall, using Proposition 11.9 of Cavallaro, Gorsky, Kreutzer, Thilikos, and Wiederrecht~\cite{CavallaroGKTW2026Optimal}. Enlarging \(f_{\ref{prop:FindGoodFlatWall}}\) by a constant factor absorbs this increase in the requested order and preserves all stated bounds.

Notice that the running time of \zcref{prop:FindGoodFlatWall} depends exponentially, in fact the dependency is in $2^{\mathbf{O}_t(r^3 \cdot \log r)}$, on the size $r$ of the requested flat wall.
This means that we cannot hope to use \zcref{prop:FindGoodFlatWall} in order to find a suspended flat wall of size $\log |G|$ efficiently.


\paragraph{Cleaning a flat mesh.}
Our goal is to make use of large flat meshes in order to insulate a small central region of such a mesh using \zcref{thm:PutTheCycleInTheAnus}.
A core property for the entire flat mesh is that its compass must be void of red vertices.
For graphs of bounded bidimensionality, this is handled by the lemma below.

\begin{proposition}[Cavallaro, Gorsky, Kreutzer, Thilikos, and Wiederrecht \cite{CavallaroGKTW2026Optimal}]\label{lemma:BlankWall}
For all integers $r \geq 2$ and $b \geq 1$ and every annotated graph $(G,R)$, if $M$ is a flat $(w \times w)$-mesh with $w \geq (3b + 1)r + 1$ in $G$ as witnessed by the sphere-rendition $\rho$, then one of the following holds:
\begin{enumerate}
 \item $\mathsf{bidim}(G,R) \geq b$, or
 \item there exists an $(r \times r)$-submesh $M'$ of $M$ such that the compass of $M'$ does not contain a vertex from $R$.
\end{enumerate}
Moreover, if $r\geq 4$ and $\rho$ witnesses that $M$ is suspended in $G$, then $\rho$ also witnesses that $M'$ is suspended in $G$.
There also exists an algorithm that, given $G$, $\rho$, and $M$ as input finds either a red $(b \times b)$-grid as a red-minor in $G$ or the mesh $M'$ as above in time $\poly(b+r)|\!|G|\!|$.
\end{proposition}

Next, we require some additional terminology.
This is because our goal is to insulate the ``centre'' of some large flat mesh, then find a large but bounded in a function depending only on $\mathsf{bidim}(G,R)$ in size suspended flat mesh $M^{\star}$, and then prepare $M^{\star}$ to yield the desired irrelevant vertex.
To make this happen, we require further definitions and some more involved machinery to be able to state the result that declares the centre of a ``homogeneous'' suspended flat mesh irrelevant for the \textsc{Disjoint Paths} problem.

\paragraph{Layers and central submeshes.}
Let $r \geq 2$ be an integer and $M$ be an $(r \times r)$-mesh.
We define the layers of $M$ recursively as follows:
The first \emph{layer} of $M$ is its perimeter.
For $i \in [2,\lfloor \nicefrac{r}{2} \rfloor]$, the \emph{$i$th layer} of $M$ is the $(i-1)$th layer of the $((r-2) \times (r-2))$-mesh obtained from $M$ by removing its perimeter and iteratively removing vertices of degree $1$.

Given an integer $q \in [3,r]$ such that $r-q$ is even, the \emph{$q$-central submesh} of $M$ is the $(q \times q)$-mesh obtained from $M$ by removing the first $\nicefrac{(r-q)}{2}$ layers of $M$ and then iteratively deleting vertices of degree at most $1$.

\paragraph{Homogeneous meshes.}
We adapt our definition of ``homogeneity'' from the work of Gorsky, Seweryn, and Wiederrecht \cite{GorskySW2026Price} in order to preserve minimal parametric dependencies and to avoid another exponential blow-up in our dependency on $\mathsf{bidim}(G,R)$.
The result is stated in the language of ``colourful graphs'' which are a generalisation of annotated graphs.
Here each vertex can carry up to $q$ distinct annotations instead of just being red.

Let $H$ be a subgraph of a graph $G$.
An \emph{$H$-bridge} in $G$ is either an edge of~$E(G)\setminus E(H)$ with both endpoints in $V(H)$, together with its endpoints, or the graph consisting of a component $D$ of $G-V(H)$, all edges joining $D$ to $H$, and their endpoints, provided that at least one such edge exists.

Let $M$ be a grounded flat mesh in $G$ witnessed by $\rho$.
Let $\Delta_M$ be the vortex-free disc bounded by the trace of the perimeter of $M$.
For every cycle $C\subseteq M$, let $\Delta_C\subseteq\Delta_M$ be the disc bounded by the trace of $C$.
Let $\mathsf{compass}_M(C)$ be the inner graph of $C$ in $\rho$ with respect to $\Delta_C$, and let $\mathsf{int}_M(C)\coloneqq\mathsf{compass}_M(C)-V(C)$.
We omit the subscript $M$ when it is clear from the context.

A \emph{$q$-colourful graph} is a pair $(G,\chi)$ where $G$ is a graph and $\chi\colon V(G) \to 2^{[q]}$ assigns to each vertex a, possibly empty, set of at most $q$ colours.
Notice that the case $q=1$ is precisely the case of annotated graphs.
We explicitly allow for $q$ to be $0$.

Given an arbitrary $q$-colourful graph $(G,\chi)$, a set $X\subseteq V(G)$ and a subgraph $H\subseteq G$, we define
\begin{align*}
 \chi(X) \coloneqq \bigcup_{v\in X} \chi(v),
\end{align*}
as well as $\chi(H) \coloneqq \chi (V(H))$.
Note that this should not be confused with the chromatic number of $H$, which is of no concern to us.
Finally, we write $(H,\chi)$ for the \emph{$q$-colourful subgraph} of $G$, where we implicitly restrict $\chi$ to the vertex set of $H$.

Let $q$ be a non-negative integer and $(G,\chi)$ be a $q$-colourful graph.
Let $r\geq 3$ be an integer and $M$ be an $(r \times r)$-mesh in $G$ such that $M$ is flat in $G$ witnessed by a sphere-rendition $\rho$.
We say that $M$ is \emph{homogeneous} in $(G,\chi)$ if for every cycle $C\subseteq M$ it holds that
\begin{align*}
 \chi( \mathsf{int} (C) ) = \chi( \mathsf{compass} (M) ).
\end{align*}

The core reason why we need to allow for more than one annotation for our homogenisation is the presence of apex vertices.
In order to be able to properly certify redundancy for some vertex in the centre of a flat mesh one needs to ensure that all possible ways to extend an instance of \textsc{Disjoint Paths} into the cells of the flatness-witnessing rendition of a flat mesh are redundant as well.
This includes the ways the apex vertices may be connected via disjoint paths deep inside those cells.
Below we introduce the technical definitions to make this intuition exact.

\paragraph{Rooted graphs.}
Let $k\geq 0$ be an integer.
A \emph{$k$-rooted graph} is a pair $(G,\mathcal R)$ where~$\mathcal R=(r_1,\dots,r_k)\in V(G)^k$.
We sometimes write $(G,r_1,\dots,r_k)$ instead of $(G,\mathcal{R})$.
The $r_i$ are called the \emph{roots} of $(G,\mathcal{R})$.
A pair $(G,\mathcal{R})$ is a \emph{rooted graph} if it is a $k$-rooted graph for some $k$.
Rooted graphs are considered up to isomorphisms preserving every root label.
The \emph{detail} of $(G,\mathcal R)$ is~$\max\bigl\{|V(G)\setminus\{r_1,\dots,r_k\}|, |E(G)|\bigr\}$.

It is important to note that annotated graphs fundamentally differ from rooted graphs.
Indeed, one may understand a rooted graph with $k$ roots as a special case of a $k$-colourful graph where each colour is carried by a single vertex.

\paragraph{The $d$-folio of a rooted graph.}
A rooted graph $(H,q_1,\dots,q_k)$ is a \emph{rooted minor} of another rooted graph $(G,r_1,\dots,r_k)$ if there exists a minor model $\varphi$ of $H$ in $G$ such that for every $i\in[k]$ it holds that $r_i \in \varphi(q_i)$.

Given an integer $d\geq 0$ and a rooted graph $(G,\mathcal{R})$, the \emph{$d$-folio} of $G$, denoted as $d\text{-}\mathsf{folio}(G,\mathcal{R})$, is the set of all rooted graphs $(H,\mathcal{Q})$ of detail at most $d$ that are rooted minors of $(G,\mathcal{R})$.

Next, we provide the definitions making exact our initial intuition.
Each cell of a flatness-witnessing rendition for some flat mesh has first of all its own boundary to take into account as roots, but it must also respect all apex vertices present ``above'' the flat mesh.

\paragraph{Augmenting cells.}
Let $a,r \in \mathbb{N}$ with $r \geq 3$.
Let $G$ be a graph, $A \subseteq V(G)$ be a set of~$a$ vertices, and let $M$ be an $(r \times r)$-mesh in $G$ that is flat in $G-A$ witnessed by the sphere rendition $\rho$.
We denote by $C(M)$ the collection of cells contained in the disc bounded by the trace of the perimeter of $M$ that avoids the unique vortex of $\rho$.

Fix an orientation of the sphere.
For each $c\in C(M)$, let $\ell\coloneqq|N(c)|$ and choose an anticlockwise ordering~$\Omega_c=\langle x_1,\dots,x_\ell\rangle$ of $N(c)$.
Let~$\lambda_c(x_i)\coloneqq i$ for each $i\in[\ell]$.
For~$\ell=0$, we use the empty ordering and labelling.
Finally, let~$\alpha_A\colon A\to[a]$ be a bijection.

For each cell $c \in C(M)$ we fix $\pi_c \colon A \cup N(c) \to [a + |N(c)|]$ such that $\pi_c(x) = \alpha_A(x)$ for all $x \in A$ and $\pi_c(x) - a = \lambda_c(x)$ for all $x \in N(c)$.
Finally, the \emph{augmented cell} $c$ is defined as the rooted graph
\begin{align*}
 \mathbf{R}_c \coloneqq (G[V(\sigma(c)) \cup A],r_1,\dots,r_{a + |N(c)|})
\end{align*}
where $r_i = \pi_c^{-1}(i)$ for all $i \in [ a + |N(c)| ]$.
We denote the graph $G[V(\sigma(c)) \cup A]$ by $R_c$.

\paragraph{Palettes of augmented cells.}
With each $c \in C(M)$ we associate a set of graphs as follows.

Let $d \geq 0$ be an integer.
We denote by $d\text{-}\mathsf{palette}(c)$ the set $d\text{-}\mathsf{folio}(\mathbf{R}_c)$ which is referred to as the \emph{$d$-palette} of $c$.

Notice that $| d\text{-}\mathsf{palette}(c) | \in 2^{\mathbf{O}((a + d)^2)}$.
In the following, the set of all rooted graphs of detail at most $d$ with at most $a + 3$\, roots is treated as the set of colours for our colourful graph.
This means, in particular, that $q \in 2^{\mathbf{O}((a + d)^2)}$.

\paragraph{A $d$-folio homogeneous wall.}
Let $a,r \in \mathbb{N}$ with $r \geq 3$.
Let $G$ be a graph, $A \subseteq V(G)$ a set of~$a$ vertices and let $M$ be an $(r \times r)$-mesh in $G$ that is flat in $G-A$ witnessed by the sphere rendition $\rho$.
We define the \emph{$(d,A,M,\rho)$-folio augmentation} of $G$, denoted by $(G^{(d,A,M,\rho)},\phi)$, as the colourful graph obtained from $G$ by introducing, for each $c \in C(M)$, a vertex $x_c$ adjacent to exactly the vertices of $N(c)$, setting $\phi(v) \coloneqq \emptyset$ for all $v\in V(G)$ and $\phi(x_c) \coloneqq d\text{-}\mathsf{palette}(c)$ for every $c \in C(M)$.
Notice that $\rho$ may be extended to a sphere rendition $\rho'$ of $G^{(d,A,M,\rho)} - A$ such that $\rho$ is the restriction of $\rho'$ to $G - A \subseteq G^{(d,A,M,\rho)} - A$ simply by setting $\sigma_{\rho'}(c) \coloneqq \sigma_{\rho}(c) + x_c + \{x_cv \mid v\in N(c)\}$ for all $c\in C(M)$ and $\sigma_{\rho'}(c) \coloneqq \sigma(c)$ for all $c \in C(\rho) \setminus C(M)$.
We call $\rho'$ the \emph{$\rho$-augmentation} for $G^{(d,A,M,\rho)}$.

In particular, $M$ is still flat in $G^{(d,A,M,\rho)} - A$ as witnessed by $\rho'$.

Now, $M$ is said to be \emph{$(d,A)$-folio homogeneous} in $G$ if $M$ is homogeneous in $(G^{(d,A,M,\rho)}-A,\phi)$.

\paragraph{Finding a folio homogeneous mesh.}
We are finally ready to state an adapted variant of the homogenisation lemma of Gorsky, Seweryn, and Wiederrecht \cite{GorskySW2026Price} to the setting of folios.

\begin{proposition}[Cavallaro, Gorsky, Kreutzer, Thilikos, and Wiederrecht \cite{CavallaroGKTW2026Optimal}]\label{prop:FolioHomoWall}
There exists a function $f_{\ref{prop:FolioHomoWall}} \colon \mathbb{N}^{3} \to \mathbb{N}$ such that for all integers $r\geq 3$ and $d \geq 0$ the following holds.
Let $G$ be a graph, $A \subseteq V(G)$ with $a \coloneqq |A|,$ and $W_0$ be a flat $f_{\ref{prop:FolioHomoWall}}(a,d,r)$-wall in $G-A$ with a sphere rendition~$\rho$ witnessing the flatness of $W_0$ in $G-A$.
Then there exists an $r$-wall $W_1 \subseteq W_0$ such that
\begin{enumerate}
 \item $W_1$ is flat in $G-A$ witnessed by $\rho,$
 \item the tangle of $W_1$ in $G-A$ is a truncation of the tangle of $W_0$ in $G-A,$ and
 \item $W_1$ is $(d,A)$-folio homogeneous in $G$.
\end{enumerate}
In particular, if $W_0$ is suspended in $G-A$ witnessed by $\rho,$ then so is $W_1$.
Moreover, $f_{\ref{prop:FolioHomoWall}}(a,d,r) \in 2^{\poly(a + d)} \cdot \poly(r)$ and there exists an algorithm that takes as input $G,$ $A,$ $W_0,$ and $\rho$ as above and computes $W_1$ in time $2^{\poly(w + a + d)} \cdot \poly(r) \cdot (|G| + |\!|G|\!|)$ where $w$ denotes the treewidth of $W_0$ in $G$.
\end{proposition}

We choose~$f_{\ref{prop:FolioHomoWall}}$ to be nondecreasing in each argument and to always take odd values at least five.

\subsubsection*{Finding a favourable configuration}
We are now ready to prove a preliminary lemma to \zcref{thm:MainIrrelevantVertex} which allows us to either determine that $\mathsf{bidim}(G,R)$ is large, the treewidth of $G$ is small, that $G$ contains a large clique minor, or to find a huge flat mesh and smaller but still sizeable suspended and folio-homogeneous flat mesh in the compass of its centre.

Before we begin, we have to make a small adaptation to catch a degenerate case in our discussion.
The issue is that we wish to preserve pieces of a given solution $S$ in a subgraph $H$ of $G$ partially determined by an apex set that resulted from an application of \zcref{lemma:GoodConfiguration} as stated below.
For this we wish to make use of the machinery for \textsc{Disjoint Paths} and folios.
However, an apex may occur internally in a piece of $S\cap H$, and its use must be recorded even when it is not an endpoint of that piece.
To record its two incidences separately, we split the apices before we start the homogenisation process.

Let $G$ be a graph and $X \subseteq V(G)$ be some vertex set.
We say that a graph $G'$ is obtained from $G$ by \emph{splitting the vertices of $X$} if $G'$ is obtained from $G$ by replacing each vertex $u \in X$ with two copies $u_1$ and $u_2$ such that $u_1$ and $u_2$ both have the same neighbourhood as $u$ in $G$ outside of $X$ and for every vertex $x \in X$ with $ux \in E(G)$ we add the edges $u_1x_1,u_1x_2,u_2x_1,u_2x_2$.

Let $d \geq 0$ be an integer.
Given a graph $G$, a set $A \subseteq V(G)$, and a mesh $M$ in $G$ such that $M$ is flat in $G-A$, we say that $M$ is \emph{strongly $(d,A)$-homogeneous} in $G$ if $M$ is $(d,A')$-folio homogeneous in $G'$ where $G'$ is obtained from $G$ by splitting the vertices in $A$ and $A' \coloneqq \{ a_1,a_2 ~ \colon ~ a \in A \}$.

We further require a subroutine for detecting whether an annotated graph has bounded bidimensionality.
Such an algorithm can be easily deduced from the reduction given by Protopapas, Thilikos, and Wiederrecht \cite{ProtopapasThilikosWiederrecht2025ColorfulMinors} that reduces the problem of checking whether an annotated graph $(H,R_H)$ is a red-minor of $(G,R)$ in polynomial time to checking for a rooted minor in a (regular) graph.
The latter can be solved in time~$2^{2^{\mathbf{poly}(|H|)}}|G|^2$ by the rooted-minor algorithm of Cavallaro, Gorsky, Kreutzer, Thilikos, and Wiederrecht~\cite{CavallaroGKTW2026Optimal}.
Since testing if $\mathsf{bidim}(G,R) \leq b$ amounts to checking if $(G,R)$ contains a red $((b + 1) \times (b + 1))$-grid as a red-minor, we obtain the following observation.

\begin{observation}\label{obs:checkBidim}
There exists an algorithm that takes as input an integer $b \geq 0$ and an annotated graph $(G,R)$ and either determines that $\mathsf{bidim}(G,R) \leq b$ or returns a red-minor-model of a red $((b + 1) \times (b + 1))$-grid in $(G,R)$ in time $2^{2^{ \mathbf{poly} (b) }} |G|^2$.
\end{observation}

\begin{lemma}\label{lemma:GoodConfiguration}
There exist functions $f_{\ref{lemma:GoodConfiguration}} \colon \mathbb{N}^5 \to \mathbb{N}$, $g_{\ref{lemma:GoodConfiguration}},w_{\ref{lemma:GoodConfiguration}} \colon \mathbb{N}^4 \to \mathbb{N}$, $h^1_{\ref{lemma:GoodConfiguration}},h^2_{\ref{lemma:GoodConfiguration}} \colon \mathbb{N} \to \mathbb{N}$ such that for all integers $t\geq 5$, $r,s \geq 3$ and $b,d \geq 0$ and every annotated graph $(G,R)$ one of the following holds.
\begin{enumerate}
    \item $G$ contains a $K_t$-minor,
    \item $\mathsf{bidim}(G,R) > b$,
    \item $\mathsf{tw}(G) \leq f_{\ref{lemma:GoodConfiguration}}(t,b,d,r,s)$, or
    \item There exist sets $A_1,A_2 \subseteq V(G)$ with $|A_i| \leq h^i_{\ref{lemma:GoodConfiguration}}(t)$ for each $i\in[2]$, a $(2r + g_{\ref{lemma:GoodConfiguration}}(t,b,d,s))$-mesh $M_1$ which is flat in $G-A_1$ and whose compass is blank, and an $s$-wall $M_2$ which is
    \begin{itemize}
        \item flat in $G-(A_1 \cup A_2)$,
        \item its compass is entirely contained within the compass of the $g_{\ref{lemma:GoodConfiguration}}(t,b,d,s)$-central submesh of $M_1$,
        \item the graph induced in $G$ by the vertices of its compass together with $A_1 \cup A_2$ has treewidth at most $w_{\ref{lemma:GoodConfiguration}}(t,b,d,s)$,
        \item $M_2$ is suspended in $G - (A_1 \cup A_2)$, and
        \item $M_2$ is strongly $(4d,A_1 \cup A_2)$-homogeneous in $G$.
    \end{itemize}
\end{enumerate}
In the last outcome, there exist sphere-renditions $\rho_1$ and $\rho_2$ witnessing the stated properties of $M_1$ and $M_2$, respectively, with the following additional property.
Let $N^{\star}$ be the $g_{\ref{lemma:GoodConfiguration}}(t,b,d,s)$-central submesh of $M_1$.
For every $\rho_1$-aligned disc $\Delta$ containing the compass of $N^{\star}$, let $G_{\Delta}$ be obtained from the crop of $G-A_1$ by $\Delta$ by restoring $A_1$ and every edge of $G$ with both endpoints in the resulting vertex set.
Then $A_2 \subseteq V(G_{\Delta})$, and the same mesh $M_2$ is flat and suspended in $G_{\Delta}-(A_1 \cup A_2)$ and strongly $(4d,A_1 \cup A_2)$-homogeneous in $G_{\Delta}$.
Moreover, these properties can be witnessed by a rendition with the same compass and the same cells inside that compass as $\rho_2$.

There exists an algorithm that takes as input an annotated graph $(G,R)$ and finds an (annotated) minor model certifying the first or second outcome, a tree-decomposition for $G$ of width at most $f_{\ref{lemma:GoodConfiguration}}(t,b,d,r,s)$, or the meshes and sets as in the last outcome together with renditions witnessing the flatness, suspension and homogeneity in time $2^{\mathbf{poly}(b + s) \cdot 2^{\mathbf{poly}(t + d)}} \cdot \mathbf{poly}(r) \cdot \mathbf{poly}(|G|)$.

    Moreover, we have
    \begin{align*}
        f_{\ref{lemma:GoodConfiguration}}(t,b,d,r,s) & \in 2^{\mathbf{poly}(t,b,d)} \cdot \mathbf{poly}(s) \cdot r,\\
        g_{\ref{lemma:GoodConfiguration}}(t,b,d,s) , w_{\ref{lemma:GoodConfiguration}}(t,b,d,s) & \in 2^{\mathbf{poly}(t,d)}\cdot \mathbf{poly}(b + s),\\
        h^1_{\ref{lemma:GoodConfiguration}}(t) & \in \mathbf{O}(t^3)\text{, and}\\
        h^2_{\ref{lemma:GoodConfiguration}}(t) & \in \mathbf{O}(t^{24}).
    \end{align*}
\end{lemma}

\begin{proof}
Let us first set up our functions properly.
For the remainder of the proof, let
\begin{align*}
    a \coloneqq g_{\ref{prop:FlatMesh}}(t),\qquad
    c \coloneqq g_{\ref{prop:FindGoodFlatWall}}(t),\qquad
    F \coloneqq \max\{1,f_{\ref{prop:FindGoodFlatWall}}(t)+g_{\ref{prop:FindGoodFlatWall}}(t)\}.
\end{align*}
We may assume that the functions in the propositions we use are nondecreasing.
Let $q$ be the smallest odd integer satisfying
\begin{align*}
    q \geq \max\big\{5,f_{\ref{prop:FolioHomoWall}}\big(2(a+c+3),4d,s\big)\big\},
\end{align*}
and let $\ell$ be the smallest odd integer such that
\begin{align*}
   \ell \geq \max\{8q+25,F \cdot (q+4)+10\}.
\end{align*}
We now set
\begin{align*}
    g_{\ref{lemma:GoodConfiguration}}(t,b,d,s)
        &\coloneqq 2\left\lceil\frac{F \cdot \ell+\ell+10}{2}\right\rceil,\\
    f_{\ref{lemma:GoodConfiguration}}(t,b,d,r,s)
        &\coloneqq f_{\ref{prop:LinearGrid}}\big(t,t^2,
        f_{\ref{prop:FlatMesh}}\big(t,(3(b+1)+1)\cdot
        (2r+g_{\ref{lemma:GoodConfiguration}}(t,b,d,s)+2)+1\big)\big),\\
    w_{\ref{lemma:GoodConfiguration}}(t,b,d,s)
        &\coloneqq F \cdot \ell+2(a+c+3),\\
    h^1_{\ref{lemma:GoodConfiguration}}(t)&\coloneqq a,\\
    h^2_{\ref{lemma:GoodConfiguration}}(t)&\coloneqq c+3.
\end{align*}
This also immediately validates the claimed bounds on these functions.
Notice in particular that $g_{\ref{lemma:GoodConfiguration}}(t,b,d,s)$ is even.

We start by using \zcref{prop:FindMinor} to check if $G$ contains a $K_t$-minor in time $2^{2^{\mathbf{poly}(t)}}\cdot |G|^2$.
If this produces a minor model $\varphi$ of $K_t$, we return $\varphi$ and stop.
Otherwise, we may now assume that $G$ is $K_t$-minor-free.

Next we make use of \zcref{prop:LinearGrid} to either find a tree-decomposition of width at most $f_{\ref{lemma:GoodConfiguration}}(t,b,d,r,s)$ for $G$, or an $x_0$-mesh $N_0$ in $G$ where
\begin{align*}
    x_0 = f_{\ref{prop:FlatMesh}}\big(t,(3(b+1)+1)\cdot
    (2r+g_{\ref{lemma:GoodConfiguration}}(t,b,d,s)+2)+1\big)
\end{align*}
In the first outcome, we return the tree-decomposition and stop.

We proceed by applying \zcref{prop:FlatMesh} while asking for a $K_t$-minor as the alternative.
Since we have already confirmed that $G$ does not contain any $K_t$-minors, we must obtain a set $A_1\subseteq V(G)$ of size at most $a$ and an $x_1$-submesh $N_1\subseteq N_0-A_1$ together with a sphere-rendition $\rho_1$ of $G-A_1$ that witnesses that $N_1$ is flat in $G-A_1$ where
\begin{align*}
    x_1 = \big((3(b+1)+1)\cdot
    (2r+g_{\ref{lemma:GoodConfiguration}}(t,b,d,s)+2)+1\big).
\end{align*}

Now we make sure that the compass of our mesh is blank.
This is achieved by applying \zcref{lemma:BlankWall} with parameter $b+1$ to $N_1$ in $(G-A_1,R\setminus A_1)$.
As a result, we obtain a $(2r+g_{\ref{lemma:GoodConfiguration}}(t,b,d,s)+2)$-submesh $N_2$ of $N_1$ whose compass is blank, or we find a red-minor model $\psi$ of the red $((b+1)\times(b+1))$-grid.
In the second outcome, $\psi$ is also a red-minor model in $(G,R)$, so we simply return that $\mathsf{bidim}(G,R) > b$ -- or $\psi$ as a witness for that fact -- and stop.
Hence, we may assume that we arrive at the first outcome.
Notice that the assertion does not require us to fully determine whether $\mathsf{bidim}(G,R)\leq b$.

Set $M_1$ to be the submesh of $N_2$ obtained by removing its outer layer.
We have found the first of our two flat meshes.
Let $N_3$ be the $g_{\ref{lemma:GoodConfiguration}}(t,b,d,s)$-central submesh of $M_1$, let $\Delta_0$ be the $\rho_1$-aligned disc defining its compass, and let $H$ be the crop of $G-A_1$ by $\Delta_0$.
Let $G_1$ be obtained from $H$ by restoring $A_1$ together with all edges of $G$ incident with $A_1$ whose other endpoint is in $V(H)\cup A_1$.
Here we use the crop itself, rather than the induced graph on its vertices, since the latter may contain additional edges drawn outside $\Delta_0$ between boundary vertices.
The $(g_{\ref{lemma:GoodConfiguration}}(t,b,d,s)-2)$-central submesh of $N_3$ is contained in $H$.
Since
\begin{align*}
    g_{\ref{lemma:GoodConfiguration}}(t,b,d,s)-3 > F \cdot \ell,
\end{align*}
we have $\mathsf{tw}(G_1-A_1) > F \cdot \ell$.

This means that, if we now apply \zcref{prop:FindGoodFlatWall} to $G_1-A_1$ with target order $\ell$, we must find a set $B\subseteq V(G_1-A_1)$ with $|B|\leq c$ and a flat $\ell$-wall $N_4$ in $G_1-A_1-B$ together with a sphere-rendition $\rho_2$ witnessing its flatness and suspension.
We use the witness supplied by the proposition with the claimed bound on its compass treewidth.
Let $K_4$ be the compass of $N_4$ in $\rho_2$ and let $P_4$ be the perimeter of $N_4$.
Then $\mathsf{tw}(K_4)\leq f_{\ref{prop:FindGoodFlatWall}}(t)\cdot\ell$.
Adding $B$ to every bag of a tree-decomposition accommodates all edges with an endpoint in $B$ and both endpoints in $V(K_4)\cup B$.
By our choice of $F$, the resulting width is at most
\begin{align*}
    f_{\ref{prop:FindGoodFlatWall}}(t)\cdot\ell+c
    \leq F\cdot\ell.
\end{align*}
Notice that this does not require the compass itself to be induced.

\begin{claim}\label{claim:ExtendToG}
There exist a set $A_2\subseteq V(H)$ with $|A_2|\leq c+3$ and a $q$-wall $N_5$ which is flat and suspended in $G-(A_1\cup A_2)$, witnessed by a sphere-rendition in which its compass is blank and entirely contained in $\Delta_0$.
Moreover, the graph induced in $G$ by the vertices of this compass together with $A_1\cup A_2$ has treewidth at most $F \cdot \ell+a+3$.
\end{claim}

\begin{claimproof}
For every cell $e$ of $\rho_1$ contained in $\Delta_0$, let $(X_e,Y_e)$ be its cell-separation, restricted to $H-B$, where
\begin{align*}
    Y_e = V(\sigma_{\rho_1}(e))\setminus B.
\end{align*}
This separation has order at most three.
We first consider the case where the tangle $\mathcal{T}$ induced by $N_4$ points into $Y_e$ for some such cell $e$.
Set $S \coloneqq N_{\rho_1}(e)$ and
\begin{align*}
    H_e\coloneqq G[V(\sigma_{\rho_1}(e))\setminus S].
\end{align*}
Notice that $|S|\leq3$.
Since $\mathcal{T}$ points into $Y_e$, the set $Y_e\setminus X_e$ contains a complete row and a complete column of $N_4$.
There are at most three rows and three columns of $N_4$ meeting $S$.
Every row avoiding $S$ meets the complete column on this side of the separation and cannot cross to the other side.
The analogous statement holds for columns.
Hence, by omitting the rows and columns meeting $S$, we find an $(\ell-3)$-submesh of $N_4$ in $H_e$.
It follows that
\begin{align*}
    \mathsf{tw}(H_e) \geq \ell-4 > F \cdot (q+4).
\end{align*}

We now apply \zcref{prop:FindGoodFlatWall} to $H_e$ with target order $q+4$.
Again, neither the clique-minor outcome nor the bounded-treewidth outcome is possible.
Thus we find a set $B_e\subseteq V(H_e)$ of size at most $c$ and a flat and suspended $(q+4)$-wall $L_e$ in $H_e-B_e$, with a witnessing rendition whose compass $K_e$ has treewidth at most $f_{\ref{prop:FindGoodFlatWall}}(t)\cdot(q+4)$.
Let $N_5$ be its $q$-central subwall, discard $B$, and set $A_2\coloneqq S\cup B_e$.

Every edge from $H_e$ to the rest of $G-A_1$ has its other endpoint in $S$.
Since $S\subseteq A_2$, we may move the entire exterior into the unique vortex of the new rendition.
Here we retain the cells inside the trace of the perimeter of the $(q+2)$-central subwall of $L_e$, rather than restricting immediately to the compass of $N_5$.
The intervening layer ensures that all cells supporting edges of $N_5$ are retained.
Indeed, two vertex-disjoint grounded cycles cannot both use edges of the same ordinary cell, since each would require two distinct boundary nodes of that cell.
Thus the support cells of $N_5$ lie on the inner side of the intervening perimeter.
This witnesses that $N_5$ is flat in $G-(A_1\cup A_2)$, with the same compass as before.
For suspension, the connectivity witnesses inside each cell are unchanged, and the disjoint paths to the perimeter of $L_e$ may be truncated at their first contacts with the perimeter of $N_5$.
These truncated paths lie in the retained region.
Its compass is contained in the cell $e$, and hence in $\Delta_0$, and is blank.

The two layers between $N_5$ and the perimeter of $L_e$ also ensure that its compass avoids that perimeter.
Every edge induced on this compass in $G$, unless it has an endpoint in $A_1\cup S\cup B_e$, therefore belongs to $K_e$.
Indeed, such an edge belongs to $H_e-B_e$, and an edge incident with a vertex strictly inside the old compass cannot be drawn outside that compass.
Restricting a tree-decomposition of $K_e$ to the vertices of the compass of $N_5$ and adding $A_1\cup S\cup B_e$ to every bag gives width at most
\begin{align*}
    f_{\ref{prop:FindGoodFlatWall}}(t)\cdot(q+4)+a+3+c
    \leq F\cdot(q+4)+a+3
    \leq F\cdot\ell+a+3.
\end{align*}
Moreover, $A_2\subseteq V(H)$ and $|A_2|\leq c+3$.
This completes the first case.
\smallskip

Hence, we may assume that the tangle $\mathcal{T}$ of $N_4$ does not point into any cell of $\rho_1$ contained in $\Delta_0$.
Let $N_4^{\circ}$ be obtained from $N_4$ by removing its two outer layers.
We first show that $N_4^{\circ}$ is grounded in $\rho_1$.

For this, suppose that some cycle $C\subseteq N_4^{\circ}$ is contained in $\sigma_{\rho_1}(e)$ for a cell $e$.
Since $N_4\subseteq H$, every edge of $C$ belongs to a cell contained in $\Delta_0$, so $e$ is such a cell.
By our assumption, the other side of its cell-separation contains a complete row and a complete column of $N_4$, both disjoint from $C$.
Let $Z\coloneqq X_e\cap Y_e$, and call a row or column of $N_4$ untouched if it avoids $Z$.
Every untouched row meets the complete column on the other side, and every untouched column meets the complete row on that side.
Hence all untouched rows and columns lie in one component $U$ of $N_4-Z$, and $U\subseteq X_e\setminus Y_e$.
In particular, $C$ avoids every untouched row and column.

Write $P_1,\ldots,P_\ell$ for the rows of $N_4$ and let
\begin{align*}
    I\coloneqq\{j\in[\ell]\colon V(C)\cap V(P_j)\neq\emptyset\}.
\end{align*}
A cycle of a mesh meets at least two distinct rows and at least two distinct columns, so $|I|\geq2$.
Since $C\subseteq N_4^{\circ}$, we also have $\min I\geq3$ and $\max I\leq\ell-2$.
There is an untouched row above all rows indexed by $I$.
Otherwise, the rows $P_1,\ldots,P_{(\min I)-1}$ and the at least two rows indexed by $I$ would give at least four distinct rows meeting $Z$, contradicting $|Z|\leq3$.
Similarly, there is an untouched row below all rows indexed by $I$.

Let $Q_L$ and $Q_R$ be the leftmost and rightmost columns meeting $C$.
These columns are distinct, and each meets $C$ in at least two distinct vertices.
Indeed, at an extremal column, every part of $C$ leaving that column must leave towards the same side.
At each row there is only one horizontal connector towards that side, so a cycle must use such connectors at two distinct rows.
This remains true when a row-column intersection is a nontrivial path, since that contact path has only one horizontal connector towards either side.
Along each of $Q_L$ and $Q_R$, we may therefore take a path from its first contact with $C$ upwards to an untouched row above $C$, and a path from its last contact with $C$ downwards to an untouched row below $C$.
These four paths are pairwise vertex-disjoint, start on $C$, and end in $U$.
Each must meet $Z$, again contradicting $|Z|\leq3$.
This argument also allows vertices of $Z$ to lie on $C$, since the four initial vertices are distinct.
Thus every cycle of $N_4^{\circ}$ is grounded in $\rho_1$.
No vertex of $N_4^{\circ}$ is drawn in a vortex, since all its edges belong to the vortex-free crop by $\Delta_0$.
Consequently, $N_4^{\circ}$ is grounded in $\rho_1$.

Choose four pairwise disjoint $(q+4)$-subwalls $T_1,\ldots,T_4$ of $N_4^{\circ}$ in a two-by-two array, with two unused rows or columns between consecutive rectangles and between the array and the perimeter of $N_4^{\circ}$.
For each $i\in[4]$, let $Q_i$ be the $q$-central subwall of $T_i$, let $C_i$ be the perimeter of $T_i$, and let $D_i$ be the perimeter of its $(q+2)$-central subwall.
Notice that our choice of $\ell$ leaves enough room for this choice, since $\ell-4 \geq 8q+21 > 2(q+4)+6$.

We make use of the following elementary consequence of groundedness.
The part of a two-connected grounded mesh carried by a cell of order at most three is a forest whose leaves are boundary nodes: a cycle would violate groundedness, and an internal leaf would violate two-connectivity of the mesh.
There is therefore at most one nontrivial tree component, with at most three leaves, and any remaining components are isolated boundary nodes.
These forests may be drawn inside their respective cells, fixing the boundary nodes and respecting the trace convention.
This gives a planar drawing of the mesh.

We also justify why this drawing has the usual rectangular embedding, including when row-column intersections are nontrivial paths.
After suppressing its degree-two vertices, a rectangular grid of order at least three is three-connected.
To see this, delete at most two non-corner grid vertices and consider the component containing all untouched rows and columns.
If the deleted vertices share a row or column, every remaining vertex belongs to that component.
Otherwise, the only additional candidates are the other two corners of the coordinate rectangle determined by the deleted vertices.
Each such vertex has a neighbour in an untouched row or column, unless it is a corner of the whole grid and both its neighbours were deleted.
Thus the only possible additional components are isolated grid corners.
Suppressing the four corners removes this exception and preserves connectivity of the remaining vertices.

In a general mesh, a nontrivial contact path replaces an interior degree-four grid vertex by an edge whose endpoints each inherit two incident grid edges, followed by subdivisions.
Boundary contact paths are singletons by the definition of a path between two vertex sets.
Splitting a vertex of a three-connected graph into two adjacent vertices, each inheriting at least two neighbours, preserves three-connectivity.
Indeed, if neither split vertex is deleted, their joining edge preserves the original connections.
If one split vertex and one other vertex are deleted, the remaining split vertex still has a neighbour in the connected graph obtained by deleting the original vertex and that other vertex.
If both split vertices are deleted, we use connectivity after deleting the original vertex.
Hence a mesh of order at least three is a subdivision of a three-connected planar graph.
Whitney's uniqueness theorem now implies that its sphere embedding is the usual rectangular one, up to a homeomorphism and reflection.
In particular, the buffered perimeters $C_1,\ldots,C_4$ bound their expected, pairwise disjoint interior discs in both renditions.

By applying this to $C_1,\ldots,C_4$ in $\rho_1$, we obtain pairwise disjoint discs $\Delta_1,\ldots,\Delta_4$, where $\Delta_i$ is bounded by the trace of $C_i$ on the side containing $Q_i$ and excluding the remainder of the buffered mesh.
At most one of these discs contains the vortex of $\rho_1$.
Choose $i\in[4]$ for which $\Delta_i$ does not contain that vortex.
Since the trace of $C_i$ lies in $\Delta_0$, whereas the vortex lies outside $\Delta_0$, we have $\Delta_i\subseteq\Delta_0$.

Let $K_i^+$ be the inner graph of $D_i$ in $\rho_2$, on its rectangular interior side, and let $K_i$ be the compass of $Q_i$ in $\rho_2$.
The buffers and the rectangular embedding imply that
\begin{align*}
    V(K_i^+)\cap V(C_i)=\emptyset,
    \qquad
    V(K_i^+)\cap V(P_4)=\emptyset.
\end{align*}
Moreover, all support cells of $Q_i$ lie on the inner side of $D_i$.
As in the first case, this follows because two vertex-disjoint grounded cycles cannot both use edges of the same ordinary cell.
We may assume that $K_i^+$ is connected.
Indeed, any component not meeting $D_i$ has no attachment across the boundary of this inner graph, whose boundary nodes belong to $D_i$, and can be placed in the exterior vortex.
This changes neither the graph nor any support cell of $Q_i$.

The separation defined by $C_i$ in $\rho_1$ now implies that $K_i^+$ is entirely contained on the side containing $Q_i$.
Indeed, its separator is contained in $V(C_i)$, which $K_i^+$ avoids, and a connected subgraph avoiding a separator cannot meet both sides.
In particular, if we let $H_{C_i}$ be the proper inner graph of $C_i$ in $\rho_1$, we obtain
\begin{align*}
    V(K_i)\subseteq V(K_i^+)\subseteq V(H_{C_i})\subseteq V(H).
\end{align*}
These vertices lie strictly on the inner side of $C_i$, and hence are not on the boundary of $\Delta_0$.
Consequently, every edge of $G-A_1$ incident with such a vertex is already present in $H$.
Notice that this argument compares two graph separations; it does not require the renditions $\rho_1$ and $\rho_2$ to agree.

Set $A_2\coloneqq B$ and $N_5\coloneqq Q_i$.
Keep the cells of $\rho_2$ on the inner side of the trace of $D_i$ and place the entire exterior in one new vortex.
This gives a sphere-rendition of $G-(A_1\cup A_2)$: no additional edge attaches to a retained interior vertex, and every attachment to the retained disc already occurs on its boundary in $\rho_2$.
The intervening layer ensures that all support cells of $Q_i$ are retained.
It therefore witnesses flatness of $N_5$, with the same compass $K_i$, which is blank and contained in $\Delta_0$.
For suspension, the connectivity witnesses inside each cell are unchanged, and disjoint paths to the old perimeter may be truncated at their first contacts with the perimeter of $N_5$.
The truncated paths lie inside the retained disc bounded by $D_i$.

Finally, $V(K_i)\cap V(P_4)=\emptyset$.
Every edge induced on $V(K_i)$ in $G-A_1-B$ belongs to $K_4$: it is present in $H-B$, and an edge incident with a vertex strictly inside the old compass cannot be drawn outside that compass.
Restricting a tree-decomposition of $K_4$ to $V(K_i)$ and adding $A_1\cup B$ to every bag therefore gives width at most
\begin{align*}
    f_{\ref{prop:FindGoodFlatWall}}(t)\cdot\ell+a+c
    \leq F\cdot\ell+a.
\end{align*}
Moreover, $A_2=B\subseteq V(H)$ and $|A_2|\leq c$.
All choices above can be made from the supplied meshes and renditions in polynomial time.
This completes the proof of the claim.
\end{claimproof}

Let $\widetilde{G}$ be obtained from $G$ by splitting the vertices in $A_1\cup A_2$ and let
\begin{align*}
    \widetilde{A}\coloneqq\{a_1,a_2\colon a\in A_1\cup A_2\}.
\end{align*}
Notice that $|\widetilde{A}|\leq2(a+c+3)$.
Moreover, splitting the apices increases the treewidth of the augmented compass by at most $a+c+3$.
Thus its treewidth in $\widetilde{G}$ is at most
\begin{align*}
    F \cdot \ell+a+3+(a+c+3)
    \leq w_{\ref{lemma:GoodConfiguration}}(t,b,d,s).
\end{align*}
Let us apply \zcref{prop:FolioHomoWall} to $N_5$ with apex set $\widetilde{A}$ and detail $4d$ in $\widetilde{G}$.
Our choice of $q$ permits this application; if necessary, we first take the central subwall of the order requested by the proposition.
As a result, we obtain an $s$-wall $M_2$ which is flat in $\widetilde{G}-\widetilde{A}$, suspended, $(4d,\widetilde{A})$-folio homogeneous, and whose compass is blank and contained in $\Delta_0$.
Notice that this means that $M_2$ is strongly $(4d,A_1\cup A_2)$-homogeneous in $G$.
Its augmented compass is contained in the augmented compass supplied by the claim, so the required treewidth bound also follows.

It remains to verify the additional assertion concerning crops.
Let $\Delta$ be a $\rho_1$-aligned disc containing the compass of $N_3$.
The compass of $M_2$ and all vertices of $A_2$ are retained in $G_{\Delta}$.
Keep the rendition on the buffered region retained in the proof of the claim and put the remainder of $G_{\Delta}-(A_1\cup A_2)$ into the exterior vortex.
This region is contained in the compass of $N_3$, and the intervening layer retains all support cells of $N_5$, and hence of $M_2$.
As in the proof of the claim, this witnesses flatness and suspension of the same mesh $M_2$.
Every augmented cell inside the compass has exactly the same induced graph together with the split apices as before the crop.
Hence all its palettes are unchanged, which proves strong homogeneity in $G_{\Delta}$.
The same argument applies after splitting vertices outside both the compass of $N_3$ and $A_1\cup A_2$, since these vertices have no neighbours in the retained compass outside $A_1\cup A_2$.
In particular, the rendition on the compass, and any vertex chosen in its central region, can be fixed before the choice of $\Delta$.

Finally, the only calls with a nonpolynomial dependence on a requested mesh order use orders at most $\ell$.
Both $q$ and $\ell$ belong to $2^{\mathbf{poly}(t+d)}\cdot\mathbf{poly}(s)$.
The clique-minor test, the two possible applications of \zcref{prop:FindGoodFlatWall}, and the homogenisation therefore take time
\begin{align*}
    2^{\mathbf{poly}(b+s)\cdot2^{\mathbf{poly}(t+d)}}
    \cdot\mathbf{poly}(r)\cdot\mathbf{poly}(|G|).
\end{align*}
All remaining steps are polynomial in the requested outer mesh order and the size of the graph.
This completes the proof.
\end{proof}

\subsection{Actually finding the irrelevant vertex}\label{subsec:NowWeDoIt}

With \zcref{lemma:RemoveCliqueMinors}, \zcref{lemma:GoodConfiguration}, and \zcref{thm:PutTheCycleInTheAnus} we now have almost all the tools we need for the proof of \zcref{thm:MainIrrelevantVertex} in place.
The only missing piece at this point is a theorem that lets us make use of suspended and homogeneous flat meshes directly.
This is handled by the following result of Cavallaro, Gorsky, Kreutzer, Thilikos, and Wiederrecht \cite{CavallaroGKTW2026Optimal}.

\paragraph{Strongly irrelevant vertices for folios.}
Let $k,d$ be non-negative integers and $(G,R)$ be an annotated graph.
We say that a vertex $v \in V(G) \setminus R$ is \emph{strongly irrelevant} for $(k,d)\text{-}\mathsf{folio}(G,R)$ if for every ordered multiset $\mathcal{R}$ consisting of $k$ roots chosen from among the vertices of $R$ it holds that $d\text{-}\mathsf{folio}(G - v,\mathcal{R}) = d\text{-}\mathsf{folio}(G,\mathcal{R})$.

\begin{proposition}[Cavallaro, Gorsky, Kreutzer, Thilikos, and Wiederrecht \cite{CavallaroGKTW2026Optimal}]\label{prop:irrelevantVertexInMesh}
There exists a function~$f_{\ref{prop:irrelevantVertexInMesh}}\colon\mathbb N^4\to\mathbb N$ such that the following holds for all non-negative integers~$b,d,k$ and odd integers~$r\geq3$.
Let $(G,R)$ be an annotated graph with~$\mathsf{bidim}(G,R)\leq b$, let $A\subseteq V(G)$, and let~$W\subseteq G-A$ be an~$f_{\ref{prop:irrelevantVertexInMesh}}(b,d,k,r)$-wall.
Suppose that $W$ is suspended in $G-A$ and~$(4d,A)$-folio homogeneous in $G$, witnessed by a sphere-rendition~$\rho$ of $G-A$, and that its compass contains no vertex of $R$.
Then, the following holds:

If $\Delta_r$ is the disc bounded by the perimeter of the $r$-central subwall $W'$ of $W$ which contains all nodes of $W'$ in $\rho$ and $$X \coloneqq \bigcup_{\substack{c \in C(\rho) \\ c \subseteq \Delta_r}} V(\sigma(c)),$$ then every $v \in X$ is strongly irrelevant for $(k,d)\text{-}\mathsf{folio}(G,R)$.

Moreover, $f_{\ref{prop:irrelevantVertexInMesh}}(b,d,k,r) \in 2^{\poly(b + d)} \cdot \poly(k + r)$.
\end{proposition}

We note that the precise wording of the homogeneity condition in the work of Cavallaro, Gorsky, Kreutzer, Thilikos, and Wiederrecht \cite{CavallaroGKTW2026Optimal} is slightly different.
However, we use the geometric interior determined by the vortex-free $C$-disc in the definition of homogeneity.
In particular, if $W$ is $(d,A)$-folio homogeneous, then for every brick $B$ of $W$ and every rooted graph $\mathbf F$ occurring in a $d$-palette of a cell in $C(W)$, there exists a cell $c'\subseteq\mathsf{int}(\Delta_B)$ such that
$\mathbf F\in d\text{-}\mathsf{palette}(c')$.
Here, $\Delta_B$ is the vortex-free disc bounded by the trace of $B$.

Indeed, the colour $\mathbf F$ occurs at a cell vertex in the geometric interior of $B$ in the folio augmentation.
This is precisely the necessary property as supplied by the main result of \cite{GorskySW2026Price} and used in the reconstruction step of the proof of
\zcref{prop:irrelevantVertexInMesh}.
This also means that this change in the definition does not interfere with the validity of \zcref{prop:FolioHomoWall} where the tools from \cite{GorskySW2026Price} are explicitly applied.
The augmented cell graphs and their apex and boundary labels are unchanged, so our homogeneity definition is indeed sufficient to make \zcref{prop:irrelevantVertexInMesh} applicable.

\irrelevantvertex*

\begin{proof}
We begin by defining our function $\tau$.
We may assume that $f_{\ref{prop:irrelevantVertexInMesh}}$ is nondecreasing in each argument and that its values with fourth argument $3$ are odd and at least five.
\begin{align*}
    t(b) &\coloneqq 6(b+1)^2+1,\\
    k(b) &\coloneqq 2\big(h^1_{\ref{lemma:GoodConfiguration}}(t(b))
        +h^2_{\ref{lemma:GoodConfiguration}}(t(b))\big)+4,\\
    s(b) &\coloneqq f_{\ref{prop:irrelevantVertexInMesh}}\big(k(b),3,k(b),3\big),\\
    r(b,n) &\coloneqq \big(2h^1_{\ref{lemma:GoodConfiguration}}(t(b))+1\big)
        \cdot\big(f_{\ref{thm:PutTheCycleInTheAnus}}(b,n)+2\big)
        +g_{\ref{thm:PutTheCycleInTheAnus}}(b,n)+2,\\
    \tau(b,n) &\coloneqq f_{\ref{lemma:GoodConfiguration}}
        \big(t(b),b,3,r(b,n),s(b)\big).
\end{align*}
It follows from this definition and the bounds from \zcref{lemma:GoodConfiguration}, \zcref{thm:PutTheCycleInTheAnus}, and \zcref{prop:irrelevantVertexInMesh} that
\begin{align*}
    \tau(b,n)\in 2^{\mathbf{poly}(b)}\cdot\log n
\end{align*}
as desired.

We begin by calling \zcref{lemma:RemoveCliqueMinors} to either correctly determine that $\mathsf{bidim}(G,R)>b$, in which case we return this outcome and stop, or find an equivalent instance $(G_0,R)$ such that $G_0$ is an induced subgraph of $G$ and does not contain a $K_{t(b)}$-minor.
This step runs in $2^{2^{\mathbf{poly}(b)}}\cdot |G|^3$ time.
If $V(G)\setminus V(G_0)\neq\emptyset$, we may stop here and return $(G_0,R)$ since, in particular, $V(G)\setminus V(G_0)$ is irrelevant.

From now on we may assume that $G=G_0$, so $G$ contains no $K_{t(b)}$-minor.
Before proceeding, we use \zcref{obs:checkBidim} to either find a red-minor model of the red $((b+1)\times(b+1))$-grid, in which case we return the first outcome and stop, or confirm that $\mathsf{bidim}(G,R)\leq b$.
This takes $2^{2^{\mathbf{poly}(b)}}\cdot |G|^2$ time and supplies the hypothesis needed for \zcref{thm:PutTheCycleInTheAnus}.
For the remainder of the proof, write $n\coloneqq |G|$.

We now call \zcref{lemma:GoodConfiguration} with parameters $t(b),b,3,r(b,n),s(b)$.
By the discussion above, its first outcome is impossible.
If it returns a red-minor model certifying $\mathsf{bidim}(G,R)>b$, we return that certificate and stop.
Similarly, if we are presented with a tree-decomposition $(T,\beta)$ for $G$ of width at most $\tau(b,n)$, we return $(T,\beta)$ and stop.

\paragraph{Preparing the annulus.}
Hence, we may assume that \zcref{lemma:GoodConfiguration} presents us -- after $2^{2^{\mathbf{poly}(b)}}\cdot\mathbf{poly}(\log n)\cdot\mathbf{poly}(n)=2^{2^{\mathbf{poly}(b)}}\cdot\mathbf{poly}(n)$ time -- with sets $A_1,A_2\subseteq V(G)$ with $|A_i|\leq h^i_{\ref{lemma:GoodConfiguration}}(t(b))$ for each $i\in[2]$, an $x_1$-mesh $M_1$ which is flat in $G-A_1$ and whose compass is blank where
\begin{align*}
    x_1 = 2r(b,n)+g_{\ref{lemma:GoodConfiguration}}(t(b),b,3,s(b)),
\end{align*}
and an $s(b)$-wall $M_2$ which is
\begin{itemize}
    \item flat in $G-(A_1\cup A_2)$,
    \item its compass is entirely contained within the compass of the $g_{\ref{lemma:GoodConfiguration}}(t(b),b,3,s(b))$-central submesh $N^{\star}$ of $M_1$,
    \item suspended in $G-(A_1\cup A_2)$, and
    \item strongly $(12,A_1\cup A_2)$-homogeneous in $G$.
\end{itemize}
Here we use the wall returned by \zcref{prop:FolioHomoWall} in the proof of \zcref{lemma:GoodConfiguration}.
Let $\rho_1$ and $\rho_2$ be the witnessing renditions supplied by that lemma, including its additional assertion concerning crops.
This means, in particular, that we may let $C_1,\dots,C_{x(b,n)}$ with $x(b,n) \coloneqq (2h^1_{\ref{lemma:GoodConfiguration}}(t(b)) + 1) \cdot (f_{\ref{thm:PutTheCycleInTheAnus}}(b,n) + 2)$ be the outermost $x(b,n)$ layers of $M_1$ where $C_1$ is the perimeter of $M_1$ and the cycles are numbered such that $C_i$ is the $i$th layer.
Removing all those layers still leaves the central $\big(2(g_{\ref{thm:PutTheCycleInTheAnus}}(b,n) + 2) + g_{\ref{lemma:GoodConfiguration}}(t(b),b,3,s(b)) \big)$-submesh $M_3$ of $M_1$ intact which means that for any $i \in[2,x(b,n)]$, $C_i$ separates $M_2$ from the perimeter of $M_1$ in $G-A_1$.
For every $i\in[2h^1_{\ref{lemma:GoodConfiguration}}(t(b)) + 1]$ let now $C^i_1,\dots,C^i_{f_{\ref{thm:PutTheCycleInTheAnus}}(b,n)}$ be the cycles $C_{(i-1)(f_{\ref{thm:PutTheCycleInTheAnus}}(b,n) + 2)+2},\dots,C_{i(f_{\ref{thm:PutTheCycleInTheAnus}}(b,n) + 2) - 1 }$ in order.
Notice also that there exist more than $g_{\ref{thm:PutTheCycleInTheAnus}}(b,n)$ rows of $M_1$ which contain rows of $M_3$ as subpaths.
Let $R_1',\dots,R_{g_{\ref{thm:PutTheCycleInTheAnus}}(b,n)}'$ be a selection of any $g_{\ref{thm:PutTheCycleInTheAnus}}(b,n)$ such rows numbered in the order they are encountered when traversing along the leftmost vertical path of $M_3$ from top to bottom.
Each of the $R_i'$ is a row of $M_1$ that contains a row of $M_3$, therefore, $R_i'$ must intersect each of the cycles $C_j$, $j \in [x(b,n)]$ precisely twice.
Indeed, if we let $U_1,\dots,U_{x(b,n)}$ be the first $x(b,n)$ columns of $M_1$, then for every $j \in [x(b,n)]$ and every $i \in [g_{\ref{thm:PutTheCycleInTheAnus}}(b,n)]$, $R_i'$ intersects $C_j$ precisely once in $U_j \cap C_j$.
For every $i \in [g_{\ref{thm:PutTheCycleInTheAnus}}(b,n)]$ let $R_i$ be the unique $U_1$-$M_3$-subpath of $R_i'$.
It follows that every $R_i$ intersects each of the $C_j$ precisely once and this intersection is a path.
Next, let us denote for each $i \in [2h^1_{\ref{lemma:GoodConfiguration}}(t(b)) + 1]$ and $j \in [g_{\ref{thm:PutTheCycleInTheAnus}}(b,n)]$ by $R^i_j$ the unique $C^i_1$-$C^i_{f_{\ref{thm:PutTheCycleInTheAnus}}(b,n)}$-subpath of $R_j$.

All of the above now gives rise to $2h^1_{\ref{lemma:GoodConfiguration}}(t(b)) + 1$ disjoint railed annuli, each of them being flat in the rendition $\rho_1$ of $G-A_1$ that witnesses the flatness of $M_1$.
Moreover, the compass of each of those annuli is blank and their domains are pairwise disjoint.
For each $i \in [2h^1_{\ref{lemma:GoodConfiguration}}(t(b)) + 1]$ let $H_i \coloneqq \bigcup_{j \in [f_{\ref{thm:PutTheCycleInTheAnus}}(b,n)]} C^i_{j} \cup \bigcup_{j \in [g_{\ref{thm:PutTheCycleInTheAnus}}(b,n)]} R^i_j$ be the resulting railed annulus with $f_{\ref{thm:PutTheCycleInTheAnus}}(b,n)$ circles and $g_{\ref{thm:PutTheCycleInTheAnus}}(b,n)$ rails.

\paragraph{Finding the right spot.}
Notice that in $G-A_1$ each of the $H_i$ is separating.
However, each vertex in $A_1$ may have an arbitrary number of neighbours in the compass of any of the $H_i$.
In order to apply \zcref{thm:PutTheCycleInTheAnus}, we require an $S$-calm annulus for some fixed solution $S$ for $(G,R)$.

Our goal is to prove that a fixed vertex in the centre of the compass of $M_2$ is irrelevant.
Let $\Delta_3$ be the vortex-free disc bounded by the trace of the perimeter of the central $3$-subwall of $M_2$ in $\rho_2$, and set
\begin{align*}
    X_0\coloneqq\bigcup_{\substack{e\in C(\rho_2)\\ e\subseteq \Delta_3}}
        V(\sigma_{\rho_2}(e)).
\end{align*}
This set is nonempty and is contained in the blank compass of $M_2$.
Fix any $v\in X_0$.
In particular, $v\notin R\cup A_1\cup A_2$, and this choice of $v$ depends only on the input graph and the renditions supplied by \zcref{lemma:GoodConfiguration}.

It suffices to show that for every solution $S_0$ of $(G,R)$ there exists a solution avoiding $v$.
This is why we created at least $2|A_1|+1$ annuli.
For any fixed solution $S_0$, at most $2|A_1|$ vertices outside $A_1$ are adjacent to $A_1$ along $S_0$.
Since the compasses of the candidate annuli are pairwise disjoint and blank, there exists $i\in[2|A_1|+1]$ such that $H_i$ is $S_0$-calm.

\paragraph{Insulating $M_2$.}
Let us now assume that $(G,R)$ is a \texttt{yes}-instance and let $S_0$ be a fixed solution.
Furthermore, let $i\in[2|A_1|+1]$ be chosen such that $H_i$ is $S_0$-calm.
We now apply \zcref{thm:PutTheCycleInTheAnus} to $(G,R)$, $S_0$, $A_1$, and $H_i$.
As a result, we obtain a new solution $S^{\star}$ that is combed in the $h_{\ref{thm:PutTheCycleInTheAnus}}(b,n)$-trimming $H^{\star}$ of $H_i$ such that $S^{\star}$ uses at most two rails of $H^{\star}$.

Notice that $H^{\star}$ has at least ten circles.
Let $\circledcirc^{\star}$ be its domain and let $\gamma$ be a $\rho_1$-aligned simple closed curve in $\operatorname{int}(\circledcirc^{\star})$ separating its two boundary components.
Let $\Delta$ be the disc bounded by $\gamma$ on the side containing $N^{\star}$, and let $G'$ be the crop of $G-A_1$ by $\Delta$.
In particular, $\Delta$ contains the compass of $N^{\star}$.
Since $S^{\star}$ uses at most two rails of $H^{\star}$, we may choose $\gamma$ so that $S^{\star}$ intersects it in at most two nodes.
Let $u_1,u_2$ be distinct vertices corresponding to nodes of $\gamma$ that include all these intersection vertices.
If there are fewer than two intersection vertices, we choose the remaining vertices arbitrarily among the other nodes of $\gamma$.

Let $G^{*}\coloneqq G[V(G')\cup A_1]$, so $G^{*}$ is precisely the graph $G_{\Delta}$ from the additional assertion of \zcref{lemma:GoodConfiguration}.
In particular, $A_2\subseteq V(G^{*})$.
Set
\begin{align*}
    A\coloneqq A_1\cup A_2,\qquad B\coloneqq A\cup\{u_1,u_2\},
\end{align*}
and let $G^{\star}$ be obtained from $G^{*}$ by splitting every vertex of $B$.
For every $x\in B$, denote its two copies by $x_1,x_2$, and put
\begin{align*}
    \widehat{A}\coloneqq\{a_1,a_2\colon a\in A\},\qquad
    R^{\star}\coloneqq\{x_1,x_2\colon x\in B\}.
\end{align*}
The vertices $u_1,u_2$ lie outside the compass of $N^{\star}$.
Moreover, the separating layers between $\gamma$ and $N^{\star}$ imply that neither has a neighbour in the compass of $M_2$ in $G-A_1$.

\paragraph{An irrelevant vertex.}
If $v\notin V(S^{\star})$, we are already done.
Hence, we may assume that $v\in V(S^{\star})$.
By construction, the annotated graph $(G^{\star},R^{\star})$ satisfies
\begin{align*}
    \mathsf{bidim}(G^{\star},R^{\star})
    \leq\sqrt{|R^{\star}|}
    \leq |R^{\star}|
    =2|A_1\cup A_2|+4
    \leq k(b).
\end{align*}
We next encode the part of $S^{\star}$ contained in $G^{*}$ by a rooted minor of detail at most three.
Every vertex of $R^{\star}$ receives its own root label.

Suppose first that $S^{\star}\subseteq G^{*}$ and that $S^{\star}$ avoids $B$.
Then $R=\emptyset$, since the part of $G^{*}$ outside $A_1$ is blank.
In this case, retain a triangle minor of $S^{\star}$ and include every vertex of $R^{\star}$ as a separate isolated rooted vertex.
The resulting rooted graph has detail three.

Otherwise, consider all edges of $S^{\star}$ that belong to $G^{*}$.
At each $x\in B$, assign these incident edges to distinct copies $x_1,x_2$.
Every nontrivial component of the resulting subgraph in $G^{\star}$ is a path with both endpoints in $R^{\star}$ and with no internal vertex in $R^{\star}$.
Indeed, every transition from $G^{*}$ to its exterior occurs either at a vertex of $A_1$ or at one of $u_1,u_2$.
Let $\mathcal{P}$ consist of these paths together with every unused vertex of $R^{\star}$ as a singleton.
The members of $\mathcal{P}$ are pairwise vertex-disjoint.

We usually contract each nontrivial path of $\mathcal{P}$ to one vertex carrying its two endpoint labels, and retain the singleton roots as separate vertices.
We make the following two exceptions.
If $S^{\star}\subseteq G^{*}$ and $S^{\star}$ meets $B$ in precisely one vertex $x$, then its split image is an $x_1$-$x_2$-path of length at least three.
For this path, retain a three-edge path with the same endpoint labels.
If $S^{\star}\subseteq G^{*}$ and $S^{\star}$ meets $B$ in precisely two vertices, at least one of the two resulting paths has length at least two.
Retain a two-edge path for this piece and contract the other piece to a vertex carrying its two endpoint labels.
In every case, we obtain a rooted graph $J$ of detail at most three: it has at most three edges and at most three unrooted vertices.

A model of $J$ supplies the required replacement pieces.
A branch set carrying two endpoint labels supplies a path between the corresponding roots, and a model of a retained two-edge or three-edge path supplies a path of at least the prescribed length.
All these paths can be chosen pairwise vertex-disjoint.
Every vertex of $R^{\star}$ carries a root label in its prescribed component of $J$, so none can occur internally in another replacement piece.
For an isolated rooted vertex, we retain just that root as a singleton.

Identifying the copies and restoring the unchanged exterior portions of $S^{\star}$ therefore preserves the cyclic connection pattern and all vertices of $R$.
The replacement paths have no internal vertices in $B$, and every vertex outside $B$ has only one copy, so no additional intersections arise upon identification.
If the entire cycle lies in $G^{*}$, the triangle, the prescribed path lengths, or the presence of at least three distinct vertices of $B$ ensures that the recovered cycle has at least three vertices.
Otherwise, every nontrivial exterior piece has length at least two, since $G^{*}$ is induced in $G$.
Thus a replacement piece cannot collapse to the same edge as an exterior piece, and restoring the exterior again yields a cycle.
Consequently, preserving this rooted minor suffices to preserve an $R$-cycle.

We now use the additional assertion of \zcref{lemma:GoodConfiguration}.
It gives a rendition $\rho_2'$ witnessing flatness and suspension of the same wall $M_2$ in $G^{*}-A$ and strong $(12,A)$-homogeneity in $G^{*}$, with exactly the same compass and the same cells inside that compass as the fixed witness $\rho_2$.
After splitting $A$, the wall is therefore $(12,\widehat{A})$-folio homogeneous.
Splitting $u_1,u_2$ leaves every augmented cell in this compass unchanged: these vertices lie outside the compass and have no neighbours in it apart from apices.
It follows that $M_2$ is flat and suspended in $G^{\star}-\widehat{A}$ and $(12,\widehat{A})$-folio homogeneous in $G^{\star}$.
Its compass contains no vertex of $R^{\star}$.
In particular, the set $X_0$ and our fixed vertex $v$ are retained in this witnessing rendition.

By the definition of $s(b)$, the order of $M_2$ is
\begin{align*}
    f_{\ref{prop:irrelevantVertexInMesh}}\big(k(b),3,k(b),3\big).
\end{align*}
Applying \zcref{prop:irrelevantVertexInMesh} with annotation bound $k(b)$, detail three, root bound $k(b)$, and apex set $\widehat{A}$ shows that every vertex of $X_0$ is strongly irrelevant for $(k(b),3)$-$\mathsf{folio}(G^{\star},R^{\star})$.
To apply this to $J$, list all vertices of $R^{\star}$ as roots and, if necessary, repeat one root until the list has length $k(b)$, giving the same extra labels to $J$.
This does not change its detail.
Hence $J$ is still a rooted minor of $G^{\star}-v$.
By the preceding reconstruction, there is an $R$-cycle in $G-v$.

Finally, notice that the wall $M_2$, its witnessing compass in $\rho_2$, and the choice of $v$ were fixed before $S_0$, the calm annulus, and $\Delta$ were chosen.
The argument therefore applies to every solution $S_0$.
If $(G,R)$ is a \texttt{yes}-instance, then so is $(G-v,R)$, and the converse follows because $G-v$ is a subgraph of $G$.
Since $v\notin R$, the vertex $v$ is irrelevant.
We return $(G-v,R)$, which is an induced subgraph containing $R$ and is still $K_{t(b)}$-minor-free, as required by the second outcome.
The vertex $v$ is found from the fixed rendition, and neither the unknown solution nor the application of annulus combing is needed by the algorithm.
All algorithmic steps have running time $2^{2^{\mathbf{poly}(b)}}\cdot\mathbf{poly}(|G|)$, which completes the proof.
\end{proof}

\section{A single-exponential parameterized algorithm for treewidth}\label{sec:BoundedTreewidth}

In this section, we give a deterministic single-exponential dynamic program for
\Tc{}.
While such an algorithm is not surprising -- indeed Gahlawat, Rathod, and Zehavi~\cite{GRZ25} state that such an algorithm follows for planar graphs from the Catalan-structure framework of Dorn, Fomin, and Thilikos~\cite{DFT12} -- we are not aware of a proof.
Dorn, Fomin, and Thilikos did not explicitly consider~\Tc.
While it is plausible that their approach indeed applies, we give a complete proof for general graphs based on the framework of Bodlaender, Cygan, Kratsch, and Nederlof~\cite{BodlaenderCyganKratschNederlof2015}, as it does not restrict the input graph to planar or~$H$-minor-free graphs, and, we believe, results in a simpler algorithm.

We start with an intuitive description of the algorithm.
As is common with algorithms based on treewidth, we will perform dynamic programming over a nice tree decomposition.
To this end, we consider partial solutions in the subgraphs associated with the bags.
The subgraph associated with a node contains the vertices and edges introduced in the subtree rooted in that node.
A partial solution for a bag~$\beta(x)$ is a linear forest (a forest of maximum degree at most two) in this subgraph such that every forgotten vertex has degree zero or two and every forgotten terminal has degree two.
We designate one edge~$e^*$ that belongs to every solution and search for a path between its endpoints that contains all terminals.
The general idea is to merge two partial solutions using matchings on the degree-one vertices in~$\beta(x)$.
Each matching describes the connectivity of these degree-one vertices, and two partial solutions can be combined if their union is a linear forest.
They can be combined into a full solution if they contain all vertices in~$R$ and form a single path between the endpoints of~$e^*$.

We next introduce some notation.
We \emph{describe} a partial solution for a bag~$\beta(x)$ by a pair~$(d,M)$, where~$d \colon \beta(x) \to \{0,1,2\}$ gives the degrees in the partial solution.
Let~$\partial d=d^{-1}(1)$ be its set of degree-one vertices, which we call the \emph{boundary}.
Every nontrivial component of the partial solution is a path with both endpoints in~$\partial d$; the matching~$M\in\pema(\partial d)$ pairs the two endpoints of each such path.
Because we are interested only in existence, partial solutions described by the same pair behave identically under every completion in the remainder of the graph.
We go one step further and do not store every feasible pair, since their number is not bounded singly exponentially in~$|\beta(x)|$.
Instead, we store only a representative family, as defined next.

We say that two matchings~$M,N\in\pema(U)$ are \emph{compatible} if the multigraph~$(U,M\uplus N)$ consists of a single cycle.
In particular, if~$|U|=2$, two identical matchings are compatible, since their union is a cycle of length two consisting of two parallel edges.
For~$U=\emptyset$, the empty matching is compatible with itself.

Let~$\mathcal{A}'\subseteq\mathcal{A}\subseteq\pema(U)$.
We say that~$\mathcal{A}'$ \emph{represents}~$\mathcal{A}$ if, for every~$N\in\pema(U)$, the existence of a matching~$M\in\mathcal{A}$ compatible with~$N$ implies the existence of a matching~$M'\in\mathcal{A}'$ compatible with~$N$.
We will use the following result due to Bodlaender, Cygan, Kratsch, and Nederlof~\cite{BodlaenderCyganKratschNederlof2015}.

\begin{lemma}
\label{lem:rank-reduction-matchings}
There exists an algorithm that given a set~${\mathcal{A} \subseteq \pema(U)}$ of perfect matchings over a set~$U$, computes a subset~$\mathcal{A'} \subseteq \mathcal{A}$ of perfect matchings such that~$\mathcal{A'}$ represents~$\mathcal{A}$.
The set~$\mathcal{A'}$ has size at most~$2^{\frac{|U|}{2}}$ and the algorithm takes~$\mathbf{O}(|\mathcal{A}|2^{|U|})$ time.
\end{lemma}

We remark that the original result can also handle weighted perfect matchings.
However, since we will only be interested in unweighted matchings, we can assign each matching a weight of 1 and then ignore the weights for ease of presentation.
Moreover, the algorithm makes use of fast matrix multiplication to improve the running time slightly, but the above running time is sufficient for our purpose.

We next explain how to merge partial solutions using a function~$\comb$.
Let~$(d_1,M_1)$ and~$(d_2,M_2)$ be two pairs representing two partial solutions and let~$B$ be the set of vertices in the intersection of the vertex sets of the two respective graphs.
Let~$d(v) = d_1(v) + d_2(v)$ for each~$v \in B$.
If~$d(v)>2$ for some~$v\in B$, then~$\comb_B(d_1,M_1,d_2,M_2)$ fails.
Otherwise, we build the multigraph~${H = \bigl(\partial d_1\cup\partial d_2,M_1\uplus M_2\bigr)}$.
If this multigraph contains a cycle, then $\comb_B(d_1,M_1,d_2,M_2)$ again fails.
Finally, if~$H$ is acyclic, then we define~$\comb_B(d_1,M_1,d_2,M_2)$ to be the pair~$(d,M')$, where~$M'$ is the perfect matching on~$\partial d$ that pairs the two endpoints of every path component of~$H$.
We next show that this pair describes the union of the two partial solutions.

\begin{lemma}
\label{lem:comb-path-forests}
Let~$G_1$ and~$G_2$ be edge-disjoint linear forests and let~$B = V(G_1) \cap V(G_2)$.
Assume that every nontrivial component of either forest has both endpoints in~$B$.
Let~$d_i(v)$ be the degree of~$v$ in~$G_i$ for each~$i\in[2]$ and each~$v\in B$ and let~$M_i$ pair two vertices of~$\partial d_i$ precisely if they are the endpoints of the same nontrivial component of~$G_i$.
Then, $G'=(V(G_1)\cup V(G_2),E(G_1)\cup E(G_2))$ is a linear forest if and only if~$\comb_B(d_1,M_1,d_2,M_2)$ does not fail.
If it does not fail, then~$\comb_B(d_1,M_1,d_2,M_2)$ describes~$E(G_1)\cup E(G_2)$.
\end{lemma}

\begin{proof}
Every vertex outside~$B$ has degree at most two in~$G'$, while the degree of~$v\in B$ is~${d(v) = d_1(v)+d_2(v)}$.
Thus, the union has maximum degree at most two precisely if the first failure condition does not apply and we assume this for the rest of the proof.
Now, replace every nontrivial component of~$G_i$ by the corresponding edge of~$M_i$ between its endpoints, and let~$H$ be the resulting multigraph.
Every vertex that has degree one in both~$G_1$ and~$G_2$ is an endpoint of an edge in~$M_1$ and in~$M_2$ and thus has degree two in~$H$.
Since each~$G_i$ is acyclic, it holds that~$G'$ contains a cycle if and only if~$H$ does, which is the second failure condition.
Hence, we have shown the first statement of the lemma.

For the second, assume that~$H$ is acyclic and let~$(d^*,M^*) = \comb_B(d_1,M_1,d_2,M_2)$.
Note that~$d^* = d$ by construction.
Since~$H$ is acyclic, each of its connected components is a path, its degree-one vertices are precisely the vertices of~$\partial d$, and two such vertices are paired by~$M^*$ precisely if they are the endpoints of the same component of~$G'$.
That is, $(d^*,M^*)$ describes~$E(G_1)\cup E(G_2)$.
\end{proof}

The last ingredient we need for our main proof is to extend the combine operation to families of perfect matchings and show that this extension preserves representation.
To this end, fix a vertex set~$B$ and two degree functions~$d_1,d_2 \colon B \to \{0,1,2\}$.
Let~$\mathcal{A}_1 \subseteq \pema(\partial d_1)$ and~$\mathcal{A}_2 \subseteq \pema(\partial d_2)$ be two sets of perfect matchings.
Let~$d(v)=d_1(v)+d_2(v)$ for all~$v \in B$.
We define
\[\Comb_B(d_1,\mathcal{A}_1,d_2,\mathcal{A}_2)
    \coloneqq \big\{M  \mid  \exists M_1\in\mathcal{A}_1.\ \exists M_2\in\mathcal{A}_2.\ \comb_B(d_1,M_1,d_2,M_2)=(d,M)\big\}.\]
If~$d(v)>2$ for some~$v\in B$, this family is empty.

\begin{lemma}
\label{lem:comb-preserves-representation}
Let~$B$ be a set, let~$d_1,d_2 \colon B \to \{0,1,2\}$ be degree functions, and let~${\mathcal{A}'_i \subseteq \mathcal{A}_i \subseteq \pema(\partial d_i)}$ such that~$\mathcal{A}'_i$ represents~$\mathcal{A}_i$ for each~$i\in[2]$.
Then, $\Comb_B(d_1,\mathcal{A}'_1,d_2,\mathcal{A}'_2)$ represents~$\Comb_B(d_1,\mathcal{A}_1,d_2,\mathcal{A}_2)$.
\end{lemma}

\begin{proof}
Let~$d(v) = d_1(v)+d_2(v)$ for each~$v \in B$.
If~$d(v)>2$ for some~$v\in B$, then both~$\Comb_B(d_1,\mathcal{A}'_1,d_2,\mathcal{A}'_2) = \emptyset$ and~$\Comb_B(d_1,\mathcal{A}_1,d_2,\mathcal{A}_2) = \emptyset$ and the former represents the latter.
Hence, we may assume that~$d(v) \leq 2$ for all~$v\in B$.
It suffices to show that replacing $\mathcal{A}_1$ by $\mathcal{A'}_1$ preserves representation while the second family is fixed.
The same argument can then be applied to the second family.

First, assume that~$\partial d=\emptyset$.
Whenever the combine operation succeeds, every vertex of~$(B,M_1 \uplus M_2)$ has degree zero or two.
Since this multigraph is also acyclic, it cannot contain any edges.
Thus, both input matchings are empty and the claim follows immediately.
So assume that~$\partial d \neq \emptyset$.
Let~$N\in\pema(\partial d)$ be any perfect matching and assume that there exists a matching~$M$ compatible with~$N$ such that~$\comb_B(d_1,M_1,d_2,M_2)=(d,M)$ for some~$M_1 \in \mathcal{A}_1$ and~$M_2\in\mathcal{A}_2$.

We will show that there is also a matching~$M'$ compatible with~$N$ such that for some~$M'_1 \in \mathcal{A'}_1$, it holds that~${\comb_B(d_1,M'_1,d_2,M_2) = (d,M')}$.
Expanding every edge of~$M$ into the corresponding path of~$M_1\uplus M_2$ shows that~$M_1\uplus M_2\uplus N$ is a single cycle.
If~$\partial d_1=\emptyset$, then~$M_1$ is the empty matching and~$\mathcal{A}_1 = \{M_1\}$.
Hence, $\mathcal{A}'_1 = \{M_1\}$ since~$\mathcal{A}'_1$ represents~$\mathcal{A}_1$.
Otherwise, deleting the edges of~$M_1$ from this cycle yields vertex-disjoint paths whose endpoints form~$\partial d_1$.
Contracting these paths gives a perfect matching~$N_1\in\pema(\partial d_1)$ compatible with~$M_1$.
Since~$\mathcal{A}'_1$ represents~$\mathcal{A}_1$, it contains a matching~$M'_1$ compatible with~$N_1$.
Expanding the edges of~$N_1$ again shows that~$M'_1\uplus M_2\uplus N$ is a single cycle.
As~$N$ is non-empty, $M'_1\uplus M_2$ is acyclic.
Thus, the combine operation succeeds and its output is~$(d,M')$, and~$M'$ is compatible with~$N$.
Replacing~$\mathcal{A}_2$ analogously completes the proof.
\end{proof}

We also use the following simple observation.
\begin{observation}
    \label{obs:transandunion}
    Let~$U$ be a set and let~$\mathcal{A}'_i \subseteq \mathcal{A}_i \subseteq \pema(U)$ such that~$\mathcal{A}'_i$ represents~$\mathcal{A}_i$ for each~$i\in[2]$.
    Then, $\mathcal{A}'_1\cup\mathcal{A}'_2$ represents~$\mathcal{A}_1\cup\mathcal{A}_2$.
    Moreover, if~$\mathcal{A}''$ represents~$\mathcal{A}'$, and~$\mathcal{A}'$ represents~$\mathcal{A}$, then~$\mathcal{A}''$ represents~$\mathcal{A}$.
\end{observation}

\begin{proof}
Fix any~$N\in\pema(U)$.
If a matching in~$\mathcal{A}_1\cup\mathcal{A}_2$ is compatible with~$N$, it belongs to one of the two families, and the corresponding representative family contains a matching compatible with~$N$.
This proves the first statement.

For the second, note that if~$\mathcal{A}''$ represents~$\mathcal{A}'$, then it holds by definition that~$\mathcal{A}'' \subseteq \mathcal{A}'$.
If a matching in~$\mathcal{A}$ is compatible with~$N$, there also exists a matching in~$\mathcal{A'}$ that is compatible with~$N$ since~$\mathcal{A'}$ represents~$\mathcal{A}$.
Since~$\mathcal{A''}$ represents~$\mathcal{A'}$, there is also a matching in~$\mathcal{A''}$ that is compatible with~$N$.
This shows that~$\mathcal{A}''$ represents~$\mathcal{A}$.
\end{proof}

We are now ready to prove \zcref{thm:t-cycle-treewidth}, which we restate here for convenience.

\treewidthdp*

\begin{proof}
Let~$(G,R)$ be an instance of \Tc.
We assume without loss of generality that~$|R| \geq 3$ as otherwise the problem can be solved trivially in~$\mathbf{O}(|\!|G|\!|)$ time by computing biconnected components.
Fix any~$t \in R$.

We first reduce the problem to finding a simple path through all terminals while designating one edge that must belong to every solution.
Construct~$G'$ by adding two new vertices~$t_1,t_2$, both adjacent to every neighbour of~$t$ in~$G$, deleting all original edges incident with~$t$, and adding the edges~$tt_1, tt_2$, and~$t_1t_2$.
Let~$R'=R \cup \{t_1,t_2\}$ and~$e^*=tt_1$.
Note that every cycle through all vertices in~$R$ in~$G$ corresponds to a cycle through all vertices in~$R'$ by replacing the two cycle edges incident with~$t$ by a path through~$t_1,t,t_2$.
The same is also true in the other direction.
Specifically, each cycle through all vertices of~$R'$ must contain the edge~$e^*$.

Let~$G^*$ be the graph obtained by deleting~$e^*$ in~$G'$.
It now suffices to find a simple~$t$-$t_1$-path in~$G^*$ containing all vertices of~$R'$.
To obtain a tree decomposition of~$G^*$, add~$t_1$ and~$t_2$ to every bag of the given decomposition that contains~$t$.
This yields a tree decomposition of width at most~$k+2$.
We use the standard~$|G^*|\cdot \poly(k)$-time procedure to turn the tree decomposition into a nice tree decomposition~$(T,\beta)$ of the same width whose root~$r$ has bag~$\beta(r)=\{t,t_1\}$.

For a node~$x$ of~$T$, let~$V_x$ be the union of the bags in the subtree rooted at~$x$, and let~$E_x$ be the set of edges introduced in that subtree.
A \emph{partial solution} at $x$ is an edge set~$F\subseteq E_x$ such that the graph~$G_F = (V_x,F)$ satisfies the following conditions:
\begin{enumerate}
    \item it is a linear forest,
    \item every forgotten terminal~$v\in(V_x\setminus \beta(x))\cap R'$ has degree two in it, and
    \item every forgotten nonterminal~$v\in(V_x\setminus \beta(x))\setminus R'$ has degree zero or two in it.
\end{enumerate}
Thus, every nontrivial component is a path with both endpoints in~$\beta(x)$.

For a degree function~$d\colon \beta(x) \rightarrow \{0,1,2\}$, let~$\mathcal{A}_x(d)$ be the family of perfect matchings~$M \in \pema(\partial d)$ such that there exists a partial solution~$F\subseteq E_x$ that is described by~$(d,M)$.
We next present the dynamic program with its recurrences (which will later be compressed into representative families).

\paragraph{Leaf nodes.}
If $x$ is a leaf, then $ \beta(x)=\{v\}$ and $E_x=\emptyset$.
Hence,
\[\mathcal{A}_x(d)=
    \begin{cases}
        \{\emptyset\}, & \text{if $d(v) = 0$,}\\
        \emptyset,     & \text{otherwise.}
    \end{cases}\]

\paragraph{Introduce-vertex nodes.}
Suppose that~$x$ introduces a vertex $v$ and has child $y$, that is, $\beta(x)=\beta(y)\cup\{v\}$.
For a degree function~$d$ on~$\beta(x)$, let~$d'(u)=d(u)$ for all~$u \in \beta(y)$.
Since no edge incident with $v$ has yet been
introduced, we get
\[\mathcal{A}_x(d)= \begin{cases}
    \mathcal{A}_y(d'), & \text{if }d(v)=0,\\
    \emptyset,         & \text{otherwise.}
  \end{cases}\]

\paragraph{Introduce-edge nodes.}
Suppose that~$x$ introduces the edge~$e=\{u,v\}$ and has child~$y$.
Note that~$\beta(x)=\beta(y)$.
If $d(u)=0$ or $d(v)=0$, then the introduced edge cannot be used and~$\mathcal{A}_x(d)=\mathcal{A}_y(d)$.
Otherwise, let~$d'(w) = d(w)$ for all~$w \in \beta(x) \setminus \{u,v\}$, $d'(u) = d(u) - 1$, and~$d'(v) = d(v) - 1$.
Moreover, let~$d''(u) = d''(v) = 1$ and~$d''(w) = 0$ for all~$w \in \beta(x) \setminus \{u,v\}$.
Then, we have
\[\mathcal{A}_x(d) = \mathcal{A}_y(d) \cup \Comb_{\beta(x)}(d',\mathcal{A}_y(d'),d'',\{\{e\}\}).\]
The first term is for the case where the edge~$e$ is not used by a solution and the second term uses~$e$ as a path with only a single edge.
It then combines this path with all suitable partial solutions for~$y$.

\paragraph{Forget-vertex nodes.}
Suppose that~$x$ forgets a vertex~$v$ and has child $y$, that is, $\beta(y)=\beta(x)\cup\{v\}$.
For a fixed degree function~$d$ on~$\beta(x)$, let~$d'(u) = d''(u) = d(u)$ for all~$u \in \beta(x)$ and let~$d'(v)=0$ and~$d''(v) = 2$.
Then, set
\[\mathcal{A}_x(d) = \begin{cases} \mathcal{A}_y(d''), & \text{if } v \in R',\\ \mathcal{A}_y(d') \cup \mathcal{A}_y(d''), & \text{if } v \notin R'.\end{cases}\]

\paragraph{Join nodes.}
Suppose that $x$ has children $y$ and $z$.
By definition, we have~$\beta(x)=\beta(y)=\beta(z)$.
Then, for every degree function~$d$ on~$\beta(x)$, we set
\[\mathcal{A}_x(d) = \bigcup_{\substack{
    d_1,d_2\colon \beta(x) \rightarrow \{0,1,2\}\\
    d_1(v)+d_2(v)=d(v) \text{ for all } v\in \beta(x)}}\Comb_{\beta(x)}(d_1,\mathcal{A}_y(d_1),d_2,\mathcal{A}_z(d_2)).
\]

The recurrences are correct by induction over~$T$.
The claims for leaf and introduce-vertex nodes follow directly from the definitions.
At an introduce-edge node, every partial solution either omits~$e$ or is the union of~$\{e\}$ and a partial solution at the child.
In the latter case, correctness follows from~\cref{lem:comb-path-forests}.
When~$v$ is forgotten, its degree must be two if it is a terminal and may be zero or two otherwise, which precisely matches the two stated cases.
Finally, at a join node, the edge sets introduced in the two child subtrees are disjoint, and the corresponding subgraphs share vertices only in $\beta(x)$.
The two degree functions sum to~$d$, and~\cref{lem:comb-path-forests} shows that the combine operation gives exactly the matchings of the resulting partial solutions.

We now compress the tables using \cref{lem:rank-reduction-matchings}.
At every node~$x$ and for every degree function~$d$, we store a family~$\widehat{\mathcal{A}}_x(d)$ that represents~$\mathcal{A}_x(d)$.
We compute the candidate family using the appropriate recurrence above, with the stored representative families in place of the full child families, and then apply~\cref{lem:rank-reduction-matchings}.
By~\cref{obs:transandunion,lem:comb-preserves-representation}, unions and combine operations preserve representation.
An induction over~$T$ therefore shows that every stored family~$\widehat{\mathcal{A}}_x(d)$ represents~$\mathcal{A}_x(d)$.

We next explain how the root table determines whether the instance is a yes-instance.
Recall that~$r$ is the root of the nice tree decomposition and~$\beta(r) = \{t,t_1\}$.
Let~$d^*(t)=d^*(t_1)=1$.
We accept if and only if~$\widehat{\mathcal{A}}_r(d^*)$ is non-empty.
Every corresponding partial solution has exactly one nontrivial component, namely a simple~$t$-$t_1$-path, and every terminal lies on this path.
Adding~$e^*$ yields a cycle through all vertices of~$R'$ in~$G'$ and hence also corresponds to a cycle through all vertices of~$R$ in~$G$.
Conversely, every solution in~$G$ yields such a path described by~$(d^*,\{e^*\})$.
Since~$\{e^*\}$ is the only perfect matching on~$\partial d^*$ and is compatible with itself, representation guarantees that~$\widehat{\mathcal{A}}_r(d^*)$ is non-empty precisely if~$\mathcal{A}_r(d^*)$ is non-empty.

It only remains to analyse the running time.
We first consider the case where the tree decomposition is given.
Let~$k'=k+3$ be an upper bound on the maximum number of vertices in any bag.
By~\cref{lem:rank-reduction-matchings}, each stored family has size at most~$2^{\frac{k'}{2}} = \sqrt{2}^{k'}$.
Note that~$\comb$ can be computed in~$\poly(k)$ time.
Since each call of~$\Comb$ is only run on reduced sets, it makes at most~$(\sqrt{2}^{k'})^2 = 2^{k'}$ calls to~$\comb$ and can therefore be computed in~$2^{k'} \poly(k)$ time.
There are at most~$3^{k'}$ degree functions for each bag and thus, the number of stored entries for each bag is at most~$(3\sqrt{2})^{k'}$.
Moreover, each set computed during the algorithm between reductions has size at most~$3^{k'} \cdot 2^{k'} = 6^{k'}$.

At a join node, there are~$6^{k'}$ possible combinations of~$d_1$ and~$d_2$ to consider: for each vertex~$v$, it can be that~$d(v) = 0$ and hence also~$d_1(v) = d_2(v) = 0$, it can be that~$d(v) = 1$ with either~$d_1(v) = 1$ and~$d_2(v) = 0$, or~$d_1(v)=0$ and~$d_2(v) = 1$, or it can be that~$d(v) = 2$.
In the last case, there are three possible cases: $d_1(v) \in \{0,1,2\}$ and~$d_2(v) = 2 - d_1(v)$.
Thus, all entries for a single node in the nice tree decomposition can be computed in~$6^{k'} \cdot 2^{k'} \cdot \poly(k') = 12^{k'}\poly(k')$ time.
Reducing the entries for a fixed node in the tree decomposition and a fixed degree function~$d$ takes time~$\mathbf{O}(6^{k'}\cdot 2^{k'})=\mathbf{O}(12^{k'})$ by~\cref{lem:rank-reduction-matchings}.
Doing this for each of the at most~$3^{k'}$ degree functions reduces all entries for a given node in the tree decomposition in~$\mathbf{O}(36^{k'})$ time.
Hence, each node can be processed in~$12^{k'}\poly(k')+\mathbf{O}(36^{k'})=\mathbf{O}(36^{k'})$ time.
Since there are~$\mathbf{O}(k'|G^*|)$ nodes, the overall running time is in~$\mathbf{O}(36^{k'}k'|G^*|) = \mathbf{O}(36^kk|G|)$.

Finally, if no tree decomposition is given as input, we can run the~$\mathbf{O}(2^{10.8k}|G|)$-time $2$-approximation algorithm for treewidth of Korhonen~\cite{Korhonen2021} with successive width guesses.
This results in a tree decomposition of width at most~$2k+1$ for~$G$ in~$\mathbf{O}(2^{10.8k}|G|) \subseteq \mathbf{O}(1783^{k}|G|)$ time.
We can then run the above algorithm with the new value~$k'' = 2k+4$.
Note that both algorithms can be run sequentially and the second algorithm runs in~$\mathbf{O}(36^{k''}k''|G|) \subseteq \mathbf{O}(1296^{k}k|G|)$ time.
Thus, the running time of the second part is absorbed by the first part and the entire algorithm runs in time~$\mathbf{O}(1783^{k}|G|)$.
This concludes the proof.
\end{proof}

\section{The complexity dichotomy}
\label{sec:Algorithm}

We now combine \zcref{thm:MainIrrelevantVertex} and \zcref{thm:t-cycle-treewidth} to obtain an \textsf{XP}-algorithm for \Tc{} parameterized by bidimensionality.
The idea is to repeatedly delete an irrelevant set until the remaining graph has treewidth at most~$f(b) \cdot \log |G|$, where~$b$ is the bidimensionality of the red set and~$f$ is a computable function that only depends on~$b$.
We then use the algorithm from \zcref{thm:t-cycle-treewidth} to get a polynomial-time algorithm whenever the bidimensionality is bounded.
We first formalise this argument and then use it afterwards to prove \zcref{thm:MainTheorem}.

\begin{theorem}
    \label{thm:mainalg}
    There exists an algorithm that takes as input an annotated graph~$(G,R)$ and solves \Tc{} in~$|G|^{2^{\mathbf{poly}(b)}}$ time, where $b \coloneqq \mathsf{bidim}(G,R)$.
\end{theorem}

\begin{proof}
    We maintain an annotated graph~$(H,R)$ and an integer $q$.
    Intuitively, $H$ is the result of removing irrelevant vertices/sets from~$G$ and~$q$ is a guess for the bidimensionality of~$(H,R)$.
    Initially, $H=G$ and~$q=1$.
    If~$H$ has at most three vertices, then we solve $(H,R)$ directly.
    Otherwise, we apply \zcref{thm:MainIrrelevantVertex} to~$(H,R)$ with parameter~$q$ and proceed according to its outcome.
    \begin{enumerate}
        \item If the algorithm determines that~$\mathsf{bidim}(H,R) > q$, then we increase $q$ by one and repeat.
        \item If the algorithm returns an induced subgraph~$H_0$ such that~$V(H) \setminus V(H_0)$ is non-empty and irrelevant, then we replace $H$ by $H_0$ and repeat.
        \item If the algorithm returns a tree-decomposition for~$H$ of width at most~$2^{\mathbf{poly}(q)}\log |H|$, then we use \zcref{thm:t-cycle-treewidth} to solve the instance~$(H,R)$.
    \end{enumerate}

    We next analyse the running time of the algorithm.
    To this end, first note that throughout the algorithm, it holds that~$R \subseteq V(H)$ and $(H,R)$ is equivalent to $(G,R)$ by \zcref{thm:MainIrrelevantVertex}.
    Moreover, $H$ is always an induced subgraph of~$G$, and hence it holds that~$\mathsf{bidim}(H,R) \leq \mathsf{bidim}(G,R) = b \leq \sqrt{|G|}$.
    Thus, throughout the algorithm we have~$q\leq\max\{1,b\}$, and the first outcome of the above case distinction can happen at most~$b$ times.
    The second outcome deletes at least one vertex and can therefore happen at most $|G|$ times.
    The last case can happen only once.
    Thus, the running time of the described algorithm is~$|G|+b+1 \leq 2|G|+1$ times the running time in \zcref{thm:MainIrrelevantVertex} (which is~$2^{2^{\mathbf{poly}(b)}} \cdot \mathbf{poly}(|G|)$) plus once the running time in \zcref{thm:t-cycle-treewidth} (which is~$\mathbf{O}(36^k \cdot k \cdot |G|)$, where~$k \in 2^{\mathbf{poly}(b)} \cdot \log |G|$).
    Thus, the running time for \zcref{thm:t-cycle-treewidth} is in~$|G|^{2^{\mathbf{poly}(b)}}$ and the overall running time for the described algorithm is in~$|G|^{2^{\mathbf{poly}(b)}}$.

    It remains to show that the algorithm is correct.
    To this end, note that the algorithm always terminates as shown above.
    Before it terminates, it always produces an equivalent instance by \zcref{thm:MainIrrelevantVertex}.
    When it terminates, it either solves a trivial instance with at most three vertices or it gives the correct answer by \zcref{thm:t-cycle-treewidth}.
    In both cases, the returned answer is correct for~$(H,R)$ and hence also for~$(G,R)$.
    This concludes the proof.
\end{proof}

With the above result at hand, we can now show our main result: \Tc{} is polynomial-time solvable on a red-minor-closed graph class~$\mathcal{C}$ if and only if there exists some constant~$c$ such that each annotated graph~$(G,R) \in \mathcal{C}$ satisfies~$\mathsf{bidim}(G,R) \leq c$ (assuming~\textsf{P}~$\neq$~\textsf{NP}).
Equivalently, \Tc{} is polynomial-time solvable when restricted to~$\mathcal{C}$ if and only if~$\mathcal{C}$ does not contain all annotated planar graphs.
We will explicitly show and use this equivalence in the following proof.

\maintheorem*

\begin{proof}
    Let $\mathcal{C}$ be a red-minor-closed class of annotated graphs.
    We first show the hardness part.
    So assume that~$\mathcal{C}$ contains all annotated planar graphs.
    We reduce from \textsc{Hamiltonian Cycle} on planar graphs, which is known to be \textsf{NP}-hard~\cite{GareyJohnsonTarjan1976PlanarHC}.
    Given an instance~$G$, we simply construct the equivalent instance~$(G,V(G))$ of \Tc.
    Since~$G$ is planar, it holds that~$(G,V(G)) \in \mathcal{C}$.
    This shows that \Tc{} is \textsf{NP}-hard restricted to~$\mathcal{C}$.

    In the other direction, assume that~$\mathcal{C}$ does not contain all annotated planar graphs.
    We will show that this is equivalent to the fact that there exists some constant~$c$ such that all annotated graphs in~$\mathcal{C}$ have bidimensionality at most~$c$.
    First, $\Gamma_{c+1}$, in which all vertices are red, is planar and has bidimensionality~$c+1$.
    In the other direction, if~$\mathcal{C}$ does not contain all annotated planar graphs, then there is a fixed annotated planar graph~$(G^*,R) \notin \mathcal{C}$.
    Then, $G^*$ is a minor of~$\Gamma_{c^*}$ for some large enough integer~$c^*$ and thus, $(G^*,R)$ is a red-minor of~$(\Gamma_{c^*},V(\Gamma_{c^*}))$.
    If any graph in~$\mathcal{C}$ has bidimensionality at least~$c^*$, then it contains an annotated graph~$(H,R')$, which contains~$(\Gamma_{c^*},V(\Gamma_{c^*}))$ as a red-minor by definition.
    Note that~$(H,R')$ also contains~$(G^*,R)$ as a red-minor, a contradiction to~$\mathcal{C}$ being closed under taking red-minors.
    Thus, every annotated graph in~$\mathcal{C}$ has bidimensionality at most~$c^*-1$.

    Finally, since the bidimensionality of all annotated graphs in~$\mathcal{C}$ is bounded by some constant~$c$, we can use \zcref{thm:mainalg} to solve any instance of \Tc{} on~$\mathcal{C}$ in $|G|^{2^{\mathbf{poly}(c)}} \subseteq \mathbf{poly}(|G|)$ time.
    This shows that \Tc{} restricted to~$\mathcal{C}$ is polynomial-time solvable and this concludes the proof.
\end{proof}

\section{Conclusion}\label{sec:Conclusion}
We established a complexity dichotomy for \Tc{} on red-minor-closed classes of annotated graphs.
The algorithmic side follows from showing that \Tc{} parameterized by bidimensionality has the weak irrelevant vertex property and that \Tc{} parameterized by treewidth can be solved in single-exponential parameterized time.
Combined with the fact that it is known that \Tc{} parameterized by bidimensionality does not have the (regular) irrelevant vertex property, this shows that there is a real distinction between the irrelevant vertex property and polynomial-time solvability for \Tc.

A natural next step is to extend this approach to \textsc{Spanning Disjoint Paths}.
Thilikos and Wiederrecht~\cite{ThilikosW2026Spanning} obtain a~$2^{2^{\mathbf{poly}(k+d)}}|G|^2$-time algorithm where $d = \mathsf{depth}_2(G,R)$.
Moreover, they show that a red-minor-closed class has the irrelevant vertex property for this problem if and only if this parameter~$d$ is bounded.
They conjectured that \textsc{Spanning Disjoint Paths} is in \textsf{XP} when parameterized by bidimensionality and the number of prescribed terminal pairs.
Our result establishes the corresponding tractability for terminal cycles, but the extension to spanning linkages requires additional ideas.
In particular, our mesh rerouting uses the freedom to change the order in which the cycle visits the red vertices.
For spanning linkages, the same operation need not preserve the prescribed connections.

Our running time also leaves open the parameterized complexity of \Tc{} with respect to bidimensionality.
Can the dependence on~$b$ in the exponent of~$|G|^{2^{\mathbf{poly}(b)}}$ be improved?
In particular, can \Tc{} parameterized by bidimensionality
be solved in~$f(b)\cdot\mathbf{poly}(|G|)$ time for some
computable function~$f$, or can one establish $\mathsf{W}[1]$-hardness?

Next, the weak irrelevant vertex property may be useful for other problems as well.
Since many problems are known to be solvable in single-exponential parameterized time when parameterized by treewidth, we believe the weak irrelevant vertex property to be a valuable addition to the algorithmic toolkit.
As a concrete open problem we pose the following challenge:
For which annotated graphs~$\mathcal{H} = (H,B)$ is the following problem \textsc{Annotated $\mathcal{H}$-Deletion} polynomial-time solvable whenever the bidimensionality of the input annotated graph~$(G,R)$ is constant?
The input to the problem is~$(G,R)$ and an integer~$k$ and the question is whether there is a set~$K$ of at most~$k$ vertices in~$G$ such that~$(G-K,R \setminus K)$ does not contain~$(H,B)$ as a red-minor.
The version of the problem without the set~$B$ is well studied and even the annotated version is known to be fixed-parameter tractable when parameterized by the solution size~$k$~\cite{ProtopapasThilikosWiederrecht2025ColorfulMinors}.
However, the argument carries unspecified and potentially very large parametric dependencies through the use of a logic-based meta-algorithm \cite{SauStamoulisThilikos2025CMSO}.
Moreover, the problem need not have the irrelevant vertex property when parameterized by bidimensionality (but \textsc{Annotated $\mathcal{H}$-Deletion} parameterized by the solution size~$k$ has it).
We conjecture that there is a large class of annotated connected graphs~$\mathcal{H}$ for which \textsc{Annotated $\mathcal{H}$-Deletion} parameterized by bidimensionality is in \textsf{XP} and we believe that the weak irrelevant vertex property might yield constructive algorithms.
For disconnected graphs, the problem is \textsf{NP}-hard for bidimensionality one even for~$H$ consisting of just three vertices and one edge (where~$B$ contains the vertex not incident to the edge).
We leave it as an explicit open problem to identify further problems that have the weak irrelevant vertex property but not the regular irrelevant vertex property.

\begin{figure}[tbp]
    \centering
    \begin{tikzpicture}
        \def\x{2.4}
        \def\y{-1.2}
                
        \foreach \i/\j/\w in {1/2/gride,1/3/guide,1/4/guile,1/6/quilt,1/7/quirt,2/1/bride,2/2/shade,2/3/shake,2/4/spake,2/5/spike,2/6/spice,2/7/quire,3/1/brede,3/3/share,3/4/spare,3/5/spire,3/7/quite,4/1/arede,4/2/serre,4/3/seare,4/4/shire,4/5/prods,4/6/brods,4/7/suite,5/1/serry,5/2/areae,5/3/woods,5/4/poods,5/5/broos,5/6/brook,5/7/groom,6/1/sorry,6/2/words,6/3/aread,6/4/tread,6/5/brood,6/6/broom,7/2/wordy,7/4/bread,7/5/broad,7/6/blood,7/7/bloom}{
            \node[rectangle,rounded corners,draw,thick,minimum height=.6cm,minimum width=1cm] at (\x*\j,\y*\i) (\w) {\w};
        }
        
        \foreach \i/\j/\w in {1/1/pride,1/5/guilt,3/2/shame,3/6/spite,6/7/gloom,7/1/worry,7/3/dread}{
            \node[rectangle,rounded corners,draw,thick,fill=red,minimum height=.6cm,minimum width=1cm] at (\x*\j,\y*\i) (\w) {\w};
        }
        
        \foreach \a/\b in {pride/gride,pride/bride,gride/guide,guide/guile,guile/guilt,guilt/quilt,quilt/quirt,quirt/quire,bride/brede,shade/shake,shade/shame,shake/spake,shake/share,spake/spike,spake/spare,spike/spice,spike/spire,spice/spite,quire/quite,brede/arede,shame/share,share/spare,share/seare,spare/spire,spire/spite,quite/suite,serre/seare,prods/brods,serry/sorry,woods/poods,broos/brook,broos/brood,brook/broom,groom/gloom,sorry/worry,words/wordy,aread/tread,aread/dread,tread/bread,brood/broom,brood/broad,gloom/bloom,worry/wordy,dread/bread,bread/broad,blood/bloom}{
            \draw[very thick] (\a) to (\b);
        }

        \foreach \a/\b in {shade/share,spike/spite,share/shire,spite/suite,arede/areae,areae/aread,aread/bread,brood/blood,broom/bloom}{
            \draw[very thick] ([xshift=-1.5pt, yshift=1.5pt]\a.south east) to ([xshift=1.5pt, yshift=-1.5pt]\b.north west);
        }

        \draw[white,line width=4pt,rounded corners] (3.45*\x,6.55*\y) -- (3.55*\x,6.45*\y);
        
        \foreach \a/\b in {gride/bride,spice/spire,spire/shire,brook/brood,tread/dread}{
            \draw[very thick] ([xshift=1.5pt, yshift=1.5pt]\a.south west) to ([xshift=-1.5pt, yshift=-1.5pt]\b.north east);
        }

        \foreach \a/\b in {spare/seare}{
            \draw[white,line width=4pt,rounded corners] (\a) -- (\b);
            \draw[very thick] ([xshift=1.5pt, yshift=1.5pt]\a.south west) to ([xshift=-1.5pt, yshift=-1.5pt]\b.north east);
        }

        \foreach \a/\b in {shake/shame,serre/serry,prods/poods,brods/broos,woods/words,groom/broom}{
            \draw[blue!15,line width=4pt,rounded corners] (\a) -- (\b);
            \draw[very thick] ([xshift=1.5pt, yshift=1.5pt]\a.south west) to ([xshift=-1.5pt, yshift=-1.5pt]\b.north east);
        }
        
        \foreach \a/\b in {spike/spite,broos/broom}{
            \draw[blue!15,line width=4pt,rounded corners] (\a) -- (\b);
            \draw[very thick] ([xshift=-1.5pt, yshift=1.5pt]\a.south east) to ([xshift=1.5pt, yshift=-1.5pt]\b.north west);
        }
        
        \foreach \i/\j/\w in {1/2/gride,1/3/guide,1/4/guile,1/6/quilt,1/7/quirt,2/1/bride,2/2/shade,2/3/shake,2/4/spake,2/5/spike,2/6/spice,2/7/quire,3/1/brede,3/3/share,3/4/spare,3/5/spire,3/7/quite,4/1/arede,4/2/serre,4/3/seare,4/4/shire,4/5/prods,4/6/brods,4/7/suite,5/1/serry,5/2/areae,5/3/woods,5/4/poods,5/5/broos,5/6/brook,5/7/groom,6/1/sorry,6/2/words,6/3/aread,6/4/tread,6/5/brood,6/6/broom,7/2/wordy,7/4/bread,7/5/broad,7/6/blood,7/7/bloom}{
            \node[rectangle,rounded corners,draw,thick,minimum height=.6cm,minimum width=1cm] at (\x*\j,\y*\i) {\w};
        }
        
        \foreach \i/\j/\w in {1/1/pride,1/5/guilt,3/2/shame,3/6/spite,6/7/gloom,7/1/worry,7/3/dread}{
            \node[rectangle,rounded corners,draw,thick,fill=red,minimum height=.6cm,minimum width=1cm] at (\x*\j,\y*\i) {\w};
        }
       
        \begin{pgfonlayer}{background}
            \begin{scope}[transform canvas={xscale=1.8}, xscale=0.55555]
                \draw[path,line width=20pt,color=blue!15] (2*\x,3*\y) -- (3*\x,2*\y) -- (5*\x,2*\y) -- (6*\x,3*\y) -- (7*\x,4*\y) -- (7*\x,1*\y) -- (1*\x,1*\y) -- (1*\x,4*\y) -- (3*\x,6*\y) -- (3*\x,7*\y) -- (5*\x,7*\y) -- (5*\x,6*\y) -- (6*\x,7*\y) -- (7*\x,7*\y) -- (7*\x,5*\y) -- (6*\x,6*\y) -- (5*\x,5*\y) -- (6*\x,4*\y) -- (5*\x,4*\y) -- (4*\x,5*\y) -- (3*\x,5*\y) -- (2*\x,6*\y) -- (2*\x,7*\y) -- (1*\x,7*\y) -- (1*\x,5*\y) -- (2*\x,4*\y) -- (3*\x,4*\y) -- (3*\x,3*\y) -- cycle;
            \end{scope}
        \end{pgfonlayer}
    \end{tikzpicture}
    \caption{A solution $R$-cycle for the instance of \textsc{Doublets} from the introduction.}
    \label{fig:sevensinssol}
\end{figure}
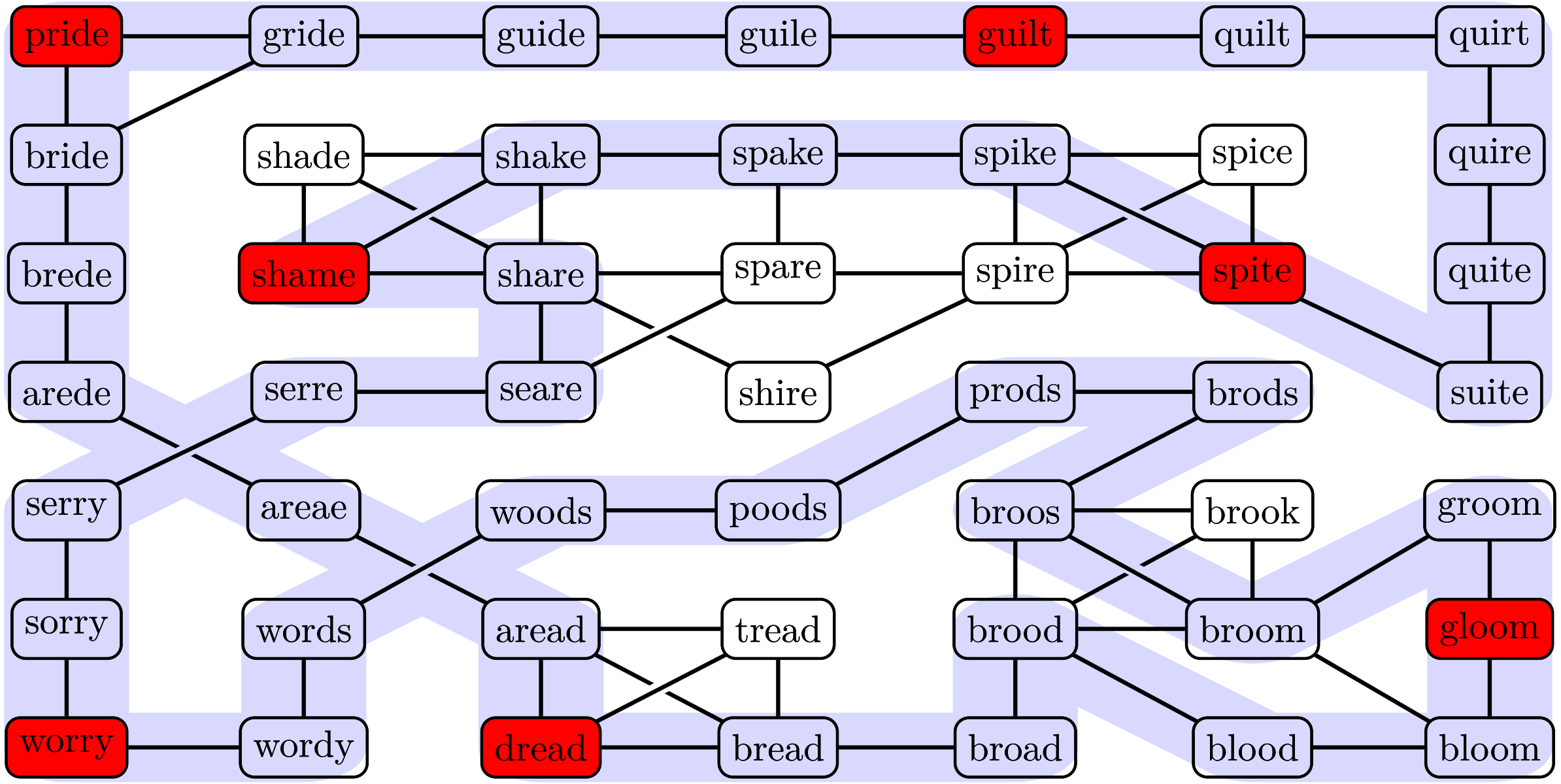

Bidimensionality is a way to measure the complexity of graph problems with terminals not only by the number of terminals, but also by the \emph{structure} of the terminals.
As bidimensionality generalises the face cover number of planar graphs and \textsc{Disjoint Paths} in planar graphs is in \textsf{XP} when parameterized by the face cover number~\cite{Schrijver91}, we wonder whether there is a directed analogue of bidimensionality that allows for similar dichotomy results for \textsc{Directed Terminal Cyclability} as we showed for \Tc.
Finally, we resolve the challenge from the introduction.
A cycle through all seven sins in the instance shown in~\cref{fig:sevensins} is shown in \cref{fig:sevensinssol}.

\paragraph{AI disclosure.}
The main results and ideas are the authors' own and have been found and developed without the use of large language models.
The authors take full responsibility for both the content and the presentation.
ChatGPT Pro Astra 6.0 was used to proofread the paper before submission, help fix some of the issues it found, and to assist the writing process.
Specifically, it was used to draft a first version of the dynamic program in \zcref{sec:BoundedTreewidth} and provided a shorter explanation for \zcref{thm:dry-village-preparation} with subsequent changes in \cref{sec:branch}.
It also drafted first versions of the introduction and the conclusion.
All these parts were subsequently substantially rewritten by the authors.

\phantomsection
\addcontentsline{toc}{section}{References}
\bibliographystyle{alphaurl}
\bibliography{literature}

@article{CyganNPPvRWJ2022SolvingConnectivity,
  author       = {Marek Cygan and
                  Jesper Nederlof and
                  Marcin Pilipczuk and
                  Micha{\l} Pilipczuk and
                  Johan M. M. van Rooij and
                  Jakub Onufry Wojtaszczyk},
  title        = {Solving Connectivity Problems Parameterized by Treewidth in Single Exponential Time},
  journal      = {{ACM} Transactions on Algorithms},
  volume       = {18},
  number       = {2},
  pages        = {17:1--17:31},
  year         = {2022},
  doi          = {10.1145/3506707}
}

@article{ThilikosW2026Spanning,
  author       = {Dimitrios M. Thilikos and
                  Sebastian Wiederrecht},
  title        = {Spanning Paths and Cycles: {S}tructural Limitations of the Irrelevant Vertex Technique},
  journal      = {CoRR},
  volume       = {abs/2607.09342},
  year         = {2026},
  doi          = {10.48550/ARXIV.2607.09342},
  eprinttype   = {arXiv}
}

@inproceedings{GorskySTW2026Catching,
  author       = {Maximilian Gorsky and
                  Giannos Stamoulis and
                  Dimitrios M. Thilikos and
                  Sebastian Wiederrecht},
  title        = {Catching Rats in {$H$}-minor-free Graphs},
  booktitle    = {Proceedings of the 2026 Annual {ACM-SIAM} Symposium on Discrete Algorithms (SODA)},
  pages        = {4644--4659},
  publisher    = {{SIAM}},
  year         = {2026},
  doi          = {10.1137/1.9781611978971.169}
}

@article{RobertsonS2003GraphMinorsXVI,
  author  = {Robertson, Neil and Seymour, Paul D.},
  title   = {Graph minors. {XVI}. {E}xcluding a non-planar graph},
  journal = {Journal of Combinatorial Theory, Series B},
  volume  = {89},
  number  = {1},
  pages   = {43--76},
  year    = {2003},
  doi     = {10.1016/S0095-8956(03)00042-X}
}

@article{RobertsonS1991GraphMinorsX,
  author  = {Robertson, Neil and Seymour, Paul D.},
  title   = {Graph minors. {X}. {O}bstructions to tree-decomposition},
  journal = {Journal of Combinatorial Theory, Series B},
  volume  = {52},
  number  = {2},
  pages   = {153--190},
  year    = {1991},
  doi     = {10.1016/0095-8956(91)90061-N}
}

@article{RobertsonS1999GraphMinorsXVII,
  author  = {Robertson, Neil and Seymour, Paul D.},
  title   = {Graph minors. {XVII}. {T}aming a vortex},
  journal = {Journal of Combinatorial Theory, Series B},
  volume  = {77},
  number  = {1},
  pages   = {162--210},
  year    = {1999},
  doi     = {10.1006/jctb.1999.1919}
}

@article{Dirac1960Abstrakten,
  author  = {Dirac, Gabriel A.},
  title   = {In abstrakten {G}raphen vorhandene vollst\"{a}ndige 4-{G}raphen und ihre {U}nterteilungen},
  journal = {Mathematische Nachrichten},
  volume  = {22},
  pages   = {61--85},
  year    = {1960},
  doi     = {10.1002/mana.19600220107}
}

@article{Gould2009Cycles,
  author  = {Gould, Ronald J.},
  title   = {A look at cycles containing specified elements of a graph},
  journal = {Discrete Mathematics},
  volume  = {309},
  number  = {21},
  pages   = {6299--6311},
  year    = {2009},
  doi     = {10.1016/j.disc.2008.04.017}
}

@article{SauST2024More,
  author  = {Sau, Ignasi and Stamoulis, Giannos and Thilikos, Dimitrios M.},
  title   = {A more accurate view of the flat wall theorem},
  journal = {Journal of Graph Theory},
  volume  = {107},
  number  = {2},
  pages   = {263--297},
  year    = {2024},
  doi     = {10.1002/jgt.23121}
}

@inproceedings{GorskySW2026Price,
  author       = {Maximilian Gorsky and
                  Michal T. Seweryn and
                  Sebastian Wiederrecht},
  title        = {The Price of Homogeneity Is Polynomial},
  booktitle    = {Proceedings of the 53rd International Colloquium on Automata, Languages, and Programming ({ICALP})},
  pages        = {100:1--100:13},
  publisher    = {Schloss Dagstuhl - Leibniz-Zentrum f{\"{u}}r Informatik},
  year         = {2026},
  doi          = {10.4230/LIPICS.ICALP.2026.100}
}

@article{GolovachST2023Combing,
  author  = {Golovach, Petr A. and Stamoulis, Giannos and Thilikos, Dimitrios M.},
  title   = {Combing a linkage in an annulus},
  journal = {SIAM Journal on Discrete Mathematics},
  volume  = {37},
  number  = {4},
  pages   = {2332--2364},
  year    = {2023},
  doi     = {10.1137/22M150914X}
}

@article{RobertsonS1986GraphMinorsV,
  author  = {Robertson, Neil and Seymour, Paul D.},
  title   = {Graph minors. {V}. {E}xcluding a planar graph},
  journal = {Journal of Combinatorial Theory, Series B},
  volume  = {41},
  number  = {1},
  pages   = {92--114},
  year    = {1986},
  doi     = {10.1016/0095-8956(86)90030-4}
}

@article{ChuzhoyT2021TowardsTight,
  author  = {Chuzhoy, Julia and Tan, Zihan},
  title   = {Towards tight(er) bounds for the excluded grid theorem},
  journal = {Journal of Combinatorial Theory, Series B},
  volume  = {146},
  pages   = {219--265},
  year    = {2021},
  doi     = {10.1016/j.jctb.2020.09.010}
}

@incollection{Reed1997Tree,
  author    = {Reed, Bruce A.},
  title     = {Tree Width and Tangles: {A} New Connectivity Measure and Some Applications},
  booktitle = {Surveys in Combinatorics},
  pages     = {87--162},
  publisher = {Cambridge University Press},
  year      = {1997},
  doi       = {10.1017/CBO9780511662119.006}
}

@incollection{Kawarabayashi2008CyclesThroughElements,
  author    = {Kawarabayashi, Ken-ichi},
  title     = {An Improved Algorithm for Finding Cycles Through Elements},
  booktitle = {Proceedings of the 13th International Conference on Integer Programming and Combinatorial Optimization ({IPCO})},
  pages     = {374--384},
  publisher = {Springer},
  year      = {2008},
  doi       = {10.1007/978-3-540-68891-4_26}
}

@article{RobertsonSeymour1995DisjointPaths,
  author  = {Robertson, Neil and Seymour, Paul D.},
  title   = {Graph Minors. {XIII}. {T}he Disjoint Paths Problem},
  journal = {Journal of Combinatorial Theory, Series B},
  volume  = {63},
  number  = {1},
  pages   = {65--110},
  year    = {1995},
  doi     = {10.1006/jctb.1995.1006}
}

@article{RobertsonSeymour2009UniqueLinkages,
  author  = {Robertson, Neil and Seymour, Paul D.},
  title   = {Graph Minors. {XXI}. {G}raphs with Unique Linkages},
  journal = {Journal of Combinatorial Theory, Series B},
  volume  = {99},
  number  = {3},
  pages   = {583--616},
  year    = {2009},
  doi     = {10.1016/j.jctb.2008.08.003}
}

@article{RobertsonSeymour2012Irrelevant,
  author  = {Robertson, Neil and Seymour, Paul D.},
  title   = {Graph Minors. {XXII}. {I}rrelevant Vertices in Linkage Problems},
  journal = {Journal of Combinatorial Theory, Series B},
  volume  = {102},
  number  = {2},
  pages   = {530--563},
  year    = {2012},
  doi     = {10.1016/j.jctb.2007.12.007}
}

@article{KawarabayashiKobayashiReed2012Quadratic,
  author  = {Kawarabayashi, Ken-ichi and Kobayashi, Yusuke and Reed, Bruce A.},
  title   = {The Disjoint Paths Problem in Quadratic Time},
  journal = {Journal of Combinatorial Theory, Series B},
  volume  = {102},
  number  = {2},
  pages   = {424--435},
  year    = {2012},
  doi     = {10.1016/j.jctb.2011.07.004}
}

@article{AdlerEtAl2017PlanarIrrelevant,
  author  = {Adler, Isolde and Kolliopoulos, Stavros G. and Krause, Philipp Klaus and Lokshtanov, Daniel and Saurabh, Saket and Thilikos, Dimitrios M.},
  title   = {Irrelevant Vertices for the Planar Disjoint Paths Problem},
  journal = {Journal of Combinatorial Theory, Series B},
  volume  = {122},
  pages   = {815--843},
  year    = {2017},
  doi     = {10.1016/j.jctb.2016.10.001}
}

@incollection{LokshtanovSaurabhZehavi2020Efficient,
  author    = {Lokshtanov, Daniel and Saurabh, Saket and Zehavi, Meirav},
  title     = {Efficient Graph Minors Theory and Parameterized Algorithms for (Planar) Disjoint Paths},
  booktitle = {Treewidth, Kernels, and Algorithms - Essays Dedicated to Hans L. Bodlaender on the Occasion of His 60th Birthday},
  pages     = {112--128},
  publisher = {Springer},
  year      = {2020},
  doi       = {10.1007/978-3-030-42071-0_9}
}

@inproceedings{KorhonenPilipczukStamoulis2024AlmostLinear,
  author    = {Korhonen, Tuukka and Pilipczuk, Micha{\l} and Stamoulis, Giannos},
  title     = {Minor Containment and Disjoint Paths in Almost-Linear Time},
  booktitle = {Proceedings of the 65th {IEEE} Annual Symposium on Foundations of Computer Science ({FOCS})},
  pages     = {53--61},
  publisher = {{IEEE} Computer Society},
  year      = {2024},
  doi       = {10.1109/FOCS61266.2024.00014}
}

@inproceedings{SauStamoulisThilikos2025CMSO,
  author    = {Sau, Ignasi and Stamoulis, Giannos and Thilikos, Dimitrios M.},
  title     = {Parameterizing the Quantification of {CMSO}: Model Checking on Minor-Closed Graph Classes},
  booktitle = {Proceedings of the 2025 Annual ACM-SIAM Symposium on Discrete Algorithms (SODA)},
  pages     = {3728--3742},
  publisher = {SIAM},
  year      = {2025},
  doi       = {10.1137/1.9781611978322.124}
}

@incollection{Karp1972Reducibility,
  author    = {Karp, Richard M.},
  title     = {Reducibility Among Combinatorial Problems},
  booktitle = {Complexity of Computer Computations},
  pages     = {85--103},
  publisher = {Plenum Press},
  year      = {1972},
  doi       = {10.1007/978-1-4684-2001-2_9}
}

@article{GareyJohnsonTarjan1976PlanarHC,
  author  = {Garey, Michael R. and Johnson, David S. and Tarjan, Robert E.},
  title   = {The Planar {H}amiltonian Circuit Problem is {NP}-Complete},
  journal = {SIAM Journal on Computing},
  volume  = {5},
  number  = {4},
  pages   = {704--714},
  year    = {1976},
  doi     = {10.1137/0205049}
}

@inproceedings{JansenSwennenhuis2024SteinerTree,
  author    = {Jansen, Bart M. P. and Swennenhuis, C{\'e}line M. F.},
  title     = {Steiner Tree Parameterized by Multiway Cut and Even Less},
  booktitle = {Proceedings of the 32nd Annual European Symposium on Algorithms ({ESA})},
  pages     = {76:1--76:16},
  publisher = {Schloss Dagstuhl -- Leibniz-Zentrum f{\"u}r Informatik},
  year      = {2024},
  doi       = {10.4230/LIPIcs.ESA.2024.76}
}

@article{CrespelleFeghaliGolovach2022Cyclability,
  author  = {Crespelle, Christophe and Golovach, Petr A.},
  title   = {Cyclability in Graph Classes},
  journal = {Discrete Applied Mathematics},
  volume  = {313},
  pages   = {147--178},
  year    = {2022},
  doi     = {10.1016/j.dam.2022.01.021}
}

@inproceedings{BjorklundHusfeldtTaslaman2012Shortest,
  author    = {Bj{\"o}rklund, Andreas and Husfeldt, Thore and Taslaman, Nina},
  title     = {Shortest Cycle Through Specified Elements},
  booktitle = {Proceedings of the 2012 Annual ACM-SIAM Symposium on Discrete Algorithms ({SODA})},
  pages     = {1747--1753},
  publisher = {SIAM},
  year      = {2012},
  doi       = {10.1137/1.9781611973099.139}
}

@inproceedings{Wahlstrom2013Tutte,
  author    = {Wahlstr{\"o}m, Magnus},
  title     = {Abusing the {T}utte Matrix: {A}n Algebraic Instance Compression for the {$K$}-Set-Cycle Problem},
  booktitle = {Proceedings of the 30th International Symposium on Theoretical Aspects of Computer Science ({STACS})},
  pages     = {341--352},
  publisher = {Schloss Dagstuhl -- Leibniz-Zentrum f{\"u}r Informatik},
  year      = {2013},
  doi       = {10.4230/LIPIcs.STACS.2013.341}
}

@article{MarxSeymourWollan2017RootedGrid,
  author  = {Marx, D{\'a}niel and Seymour, Paul D. and Wollan, Paul},
  title   = {Rooted Grid Minors},
  journal = {Journal of Combinatorial Theory, Series B},
  volume  = {122},
  pages   = {428--437},
  year    = {2017},
  doi     = {10.1016/j.jctb.2016.07.003}
}

@article{HodorLaMicekRambaud2026ApexForest,
  author  = {Hodor, J{\k{e}}drzej and La, Hoang and Micek, Piotr and Rambaud, Cl{\'e}ment},
  title   = {Quickly Excluding an Apex-Forest},
  journal = {SIAM Journal on Discrete Mathematics},
  volume  = {40},
  number  = {1},
  pages   = {282--307},
  year    = {2026},
  doi     = {10.1137/24M1659248}
}

@article{ThilikosWiederrecht2025Bidimensionality,
  author        = {Thilikos, Dimitrios M. and Wiederrecht, Sebastian},
  title         = {The Graph Minor Structure Theorem Through Bidimensionality},
  journal       = {CoRR},
  volume        = {abs/2306.01724},
  year          = {2025},
  eprint        = {2306.01724},
  archivePrefix = {arXiv},
  primaryClass  = {math.CO},
  doi           = {10.48550/arXiv.2306.01724}
}

@inproceedings{ProtopapasThilikosWiederrecht2025ColorfulMinors,
  author       = {Evangelos Protopapas and
                  Dimitrios M. Thilikos and
                  Sebastian Wiederrecht},
  title        = {Colorful Minors},
  booktitle    = {Proceedings of the 53rd International Colloquium on Automata, Languages, and Programming ({ICALP})},
  pages        = {149:1--149:17},
  publisher    = {Schloss Dagstuhl - Leibniz-Zentrum f{\"{u}}r Informatik},
  year         = {2026},
  doi          = {10.4230/LIPICS.ICALP.2026.149}
}

@article{CavallaroGKTW2026Optimal,
  author        = {Cavallaro, Dario and Gorsky, Maximilian and Kreutzer, Stephan and Thilikos, Dimitrios M. and Wiederrecht, Sebastian},
  title         = {Optimal Bounds for the {$k$}-Disjoint Paths Problem},
  journal       = {CoRR},
  volume        = {abs/2605.14902},
  year          = {2026},
  eprint        = {2605.14902},
  archivePrefix = {arXiv},
  primaryClass  = {math.CO},
  doi           = {10.48550/arXiv.2605.14902}
}

@article{KawarabayashiTW2021Quickly,
  author        = {Kawarabayashi, Ken-ichi and Thomas, Robin and Wollan, Paul},
  title         = {Quickly Excluding a Non-Planar Graph},
  journal       = {CoRR},
  volume        = {abs/2010.12397},
  year          = {2020},
  eprint        = {2010.12397},
  archivePrefix = {arXiv},
  primaryClass  = {math.CO},
  doi           = {10.48550/arXiv.2010.12397}
}

@article{ThilikosW2024Excluding,
  author        = {Thilikos, Dimitrios M. and Wiederrecht, Sebastian},
  title         = {Excluding Surfaces as Minors in Graphs},
  journal       = {CoRR},
  volume        = {abs/2601.19230},
  year          = {2026},
  eprint        = {2601.19230},
  archivePrefix = {arXiv},
  primaryClass  = {math.CO},
  doi           = {10.48550/arXiv.2601.19230}
}

@inproceedings{PaulPTW2024Obstructionsa,
  author    = {Paul, Christophe and Protopapas, Evangelos and Thilikos, Dimitrios M. and Wiederrecht, Sebastian},
  title     = {Obstructions to {E}rd{\H{o}}s--{P}{\'o}sa Dualities for Minors},
  booktitle = {Proceedings of the 65th {IEEE} Annual Symposium on Foundations of Computer Science ({FOCS})},
  pages     = {31--52},
  publisher = {{IEEE} Computer Society},
  year      = {2024},
  doi       = {10.1109/FOCS61266.2024.00013}
}

@article{PaulPTW2025Local,
  author        = {Paul, Christophe and Protopapas, Evangelos and Thilikos, Dimitrios M. and Wiederrecht, Sebastian},
  title         = {The Local Structure Theorem for Graph Minors with Finite Index},
  journal       = {CoRR},
  volume        = {abs/2507.02769},
  year          = {2025},
  eprint        = {2507.02769},
  archivePrefix = {arXiv},
  primaryClass  = {math.CO},
  doi           = {10.48550/arXiv.2507.02769}
}

@inproceedings{GorskySW2025Polynomial,
  author    = {Gorsky, Maximilian and Seweryn, Micha{\l} T. and Wiederrecht, Sebastian},
  title     = {Polynomial Bounds for the Graph Minor Structure Theorem},
  booktitle = {Proceedings of the 66th {IEEE} Annual Symposium on Foundations of Computer Science ({FOCS})},
  pages     = {1961--1978},
  publisher = {{IEEE} Computer Society},
  year      = {2025},
  doi       = {10.1109/FOCS63196.2025.00104}
}

@inproceedings{GorskyPW2026Quickly,
  author       = {Maximilian Gorsky and
                  Evangelos Protopapas and
                  Sebastian Wiederrecht},
  title        = {Quickly Excluding an Annotated Planar Graph},
  booktitle    = {Proceedings of the 53rd International Colloquium on Automata, Languages, and Programming ({ICALP})},
  pages        = {99:1--99:17},
  publisher    = {Schloss Dagstuhl - Leibniz-Zentrum f{\"{u}}r Informatik},
  year         = {2026},
  doi          = {10.4230/LIPICS.ICALP.2026.99}
}

@article{BodlaenderCyganKratschNederlof2015,
  author  = {Hans L. Bodlaender and Marek Cygan and
             Stefan Kratsch and Jesper Nederlof},
  title   = {Deterministic Single Exponential Time Algorithms for
             Connectivity Problems Parameterized by Treewidth},
  journal = {Information and Computation},
  volume  = {243},
  pages   = {86--111},
  year    = {2015},
  doi     = {10.1016/j.ic.2014.12.008}
}

@article{Korhonen2021,
  title={A single-exponential time 2-approximation algorithm for treewidth},
  author={Korhonen, Tuukka},
  journal={SIAM Journal on Computing},
  pages={FOCS21-174--FOCS21-194},
  year={2023},
  publisher={SIAM},
  doi={10.1137/22M147551X}
}

@article{DFT12,
  author       = {Frederic Dorn and
                  Fedor V. Fomin and
                  Dimitrios M. Thilikos},
  title        = {Catalan structures and dynamic programming in {$H$}-minor-free graphs},
  journal      = {Journal of Computer and System Sciences},
  volume       = {78},
  number       = {5},
  pages        = {1606--1622},
  year         = {2012},
  doi          = {10.1016/J.JCSS.2012.02.004}
}

@inproceedings{GRZ25,
  author       = {Harmender Gahlawat and
                  Abhishek Rathod and
                  Meirav Zehavi},
  title        = {({A}lmost-)Optimal {FPT} Algorithm and Kernel for {$T$}-Cycle on Planar Graphs},
  booktitle    = {Proceedings of the 52nd International Colloquium on Automata, Languages, and Programming {(ICALP)}},
  volume       = {334},
  pages        = {82:1--82:18},
  publisher    = {Schloss Dagstuhl - Leibniz-Zentrum f{\"{u}}r Informatik},
  year         = {2025},
  doi          = {10.4230/LIPICS.ICALP.2025.82},
}

@article{BNRZ23,
  author       = {Matthias Bentert and
                  Andr{\'{e}} Nichterlein and
                  Malte Renken and
                  Philipp Zschoche},
  title        = {Using a Geometric Lens to Find $k$-Disjoint Shortest Paths},
  journal      = {{SIAM} Journal on Discrete Mathematics},
  volume       = {37},
  number       = {3},
  pages        = {1674--1703},
  year         = {2023},
  doi          = {10.1137/22M1527398}
}

@inproceedings{Loc21,
	Author = {Lochet, William},
	Booktitle = {Proceedings of the 2021 ACM-SIAM Symposium on Discrete Algorithms (SODA)},
	Organization = {SIAM},
	Pages = {169--178},
	Title = {A Polynomial Time Algorithm for the $k$-Disjoint Shortest Paths Problem},
    doi = {10.1137/1.9781611976465.12},
	Year = {2021}
}

@article{Eilam-Tzoreff98,
  author       = {Tali Eilam{-}Tzoreff},
  title        = {The Disjoint Shortest Paths Problem},
  journal      = {Discrete Applied Mathematics},
  volume       = {85},
  number       = {2},
  pages        = {113--138},
  year         = {1998},
  doi          = {10.1016/S0166-218X(97)00121-2}
}

@article{BFG26,
  author       = {Matthias Bentert and
                  Fedor V. Fomin and
                  Petr A. Golovach},
  title        = {Tight Approximation and Kernelization Bounds for Vertex-Disjoint Shortest Paths},
  journal      = {Theory of Computing Systems},
  volume       = {70},
  number       = {1},
  pages        = {5},
  year         = {2026},
  doi          = {10.1007/S00224-025-10252-9}
}

@inproceedings{BCHKKS26,
  author       = {Matthias Bentert and
                  Dario Cavallaro and
                  Amelie Heindl and
                  Ken{-}ichi Kawarabayashi and
                  Stephan Kreutzer and
                  Johannes Schr{\"{o}}der},
  title        = {The Directed Disjoint Paths Problem with Congestion},
  booktitle    = {Proceedings of the 2026 Annual ACM-SIAM Symposium on Discrete Algorithms (SODA)},
  pages        = {2291--2315},
  publisher    = {{SIAM}},
  year         = {2026},
  doi          = {10.1137/1.9781611978971.82}
}

@article{FHW80,
  author       = {Steven Fortune and
                  John E. Hopcroft and
                  James Wyllie},
  title        = {The Directed Subgraph Homeomorphism Problem},
  journal      = {Theoretical Computer Science},
  volume       = {10},
  pages        = {111--121},
  year         = {1980},
  doi          = {10.1016/0304-3975(80)90009-2}
}

@article{DumasHMMRS2026Invitation,
  author       = {Dumas, Ma{\"e}l and Hodor, J{\k{e}}drzej and Mazur, Tomasz and Micek, Piotr and Rambaud, Cl{\'e}ment and Stamoulis, Giannos},
  title        = {Invitation to torso parameters of annotated graphs},
  journal      = {CoRR},
  volume       = {abs/2610.25905},
  year         = {2026},
  doi          = {10.48550/arXiv.2610.25905},
  eprinttype   = {arXiv}
}

@inproceedings{CavallaroKK2024Eulerian,
  author       = {Dario Giuliano Cavallaro and
                  Ken{-}ichi Kawarabayashi and
                  Stephan Kreutzer},
  title        = {Edge-Disjoint Paths in {E}ulerian Digraphs},
  booktitle    = {Proceedings of the 56th Annual {ACM} Symposium on Theory of Computing ({STOC})},
  pages        = {704--715},
  publisher    = {{ACM}},
  year         = {2024},
  doi          = {10.1145/3618260.3649758},
}

@inproceedings{CyganMPP2013PlanarDigraph,
  author       = {Marek Cygan and
                  D{\'{a}}niel Marx and
                  Marcin Pilipczuk and
                  Michal Pilipczuk},
  title        = {The Planar Directed $k$-Vertex-Disjoint Paths Problem Is Fixed-Parameter Tractable},
  booktitle    = {Proceedings of the 54th Annual {IEEE} Symposium on Foundations of Computer Science ({FOCS})},
  pages        = {197--206},
  publisher    = {{IEEE} Computer Society},
  year         = {2013},
  doi          = {10.1109/FOCS.2013.29},
}

@article{Schrijver91,
  author       = {Alexander Schrijver},
  title        = {Disjoint Homotopic Paths and Trees in a Planar Graph},
  journal      = {Discrete \& Computational Geometry},
  volume       = {6},
  pages        = {527--574},
  year         = {1991},
  doi          = {10.1007/BF02574704},
}

@article{BruhnUlmer2022Packing,
  author  = {Henning Bruhn and Arthur Ulmer},
  title   = {Packing {$A$}-paths of length zero modulo four},
  journal = {European Journal of Combinatorics},
  volume  = {99},
  pages   = {103422},
  year    = {2022},
  doi     = {10.1016/j.ejc.2021.103422}
}

@booklet{Car78,
  author  = {Lewis Carroll},
  title   = {{Word-Links}: {A} Game for Two Players, or a Round Game},
  address = {Oxford},
  year    = {1878},
  url     = {https://schnark.github.io/lewis-carroll/html/texts/word-links-printed.html}
}

\end{document}